\documentclass[11pt,a4paper]{article}

\usepackage[margin=2cm]{geometry}
\usepackage{amsfonts,amsmath}
\usepackage{setspace}
\usepackage{cite}
\usepackage{rotating}
\usepackage{multirow}
\usepackage{booktabs,caption,fixltx2e}
\usepackage[flushleft]{threeparttable}
\usepackage[colorlinks=true,
            linkcolor=red,
            urlcolor=blue,
            citecolor=gray]{hyperref}
\usepackage{epstopdf}
\usepackage{upgreek}
\usepackage{verbatim}
\usepackage{multirow}
\usepackage{enumitem}
\usepackage{gensymb}
\usepackage[pdftex,dvipsnames]{xcolor}  

\newcommand{\x}{{\mathbf x}}

\newcommand{\y}{{\mathbf y}}

\newcommand{\I}{{\mathbf I}}

\newcommand{\IG}{\includegraphics}
\newcommand{\Real}{\mathbb R}

\newcommand{\f}{{\mathbf f}}

\begin{document}

\title{Gaussian Processes Retrieval of LAI from Sentinel-2 Top-of-Atmosphere Radiance Data}

\author{Jose Estevez$^{1,}$, Jorge Vicent$^{2,}$, Juan Pablo Rivera-Caicedo$^{3,}$,\\ 
Pablo Morcillo-Pallar\'es$^{1,}$, Francesco Vuolo$^{4,}$, Neus Sabater$^{1,5}$, \\
Gustau Camps-Valls$^{1,}$, Jos\'e Moreno$^{1}$ and Jochem Verrelst$^{1,}$*\\
$^{1}$ Image Processing Laboratory (IPL), Universitat de Val\`encia, Val\`encia, Spain; \\
$^{2}$ Magellium, Toulouse, France;\\
$^{3}$ CONACyT-UAN,  Universidad Aut\'onoma de Nayarit, Nayarit, Mexico;\\
$^{4}$ University of Natural Resources and Life Sciences, Vienna (BOKU), Vienna, Austria\\
$^{5}$ Finnish Meteorological Institute, Helsinki, Finland.
}

\date{\bf Preprint of paper published in ISPRS Journal of Photogrammetry and Remote Sensing
Volume 167, September 2020, Pages 289-304. https://doi.org/10.1016/j.isprsjprs.2020.07.004}

\maketitle

\begin{abstract}
{Retrieval of vegetation properties from satellite and airborne optical data usually takes place after atmospheric correction, yet it is also possible to develop retrieval algorithms directly from top-of-atmosphere (TOA) radiance data.  One of the key vegetation variables that can be retrieved from at-sensor TOA radiance data is leaf area index (LAI) if algorithms account for variability in atmosphere. We demonstrate the feasibility of LAI retrieval from Sentinel-2 (S2) TOA radiance data (L1C product) in a hybrid machine learning framework.
To achieve this, the coupled leaf-canopy-atmosphere radiative transfer models PROSAIL-6S were used to simulate a look-up table (LUT) of TOA radiance data and associated input variables. This LUT was then used to train the Bayesian machine learning algorithms Gaussian processes regression (GPR) and variational heteroscedastic GPR (VHGPR).  PROSAIL simulations were also used to train GPR and VHGPR models for LAI retrieval from S2 images at bottom-of-atmosphere (BOA) level (L2A product) for comparison purposes. The BOA and TOA LAI products were consistently validated against a field dataset with GPR ($R^2$ of 0.78) and with VHGPR ($R^2$ of 0.80) and for both cases a slightly lower RMSE for the TOA LAI product. Because of delivering superior accuracies and lower uncertainties, VHGPR was further applied for LAI mapping using S2 acquisitions over the agricultural sites Marchfeld (Austria) and Barrax (Spain). The VHGPR models led to consistent LAI maps at BOA and TOA scale. The LAI maps were also compared against LAI maps as generated by the SNAP toolbox, which is based on a neural network (NN). Maps were again consistent, however  the SNAP NN algorithm tend to overestimate over dense vegetation cover. Overall, this study demonstrated that hybrid LAI retrieval algorithms can be developed from TOA radiance data given a cloud-free sky, thus without the need of atmospheric correction. To the benefit of the community, the development of such hybrid algorithms for the retrieval vegetation properties from BOA or TOA images has been streamlined in the freely downloadable ALG-ARTMO software framework.
}
\end{abstract}


\section{Introduction}\label{sec:intro}

The estimation of vegetation biophysical variables is key for a wide range of ecological and agricultural applications \cite{Verrelst2015b, Verrelst2019}. 
Particularly leaf area index (LAI) has been proven to be a successful variable retrievable from optical sensors mounted on Earth observing satellites \cite{Verrelst2015,yan2019}.  
Copernicus' flagship for terrestrial earth observation (EO), i.e. the Sentinel-2 (S2) constellation, provides free, full and open access optical data with very short revisit times (2-3 days in mid-latitudes), high spatial resolution (<30 m), and good spectral resolution (10-180 nm) \cite{Drusch2012,Malenovsky2012}. 
This vast data stream has proven to be convenient for the quantification and monitoring of vegetation characteristics, with LAI as the most successful indicator of vegetation density \cite{Verrelst2015b,Verrelst2015}.  

However, optical missions do not measure vegetation properties directly, and some essential pre-processing steps are required to transform at-sensor reflected radiation into interpretable surface reflectance measures, i.e. radiometric calibration and correction, geometric correction, and ultimately atmospheric correction. 
The retrieval of biophysical variables takes place typically after the atmospheric correction step where radiometrically and spectrally calibrated top-of-atmosphere (TOA) radiance is converted into bottom-of-atmosphere (BOA) reflectance. 
Consequently, the pre-processing from TOA to BOA data is a critical step, and determines the success of the subsequent retrieval process \cite{Laurent2011a}. Nevertheless, the TOA radiance to BOA reflectance conversion is not so straightforward. Typically, the atmospheric correction is based on the inversion of an atmospheric radiative transfer model (RTM) commonly through interpolation of pre-computed look-up tables (LUT). Together with the intrinsic errors of LUT interpolation, the ill-posedness of the inversion of atmospheric characteristics introduces important uncertainties in atmospheric correction \cite{guanter2009}. Also, the atmospheric correction generally makes the assumption that the surface is Lambertian. Other steps that introduce errors are the corrections for topographic, adjacency, and bi-directional surface reflectance effects. These corrections are applied sequentially and independently, potentially accumulating errors into the BOA reflectance data \cite{gao2009}.

In an attempt to overcome this limitation, it has earlier been suggested and successfully tested to retrieve biophysical variables directly from at-sensor TOA radiance, thus without the necessity to go through the atmospheric correction process \cite{Laurent2011a,Laurent2011b,Laurent2013,Mousivand2015,Shi2017,Verrelst2019TOC2TOA}. The possibility of retrieving LAI directly from TOA radiance data was recently theoretically confirmed by calculating a global sensitivity analysis, thereby varying all leaf-canopy-atmosphere RTM variables. At TOA radiance, LAI proved to be a dominant variable along the spectral range, and at multiple spectral regions it is not influenced by the atmospheric variables, especially in the short wave infrared (SWIR) \cite{Verrelst2019TOC2TOA}. 
An advantage of retrieving vegetation properties directly from TOA is that the combined model simulates TOA radiances, which is the physical variable measured by the sensor. This means that the simulated quantities can be directly compared with the reference measurements, unlike the BOA approach where a mismatch may exist due to the approximations and assumptions in the atmospheric correction step \cite{Laurent2011a}. The downside of these approaches, however, is that they require a sound physical understanding on the factors determining the at-sensor spectral TOA radiance, e.g. as studied in \cite{faurtyot1997,verhoef2003,verhoef2007}.
Although for operational products such as the S2 atmospherically-corrected BOA reflectance products are freely provided, for experimental missions or airborne campaign atmospheric corrections are still a mandatory pre-processing step.
It was also with such kinds of experimental data that the aforementioned TOA retrieval approaches were presented.  

When it comes to the retrieval of vegetation properties from optical EO data, such as LAI, four principal families of retrieval methods can be identified: (1) parametric regression, (2) nonparametric regression, (3) inversion of RTMs, and (4) hybrid or combined method \cite{Verrelst2015b, Verrelst2019}. Regarding the retrieval of LAI from TOA radiance data, earlier retrieval approaches are to be found into the third category of methods, i.e., inversion of leaf a coupled canopy-atmosphere RTMs by making use of predefined look-up tables (LUT). The main drawback of this method is that it takes a long computational time, i.e. for each pixel querying and interpolating the LUT for inversion through a minimization function \cite{Verrelst2015b, Verrelst2019}. In this regard, hybrid retrieval methods has become an appealing alternative due to the fast progress in machine learning methods, for the last few years hybrid retrieval methods became an appealing alternative. 
Hybrids methods establish a statistical relationship between simulated spectra and a biophysical variable. These type of methods have been particularly successful in operational processing of EO data because they exploit the generic properties of physically-based methods combined with the flexibility and computational efficiency of machine learning regression algorithms (MLRAs) \cite{Verrelst2015b, Verrelst2019}. One of the major advantages of these methods is that, once the MLRA is trained it can process an image into a vegetation product quasi-instantly. 

Hybrid retrieval implementations in an operational context has long been restricted to artificial neural networks (NNs). 
The combination of artificial NNs with the radiative transfer model (RTM) PROSAIL has long been used  in operational applications \cite{Bacour06,Baret2013} and kept on being used, e.g for the processing of S2 images.  For instance, the retrieval algorithm has been implemented into the biophysical processor tool of the Sentinel Application Platform (SNAP) \cite{weiss2016}. 
At the same time, related studies reveal advantages in the application of alternative MLRAs over conventional NNs techniques \cite{upreti2019}. 
Especially the MLRA families of decision trees and kernel-based methods \cite{CampsValls09wiley} proved to be successful \cite{Verrelst2015b, Verrelst2019}. 
These methods tend to be simpler to train and can perform more robust than NNs while maintaining competitive accuracies \cite{Verrelst12a,Verrelst2015}. From the kernel-based MLRAs family noteworthy are the algorithms kernel ridge regression \cite{Suykens1999} because of its simplicity and therefore fast run-time, and Gaussian Process Regression (GPR) \cite{rasmussen06} because of its ability to provide additional information such as ranking of relevant bands as well as associated uncertainties \cite{Verrelst2013b,Verrelst2015}.

Given the progress made by the MLRAs, new opportunities emerged to develop retrieval algorithms directly from TOA radiance data. For instance, of interest to implement are hybrid algorithms with advanced MLRAs that at the same time provides associated uncertainty estimates \cite{Verrelst2019TOC2TOA}. 
Yet, what makes the implementation of TOA approach challenging is the atmospheric part of the coupled model to account for variability in atmospheric effects. Although multiple atmospheric RTMs have been developed, these models are often difficult to configure for generating a large amount of simulations. Widely used atmospheric RTMs include 6SV \cite{vermote1997}, libRadtran \cite{mayer2005} and MODTRAN \cite{Berk2006}. 
To overcome this limitation, the Atmospheric Look-up table Generator (ALG) is one of the few software packages that enables executing atmospheric RTMs with a friendly graphical user interface \cite{vicent2019}. In addition, the few software packages available to automate the retrieval from TOA data are still experimental. To the best of our knowledge, only the automated radiative transfer models Operator (ARTMO) scientific software framework is able to provide these steps into a streamlined and quasi-automatic way \cite{Verrelst12c}.  
ARTMO not only runs leaf-canopy models, its recent TOC2TOA toolbox allows coupling a leaf-canopy LUT with an ALG-generated atmospheric LUT to upscale the data to TOA radiance level given the assumption of a Lambertian surface \cite{Verrelst2019TOC2TOA}. 
At the same time, with ARTMO's MLRA toolbox we can train retrieval algorithms and generate the biophysical maps from remote sensing data from BOA reflectance or from TOA radiance, as initially explored in \cite{Verrelst2019TOC2TOA}.

Building on experience from above studies, the main objective of this work was to develop, optimize and validate a hybrid LAI retrieval algorithm applicable directly to S2 TOA radiance data. To demonstrate its validity, the developed algorithm was compared against a hybrid retrieval algorithm applicable to S2 BOA reflectance data. 
The pursued approach was as follows. First, a coupled atmosphere-canopy RT model was used to simulate a database of TOA radiance data with associated input variables. This database was then used to train a GPR model for LAI retrieval from a S2 TOA radiance image (L1C product) over the agricultural region Barrax, Spain. To do so, an atmosphere 6SV database was generated with the ALG toolbox, which was then coupled  with PROSAIL simulations to generate TOA radiance data in ARTMO's TOC2TOA toolbox. The TOA radiance data was then used by the MLRA toolbox for developing the retrieval algorithm. Similarly, PROSAIL simulations were used to train GPR model for LAI retrieval from S2 images at BOA level (L2A product) for comparison purposes. Finally, the obtained maps were validated against LAI maps as generated by the SNAP toolbox, which is based on a NN algorithm \cite{weiss2016}.

\section{Theoretical framework top-of-canopy and top-of-atmosphere simulations for retrieval}\label{sec:TOAtheory}

\subsection{Leaf \& canopy RTMs: PROSAIL}

In hybrid biophysical variable retrieval models, the model development is based on simulated data coming from RTMs. When aiming to develop hybrid retrieval models applicable to at-sensor TOA radiance data, then the simulated data come from coupled vegetation surface and atmosphere RTMs. 
Here, we demonstrate the feasibility of this approach with the most standard leaf-canopy-atmosphere RTMs, as they are fast and freely available to the community. Vegetation top-of-canopy (TOC) reflectance simulations come from the combination of the leaf RTM PROSPECT-4 \cite{Feret08} with the canopy RTM SAIL \cite{Verhoef1984}, also known as PROSAIL \cite{Jacquemoud2009}. 
PROSPECT-4 is  one of the most widely used RTMs that simulates leaf optical properties. It calculates directional-hemispherical reflectance and transmittance measured from 400 nm to 2500 nm at 1 nm spectral sampling. SAIL solves the radiative transfer equation for scattering and absorption of four upward/downward fluxes at the canopy scale. 
The leaf reflectance ($\rho_l$) and transmittance ($\tau_l$) outputs of PROSPECT are entered into SAIL model to simulate the top-of-canopy (TOC)  reflectance ($\rho_c$) in the 400-2500 nm spectral range at 1 nm sampling. The soil spectral reflectance is another important input of SAIL. Generally, field radiometric data is used, but also spectra from images have been successfully used \cite{Verrelst2019}. 
 
By varying the RTM input variables, multiple model realization are run and both inputs and output spectra are stored in databases. The databases can subsequently be used for further processing such as mapping applications, e.g. by means of applying inversion strategies through minimization functions \cite{Rivera13a,Verrelst14}, or by means of using these databases for training a hybrid retrieval strategy \cite{Caicedo2014,Verrelst2016AL}.  However, with PROSAIL only TOC reflectance simulations are generated, which means these data can solely be used to images after atmospheric correction. An additional step is thus required when developing retrieval strategies directly from TOA radiance data, i.e. the coupling with an atmospheric RTM.

\subsection{Coupling with the atmospheric RTM: 6SV}

In order to enable extracting biophysical variables directly from TOA radiance data, it is necessary to upscale the PROSAIL-simulated TOC database to TOA radiance level. 
This is achieved by means of coupling the database with simulations from an atmospheric RTM. 
Among the multiple atmospheric RTMs available, the 6SV (Second Simulation of the Satellite Signal in the Solar Spectrum) \cite{vermote1997} is probably one of the most widely used computer code that simulated the propagation of radiation through the atmosphere. 6SV is an improved version of 5S code, developed by the Laboratoire d'Optique Atmospherique. It takes into account the main atmospheric effects like gaseous absorption by water vapor, carbon dioxide, oxygen and ozone; scattering by molecules and aerosols. The computational accuracy for Rayleigh and aerosol scattering effects is based on the use of state-of-the-art approximations and implementation of the successive order of scattering (SOS) algorithm \cite{LENOBLE2007479}. Just like PROSAIL, the simulations of 6SV are in the 400-2500 nm spectral range but at a spectral resolution of 2.5 nm.

The output of 6SV are the following atmospheric transfer functions for each combination of key input parameters:

\begin{itemize}[noitemsep,topsep=0pt]
\item $\rho_0$: Intrinsic atmospheric reflectance (unitless).
\item $T_{gas}$: Total gas transmittance (unitless).
\item $T_{dwn}$ and $T_{up}$: Total downwards and upwards transmittance due to scattering (unitless)
\item $S$: Spherical albedo (unitless).
\item $I_0$: Extraterrestrial solar irradiance in [mW$\cdot$m$^{-2}\cdot$nm$^{-1}$].
\end{itemize}

\begin{figure}[!ht]
	\centering
	\IG[trim={0cm 0cm 0cm 0cm},clip,width=17cm]{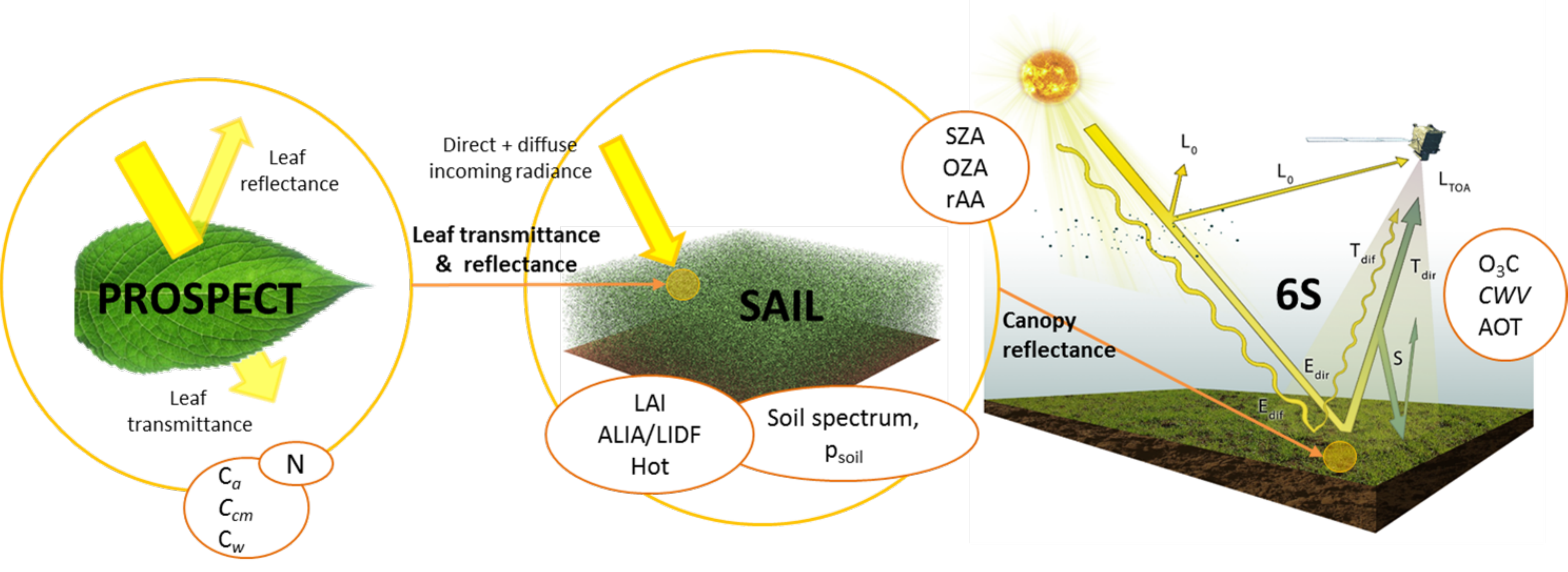}
	\caption{Schematic illustration of the coupled PROSAIL with 6S. The PROSAIL part is with permission adapted from \cite{Berger2018}. For 6S only the dominant continuous variables are given. }  
	\label{PROSAILscheme}
\end{figure}

TOA radiance spectra (L) is calculated by coupling the generated atmospheric transfer functions from 6SV with the Lambertian surface reflectance ($\rho$) from PROSAIL following the equation:
\begin{align} \label{eqn:ltoa}
    L = \frac{I_0\mu_{il}}{\pi}T_{gas}\left[\rho_0 + \frac{T_{dwn}T_{up}\rho}{1-S\rho}\right]
\end{align}
where $\mu_{il}=\cos\theta_{il}$ being $\theta_{il}$ the solar zenith angle.  For the sake of simplicity, the spectral dependency of all terms in the equation \ref{eqn:ltoa} has been omitted. A schematic overview of the coupling of PROSAIL with 6S is provided in Figure \ref{PROSAILscheme}.

\subsection{Gaussian Process Regression}
 
We included Gaussian Process Regression (GPR)~\cite{rasmussen06} in the hybrid retrieval scheme because it has proven good retrieval performance in variable retrieval \cite{Verrelst12b,Verrelst13a} and model emulation in general \cite{GRSM_2016,Vicent2018,Svendsen19amogape,CampsValls19nsr}, and when is applied to S2 and Sentinel-3 data in particular  \cite{Verrelst12a,Verrelst2013b,Verrelst2015,upreti2019}.   See also reviews of \cite{Verrelst2015b,Verrelst2019} for a rationale of using GPR as opposed to alternative MLRAs.

Notationally, the GPR model establishes a relation between the input ($B$-bands spectra) $\x\in \Real^B$ and the output variable (canopy parameter to be retrieved) $y\in \Real$ of the form:
\begin{equation}
\hat y = f(\x) = \sum_{i=1}^N\alpha_i K(\x_i,\x),
\end{equation}
where $\{\x_i\}_{i=1}^N$ are the spectra used in the training phase, $\alpha_i\in \Real$ is the weight assigned to each one of them, 
and $K$ is a function evaluating the similarity between the test spectrum $\x$ and all $N$ training spectra, $\x_i=[x_i^1,x_i^2,\ldots,x_i^B]^\top$, $i=1,\ldots,N$. We used a scaled Gaussian kernel function, 
\begin{equation}
K(\x_i,\x_j) = \nu \exp\bigg(-\sum_{b=1}^B \dfrac{{(x_i^{b}-x_j^{b})}^2}{2\sigma_b^2}\bigg) + \delta_{ij}\cdot \sigma_n^2, 
\end{equation}
where $\nu$ is a scaling factor, $B$ is the number of bands, $\sigma_b$ is a dedicated parameter controlling the spread of the relations for each particular spectral band $b$, $\sigma_n$ is the noise standard deviation and $\delta_{ij}$ is the Kronecker's symbol. The kernel is thus parametrized by signal ($\nu$, $\sigma_b$) and noise ($\sigma_n$) hyperparameters, collectively denoted as $\boldsymbol{\theta}=\{\nu,\sigma_b,\sigma_n\}$.

For training purposes, we assume that the observed variable is formed by noisy observations of the true underlying function $y = f(\x) + \epsilon$. Moreover we assume the noise to be additive independently identically Gaussian distributed with zero mean and variance $\sigma_n$. Let us define the stacked output values $\y = (y_1,\ldots,y_n)^\top$, the covariance terms of the test point ${\bf k}_\ast=[k(\x_\ast, \x_1),\ldots, k(\x_\ast,\x_n)]^\top$, and $k_{\ast\ast} = k(\x_\ast,\x_\ast)$ represents the self-similarity of $\x_\ast$. From the previous model assumption, the output values are distributed according to:
\begin{eqnarray}
\bigg(
\begin{array}{c}
{\bf y}       \\
f(x_\ast)
\end{array}
\bigg)
\sim {\mathcal N} \bigg({\bf 0},\bigg(
\begin{array}{cc}
{\bf K} + \sigma_n^2 {\bf I}      &      {\bf k}_\ast \\
{\bf k}_\ast^\top   &      k_{\ast\ast} 
\end{array}
\bigg) \bigg).
\end{eqnarray}
For prediction purposes, the GPR is obtained by computing the posterior distribution over the unknown output $\y_\ast$, $p(\y_\ast|{\bf x}_\ast, {\mathcal D})$, where ${\mathcal D}\equiv\{{\bf x}_n,y_n|n=1,\ldots,N\}$ is the training dataset. Interestingly, this posterior can be shown to be a Gaussian distribution, $p(y_\ast|{\bf x}_\ast,{\mathcal D})$ = ${\mathcal N}(y_\ast|\mu_{\text{GP}*},\sigma_{\text{GP}*}^2)$, for which one can estimate the {\em predictive mean} (point-wise predictions):
\begin{eqnarray}
\mu_{\text{GP}*} = k_\ast^\top({\bf K} + \sigma_n^2{\bf I})^{-1}{\bf y},
\end{eqnarray} 
and the {\em predictive variance} (confidence intervals):
\begin{eqnarray}
\sigma_{\text{GP}*}^2 = k_{\ast\ast} - {\bf k}_\ast^\top({\bf K} + \sigma_n^2{\bf I})^{-1}{\bf k}_\ast.
\end{eqnarray}
The corresponding hyperparameters $\boldsymbol{\theta}$ are typically selected by Type-II Maximum Likelihood, using the marginal likelihood (also called {\em evidence}) of the observations, which is also analytical. When the derivatives of the log-evidence are also analytical, which is often the case, conjugated gradient ascent is typically used for optimization (see \cite{rasmussen06} for further details).  
A more detailed survey on GPR properties in remote sensing is provided in~\cite{GRSM_2016}, and a perspective outlook in \cite{CampsValls19nsr}.

With respect to EO mapping applications, GPR is simple to train and works well with a relative small data set, as opposed to other methods like neural networks. The use of the ARD kernel function makes the GPR model quite flexible, and often outperforms other non-parametric regression methods in remote sensing applications. Furthermore, GPR provides information about the level of uncertainty (or confidence intervals) for prediction, e.g. confidence map that provides insight in the robustness of the retrieval \cite{Verrelst2013b}, and about the relevance of bands, which can be used for identifying the sensitive spectral regions~\cite{Verrelst2016GPR,GRSM_2016,CampsValls19nsr}.

\subsection{Heteroscedastic Gaussian Process Regression}

Despite the great advantages for modeling, an important challenge in the practical use of GPR in EO mapping problems comes from the fact that very often signal and noise are often correlated. As seen before, the standard GP modeling assumes that the variance of the noise process $\sigma_n$ is independent of the signal, which does not hold in most of EO applications. This strong assumption of homoscedasticity is generally broken in many biophysical retrieval problems because the acquisition process is typically affected by noise in different amounts depending on the measured range of the variable. In order to deal with input-dependent noise variance, heteroscedastic GPs let noise power vary smoothly throughout input space,  $\sigma_n(\x)$. This, however, does not lead to closed-form solutions, and several approximations have been proposed in the literature. Among them, the marginalized variational approximation yields a richer and more flexible heteroscedastic GP model~\cite{Lazaro-GredillaT11}, which has yielded very good results in biophysical retrieval from EO data  \cite{lazaro2013,GRSM_2016}.  
\subsection{Satellite measurements (Sentinel-2)}\label{Sentinel-2}

ESA's Sentinel-2 (S2) is a polar-orbiting, super-spectral and high spatial resolution mission integrated by a pair of satellites (Sentinel-2A and Sentinel-2B) that enables a global revisit time below 5 days. The S2 mission deliver data from all land surfaces and coastal areas for supporting agro-ecosystems application within the European Commission's Copernicus programme. Each S2 satellite carries a Multi-Spectral Imager (MSI) which have 13 spectral bands covering from the visible and NIR (VNIR) to SWIR spectral domains. MSI ranges from 400 to 2400 nm, with pixel sizes of 10, 20, or 60 m, depending on the spectral band \cite{Drusch2012}. Three of these bands are located in the red-edge (centered at 705, 740 and 783 nm), an important region for vegetation study. Other band configuration details of MSI are included in Table \ref{MSI_bands}.The super-spectral resolution, the inclusion of the red edge region of the spectrum, the high revisit frequency and the high radiometric quality \cite{Gascon2017} make S2 optical data very convenient to estimate the biophysical variables.

Two reflectance products from S2 MSI are available at different processing levels: Level-1C and Level-2A. L1C product provides TOA reflectances (i.e., TOA radiance normalized by incident solar irradiance). The processing chain for this product involves radiometric calibration, geometric calibration and orthorectification. L2A product provides BOA reflectance from L1C product. Sentinel-2 Atmospheric Correction (S2AC) is achieved by means of the Sen2Cor atmospheric correction scheme \cite{MainKnorn2017}. The baseline process of Sen2Cor is the cirrus/haze detection and removal \cite{sen2corCirrus,sen2corCloud}. The Aerosol Optical Thickness (AOT) can be preferably derived from Dark Dense Vegetation (DDV) targets and water bodies \cite{sen2corDDV}. Water vapour retrieval over land is performed using the Atmospheric Pre-corrected Differential Absorption (ADPA) \cite{SCHLAPFER1998353}. Sen2Cor is based on the libRadtran radiative transfer model \cite{emde2015} which simulates a wide variety of atmospheric conditions, solar geometries and ground elevations. A pre-computed LUT based on this atmospheric model is used to invert surface reflectance, using LUT interpolation to fill gaps in the simulation. Other optional pre-processing steps are performed such as correction for adjacency, topography and BRDF effects. Additionally, the Lambertian surface assumption is applied \cite{Richter2011ATDB}. All the errors of L2A product related to measurements, modelling and assumptions, in combination with error propagation, results in non-negligible uncertainties that can impact a further biophysical retrieval. 
L1C and L2A products are made available to users via the Copernicus Open Access Hub (SciHub). L2A can also be generated by the user from the L1C product using the Sentinel-2 Toolbox or the standalone version of the Sen2Cor processor. This offline processing allows the user to set certain input parameters \cite{Muller-Wilm2018} e.g. the type of aerosol (rural and maritime), the type of atmosphere (mid latitude summer and mid latitude winter) and the ozone concentration. In addition, there is the option to enable the cirrus correction and the BRDF correction, which are disabled by default. Whereas this mode allows the user to adjust certain parameters of the atmospheric correction to the local environment conditions, the product offered by ESA core through Scihub is processed with the default values defined in the Sen2cor algorithm \cite{Clerc2020}.

\begin{table}[!ht] 
\small
\begin{center}
\caption{Sentinel-2 MSI band setting (bands used in this experiment are indicated in bold).}  
\vspace{-0.2cm}
\resizebox{1.0\textwidth}{!}{ 
\begin{tabular}{l*{13}{c}}
\hline
Band \# & B1 &  \bf{B2} & \bf{B3} & \bf{B4} & \bf{B5} & \bf{B6} &  \bf{B7} & \bf{B8} & \bf{B8a} & B9 & B10 & \bf{B11} & \bf{B12} \\
Band center (nm) & 443 & \bf{490} & \bf{560} & \bf{665} & \bf{705} & \bf{740} & \bf{783} & \bf{842} & \bf{865} & 945 & 1375 & \bf{1610} & \bf{2190} \\
Band width (nm) & 20 & \bf{65} & \bf{35} & \bf{30} & \bf{15} & \bf{15} & \bf{20} & \bf{115} & \bf{20} & 20 & 30 &  \bf{9} & \bf{180} \\
Spatial resolution (m) & 60 & \bf{10} & \bf{10} & \bf{10} & \bf{20} & \bf{20} &  \bf{20} & \bf{10} & \bf{20} & 60 & 60 &  \bf{20} & \bf{20} \\
\hline
\label{MSI_bands}
\end{tabular}}
\end{center}
\end{table}

\section{Materials and Methods}\label{sec:MaterialsMethods}

\begin{figure}[!ht]
	\centering
	\IG[trim={0cm 7.5cm 0cm 0cm},clip,width=15cm]{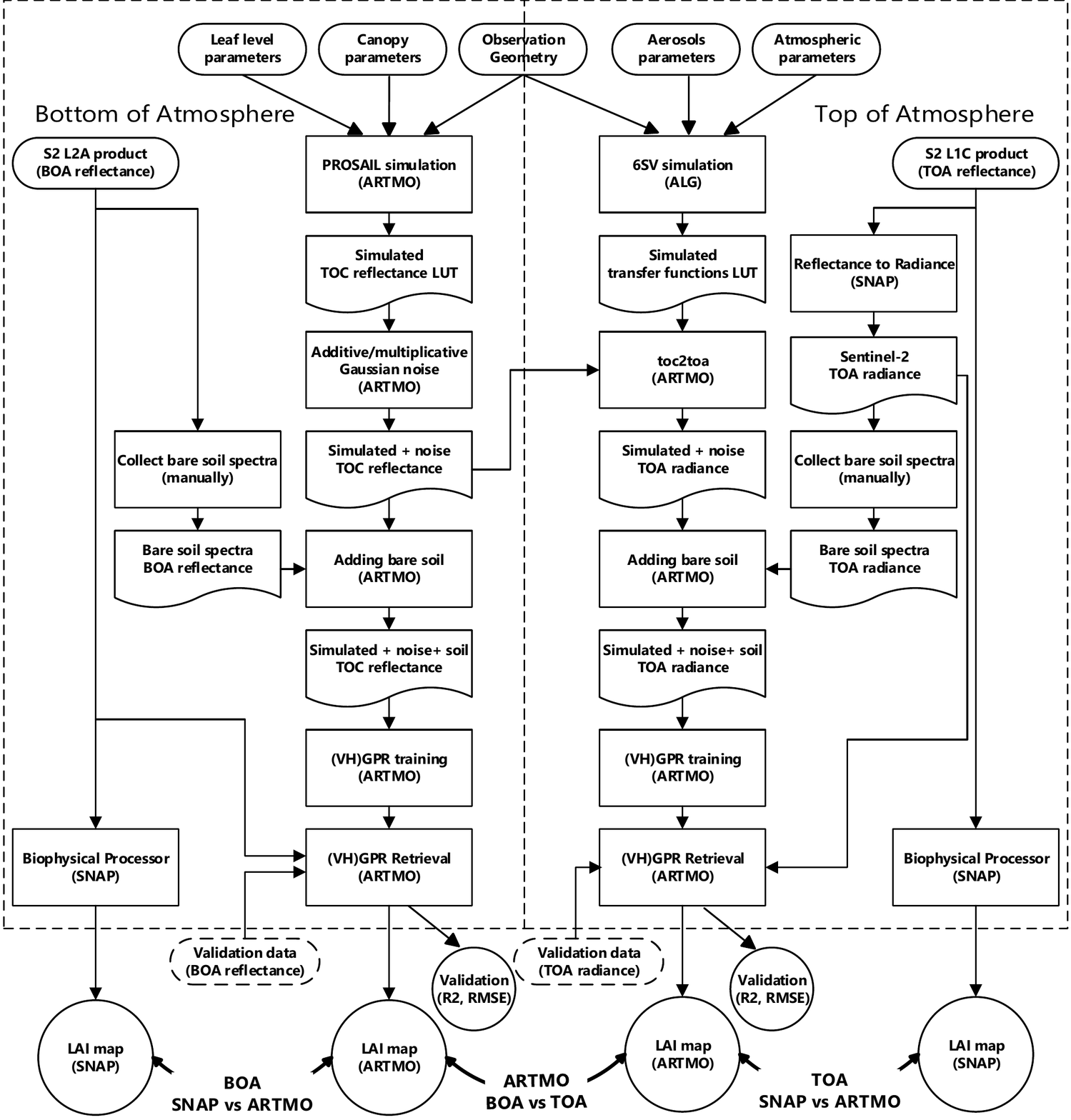}
	\caption{Flowchart of the pursued work-flow. Divided into two levels: bottom-of-atmosphere (left) and top-of-canopy (right). A rounded rectangle represents an input data/parameters, a normal rectangle represents a task/process, a rectangle with a curved bottom represents an intermediate output and a circle represents a final output/result. } 
	\label{flowchart}
\end{figure}

Regarding the retrieval of LAI from S2 BOA and TOA data, we followed the recently proposed methodology by Verrelst et al., (2019) \cite{Verrelst2019TOC2TOA}. 
In short, the hybrid retrieval strategy relies on GPR and VHGPR models trained by simulations from PROSAIL at the canopy scale and from PROSAIL-6SV models at the atmosphere scale. Trained models are then applied S2 to L2A (BOA) and L1C (TOA) data for LAI mapping and validation with ground measurements. 
The following steps were necessary to conduct the methodology: (1) generation of simulations with the models PROSAIL and 6SV; (2) coupling PROSAIL and 6SV to upscale at TOA level; (3) training GPR and VHGPR with model simulations and cross-validation; (4) validation with S2 data and ground measurements; (5) mapping variables; and (6) comparison with LAI product coming from the SNAP Biophysical Processor. A schematic overview of the  method is provided in Figure \ref{flowchart}, and key steps are detailed in the following sections, starting with a description of the used Sentinel-2 data.  

\subsection{PROSAIL simulations}\label{prosail_simulations}

The PROSAIL simulations were produced by coupling PROSPECT-4 with SAIL within the ARTMO framework. The PROSPECT-4 and 4SAIL input variables with their sampling range and distribution are shown in Table \ref{prosail_conf}. This parametrization is based on the measurements campaigns and/or other studies which used the same crops \cite{Rivera13a,Verrelst14,Verrelst2015}. Hot spot, sky viewing factor and solar/viewing angles were fixed. LAI and LCC were 100 time sampled with Gaussian distribution and the rest of variables were 10 times sampled with uniform distribution.  
LAI and LCC required to put more emphasis on the values at the actual growth stages of the crops. The selected values of illumination and viewing conditions agree with the satellite overpass conditions.  
The combination of all input variables values would produce an unrealistic number of simulations of 1 billion. For that reason, a smaller subset was randomly chosen with 1000 reflectance realizations. Since PROSAIL simulates reflectance from 400 nm to 2500 nm with a 1 nm spectral resolution the output spectra were resampled to the band settings of S2 (Table \ref{MSI_bands}) using the spectral response function provided by the Ground Segment as from 15 January 2018~\cite{s2isrf}.

\begin{table}[!ht]
\small
\begin{center}
    \caption{Ranges, values and  distributions of input parameters used to establish the synthetic canopy reflectance database for use in the LUT. ${\bar x}$: mean, SD: standard deviation.}
\begin{tabular}{llccc@{}}
\hline
       \multicolumn{2}{c}{\bf Model Parameters} & {\bf Units} & {\bf Range} & {\bf Distribution}\\
\hline			
\multicolumn{5}{l}{$Leaf$ $parameters$: PROSPECT-4} \\
$N$ &  Leaf structure index & unitless & 1.3-2.5 & Uniform \\
LCC & Leaf chlorophyll content & [${\mu}$g/cm$^{2}$]  & 5-75 & Gaussian (${\bar x}$: 35, SD: 30) \\
$C_m$ &  Leaf dry matter content & [g/cm$^{2}$] & 0.001-0.03 & Uniform\\
$C_w$ &  Leaf water content & [cm] & 0.002-0.05 & Uniform\\
\\
\multicolumn{5}{l}{$Canopy$ $variables$: 4SAIL}\\
LAI & Leaf area index & [m$^{2}$/m$^{2}$] & 0.1-7 & Gaussian (${\bar x}$: 3, SD: 2)\\
$\alpha_{soil}$ & Soil scaling factor & unitless & 0-1 & Uniform \\
ALA & Average leaf angle & [$\degree$] & 40-70 & Uniform\\
HotS & Hot spot parameter & [m/m] & 0.01 & -\\
skyl & Diffuse incoming solar radiation & [fraction] & 0.05 & - \\     
$\theta_s$ & Sun zenith angle & [$\degree$] & 30 & - \\
$\theta_v$ & View zenith angle & [$\degree$] & 0 & - \\
$\phi$ & Sun-sensor azimuth angle & [$\degree$] & 0 & - \\ 
\hline
	   \end{tabular}
	   \label{prosail_conf}
\end{center}
\end{table}

\subsection{Noise model and added soil spectra}\label{noise_model}

In order to improve the performance of the retrieval, ideally the training database should be as similar as possible to real Sentinel-2 data. It implies to consider some uncertainties associate with sensor measurement accuracy and data processing including radiometric calibration, atmospheric and geometric corrections. These different source of uncertainties can introduce additive and multiplicative errors which can be band dependent (applied to a single band) and band independent (applied to all bands) \cite{Verger2011}. All these variabilities and uncertainties were introduced in the simulated LUT based on white Gaussian noise, according to the noise model provided in Equation \ref{eqn:gaussian_noise} \cite{weiss2016}:

\begin{align} \label{eqn:gaussian_noise}
    R^*(\lambda) = R(\lambda).\left(1+\frac{MD(\lambda)+MI}{100}\right)+AD(\lambda)+AI
\end{align}

where R($\lambda$) and R{*}($\lambda$) represent respectively the raw simulated reflectance for band $\lambda$ and the reflectance with uncertainties for band $\lambda$. MD and MI are the multiplicative wavelength dependent noise and the multiplicative wavelength independent noise, respectively. AD and AI are the additive wavelength dependent noise and the additive wavelength independent noise. After some testing of additive and multiplicative noise, a value of 0.01 for AD and AI, and a value of 4\% for MD and MI were used for all the bands. Similar noise levels were successfully used in recent works \cite{upreti2019,Verrelst2019TOC2TOA}, trying to reduce the over-fitting on the MLRA training database.

Further, because PROSAIL is a vegetation canopy model and not prepared to simulate variability in soil cover, reflectance spectra from bare soil pixels were added to the PROSAIL simulations \cite{Verrelst2019TOC2TOA}. A dataset of 30 distinct  soil samples were visually identified from the S2  L2A and L1C products (BOA and TOA), trying to collect the more representative soil spectral signatures. Because these spectra came from the image itself, no additional noise was added to it.

\subsection{Atmospheric simulations and coupling}\label{atmospheric_simulations}

The role of the atmosphere was simulated using the 6S code (6SV2.1) \cite{Kotchenova2006,Kotchenova2007} within the ALG framework. The model input variables (Table \ref{6S_conf}) were set considering the experimental data conditions, similar to \cite{Verrelst2019TOC2TOA}. The distribution of the variables follows the Latin Hypercube Sampling (LHS) method. The atmospheric profile mode used was Mid-Latitude Summer and the aerosol model selected was Continental. The geometric conditions of the canopy model PROSAIL were preserved. Finally, this atmospheric simulation generates an output of 1000 samples which is stored in a database. This database consists on pairs of transfer functions and atmospheric parameters. The spectral range was limited to the S2 MSI spectral configuration (Table \ref{MSI_bands}), matching with PROSAIL simulations.  

\begin{table}[!ht] 
\begin{center}
    \caption{Range of 6S input variables used for the simulations of the atmospheric transfer functions.}
     \begin{tabular}{llccc@{}}
     \\
\hline
       \multicolumn{2}{c}{\bf Model variables} & {\bf Units} & {\bf Minimum} & {\bf Maximum}\\
\hline			
$O3C$ &  O$_3$ Column concentration & [amt-cm] & 0.25 & 0.35 \\
$CWV$ & Columnar Water Vapor & [$g.cm^{-2}$]  & 0.4 & 4.5 \\
$AOT$ &  Aerosol Optical Thickness & unitless & 0.05 & 0.5\\   
$\theta_s$ & Sun zenith angle & [$\degree$] & 30 & - \\
$\theta_v$ & View zenith angle & [$\degree$] & 0 & - \\
$\phi$ & Sun-sensor azimuth angle & [$\degree$] & 0 & - \\ 
\hline
	   \end{tabular}
	   \label{6S_conf}
\end{center}
\end{table}
 
The original 1 nm sampling of PROSAIL surface simulations were resampled by spline interpolation to the 2.5 nm sampling 6SV atmospheric simulations. With this common spectral sampling, the surface and atmospheric simulations were randomly combined and propagated to TOA radiance, following the equation \ref{eqn:ltoa} with the Lambertian surface assumption. This step was conducted using ARTMO's TOC2TOA toolbox, which generate the TOA database consisting of pairs of radiance spectra and associated vegetation-atmosphere parameters. Finally, ARTMO's TOC2TOA toolbox applied a convolution of this high-spectral resolution TOA radiance by the S2 spectral response function.

\subsection{Training the MLRA, retrieval and cross-validation}\label{training}

The TOC database simulated with model PROSAIL and the TOA LUT simulated with the combined PROSAIL-6SV models were used to train GPR into LAI retrieval models applicable to S2 at the corresponding BOA (L2A) and TOA (L1C) level. In order to assess the theoretical GPR retrieval performance, a 5-fold cross-validation was performed.
Three goodness-of-fit metrics were calculated, being: (1) the root-mean-squared error (RMSE), (2) the normalized RMSE (NRMSE in \%) and (3) the coefficient of determination ($R^2$).

\subsection{Validation with ground measurements and Sentinel-2 data}\label{validation}

\subsubsection{Marchfeld site}\label{Marchfeld}

Validation data was collected from an area located east of Vienna in Lower Austria (Lat. 48$\degree$N, Long. 17$\degree$E). The site is a major agricultural production area in Austria with cropland occupying 60,000ha, of which about 21,000ha are irrigated regularly with groundwater throughout the growing season \cite{Neugebauer2014}. The region is characterized by a semi-arid climate representing the driest region of Austria. 

The field campaign took place from April to August 2016. Eight different crop types, distributed over 72 parcels, were monitored to represent the prevailing crop types in the  study area \cite{Vuolo2018}. The parcels include 33 ordinary fields and 39 one-hectare experimental plots. 
In-situ LAI measurements where collected with a Li-Cor LAI-2200 Plant Canopy Analyzer \cite{li1992lai}. The LAI-2200's sensor operates a non-destructive method and is sensitive to all light blocking objects in its view. It estimates the LAI from the values of canopy transmittance by identifying the attenuation of the radiation as it passes through the canopy \cite{li1992lai}. Therefore, measurements were taken above- (A) and below-canopy (B). LAI estimates represent the effective Plant Area Index (PAIe), because the optical sensor does not distinguish between photosynthetically active leaves and inactive parts of the plants such as senescent leaves or stems. Care was taken to measure LAI only on photosynthetically active vegetation. 
For example, measurements were interrupted on winter cereals as soon as the first signs of senescence started to appear. The LAI-2200 was deployed in an elementary sampling units (ESUs) using a radius of 5-10 m of a georeferenced point (accuracy of $\pm$3-5m). Each unit represented a homogeneous area with a single crop type. 
The ESUs were randomly chosen from the study area, with the only restriction being the fields' accessibility for time restraints. For ordinary fields the centres of the ESUs were placed in a corner of a squared area of 60 m by 60 m within the field and measured from the field border. It was imperative that the field conditions were relatively homogeneous in terms of crop development. 
The ESUs located in the experimental plots, part of a larger experimental setup, were located in the centre of each one-hectare plot. Winter cereal, onion, and potato were assessed through three replications of one A and eight B measurements, randomly distributed in the ESU. Row crops like maize, carrot and sugar beet were estimated with four replications of one A and six B measurements, for a total of 24 single measurements to generate a single LAI value per ESU. The final dataset of in-situ collected LAI consists of 114 measurements and complementary L1C radiance and L2A reflectance values extracted from the satellite image data. The SNAP "Reflectance to Radiance" tool was used to convert L1C reflectance to radiance. Table \ref{Dates} lists the satellite images and the corresponding dates of field measurements which were used for the analysis.

\begin{table}[!h] 
\begin{center}
    \caption{Satellite acquisition and ground measurements dates from April to September 2016 for Marchfeld campaign.}
     \begin{tabular}{llccc@{}}
     \\
\hline
       \multicolumn{1}{c}{\bf Date-Ground Measurements} & {\bf Date-Satellite Acquisition} & {\bf Difference (Days)}\\
\hline			
13 April & 13 April & 0\\
18 April & 13 April  & 5\\
25 April & 26 April & 1\\   
2 May & 6 May & 4\\
9 May & 6 May & 3\\
27-28 June & 25 June & 3\\
4-5 July & 2 July & 3\\
16 August & 14 August & 2\\
31 August & 31 August & 3\\
12 September & 10 September & 3\\
\hline
	   \end{tabular}
	   \label{Dates}
\end{center}
\end{table}

\subsubsection{Barrax site}\label{Barrax}

As second test, the agricultural area Barrax, Spain, was chosen (Lat. 39$\degree$N, Long. -2$\degree$E). 
Although no field campaign has been conducted for the last few years, this site was long used as reference for ESA field campaigns in support of satellite missions (e.g., SPARC, SEN3EXP). The Barrax agricultural area has a rectangular form and an extent of 5 km by 10 km, and is characterized by a flat morphology and large, uniform land-use units. The region consists of approximately 65\% dry land and 35\% irrigated land, mainly by center pivot irrigation systems. It leads to a patchy landscape with large circular fields. Cultivated crops include garlic, alfalfa, onion, sunflower, corn, potato, sugar beet,
vineyard and wheat. The annual rainfall average is about 400 mm.

\subsection{Mapping and comparison with SNAP Biophysical Processor}\label{mapping}

\begin{figure}[!h]
	\centering
	\IG[trim={0cm 17cm 0cm 0cm},clip,width=8cm]{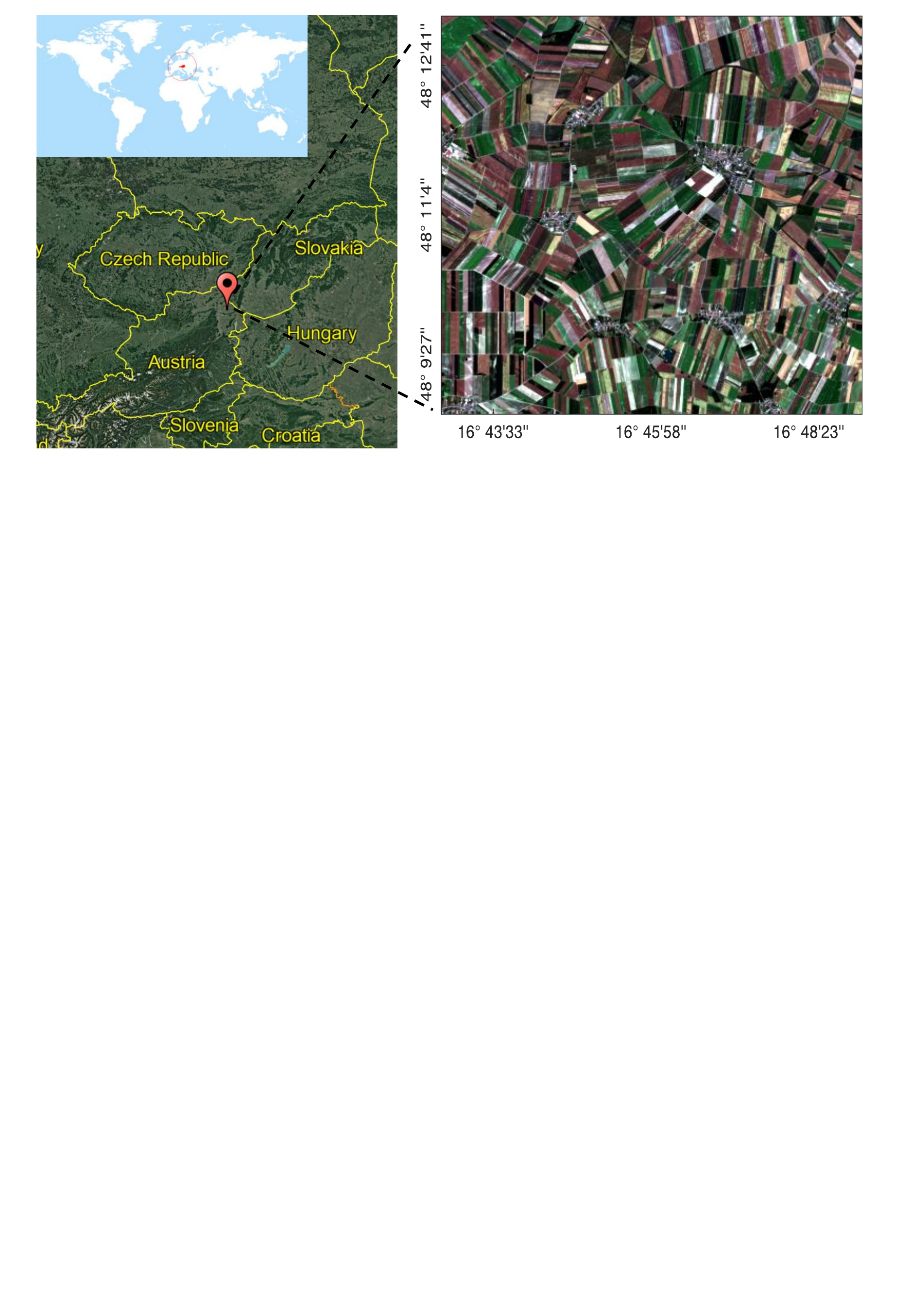} \\
	\IG[trim={0cm 17cm 0cm 0cm},clip,width=8cm]{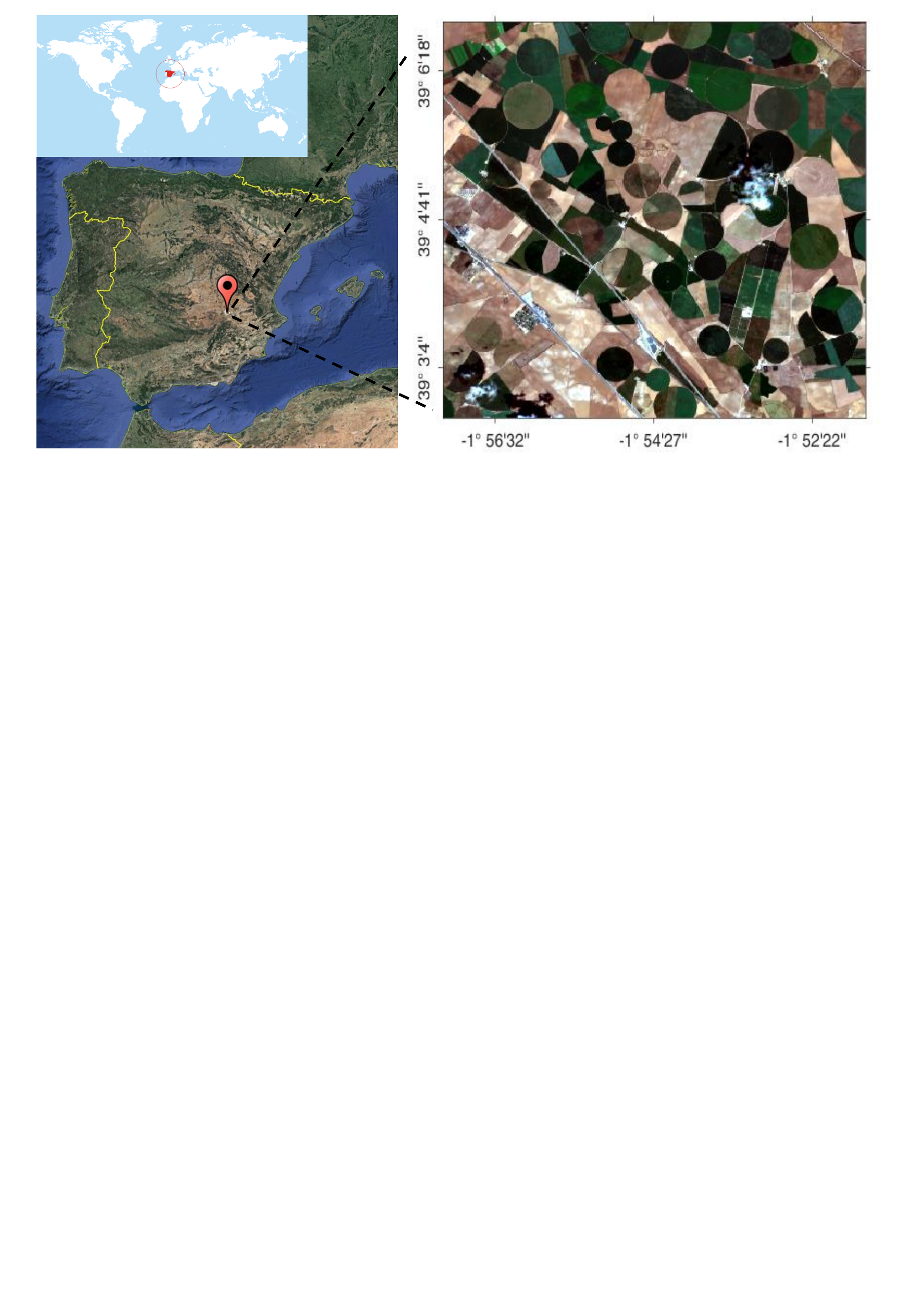} \\
	\caption{RGB composition of Sentinel-2 MSI over Marchfeld, Austria (left) and Barrax, Spain (right).} 
	\label{RGB_Barrax}
\end{figure}

As a final step, in order to evaluate the capability of the GPR and VHGPR models to generate LAI maps, we used the trained models to generate LAI maps from S2 images using ARTMO's MLRA toolbox \cite{Caicedo2014}. To do so, for both the Marchfeld and Barrax sites, a cloud-free spatial subset of S2 L1C (TOA) and L2A (BOA) imagery was used to evaluate the retrieval performance at both TOA and BOA scale. For Marchfeld an acquisition during the field campaign (2 July 2016) was used, while for Barrax a more recent acquisition was used (5 June 2017). The average SZA values were 30$^\circ$ and 20$^\circ$ respectively for Marchfeld and Barrax. A RGB of the subsets are shown in Figure \ref{RGB_Barrax}. For Barrax site the L2A product was directly downloaded from Scihub. For Marchfeld site, since the image is not available at L2A for this acquisition date in the Scihub, the L1C subset was processed offline using Sen2Cor Atmospheric Correction Processor (version 2.5.5) to perform the L2A reflectance. For this atmospheric correction the default parameters of the Sen2Cor algorithm were used \cite{Clerc2020}.
Only the 10 m and 20 m bands were used from S2 images, being the bands 2 to 8, 8a, 11, and 12. The  images were resampled to 20 m and for  both sites a spatial subset of 400 by 400 pixels was selected.

Lastly, the generated LAI maps were compared against the maps generated with the Biophysical Processor of SNAP software \cite{weiss2016}. In SNAP, the retrieval of LAI is based on a neural network (NN) algorithm and is built on earlier experience for other ESA land missions \cite{weiss2016}. 
This algorithm is also trained with a PROSAIL LUT. The variables range and distributions established for the LUT were similar to the ones we assumed in PROSAIL (Table \ref{prosail_conf}). Although ESA recommends to use this tool with TOC reflectance data, we also use TOA reflectance data to generate the maps for comparison purposes.


\section{Results}\label{sec:results}

\subsection{GPR LAI models cross-validation and validation against field data} \label{validation_results}

\begin{figure*}[!h]
\centering
\small
\begin{tabular}{cccc}
TOC (BOA) reflectance & TOA radiance\\
\raisebox{-0.5\height}{\IG[height=3cm,trim={5cm 21.6cm 5cm 0.5cm},clip]{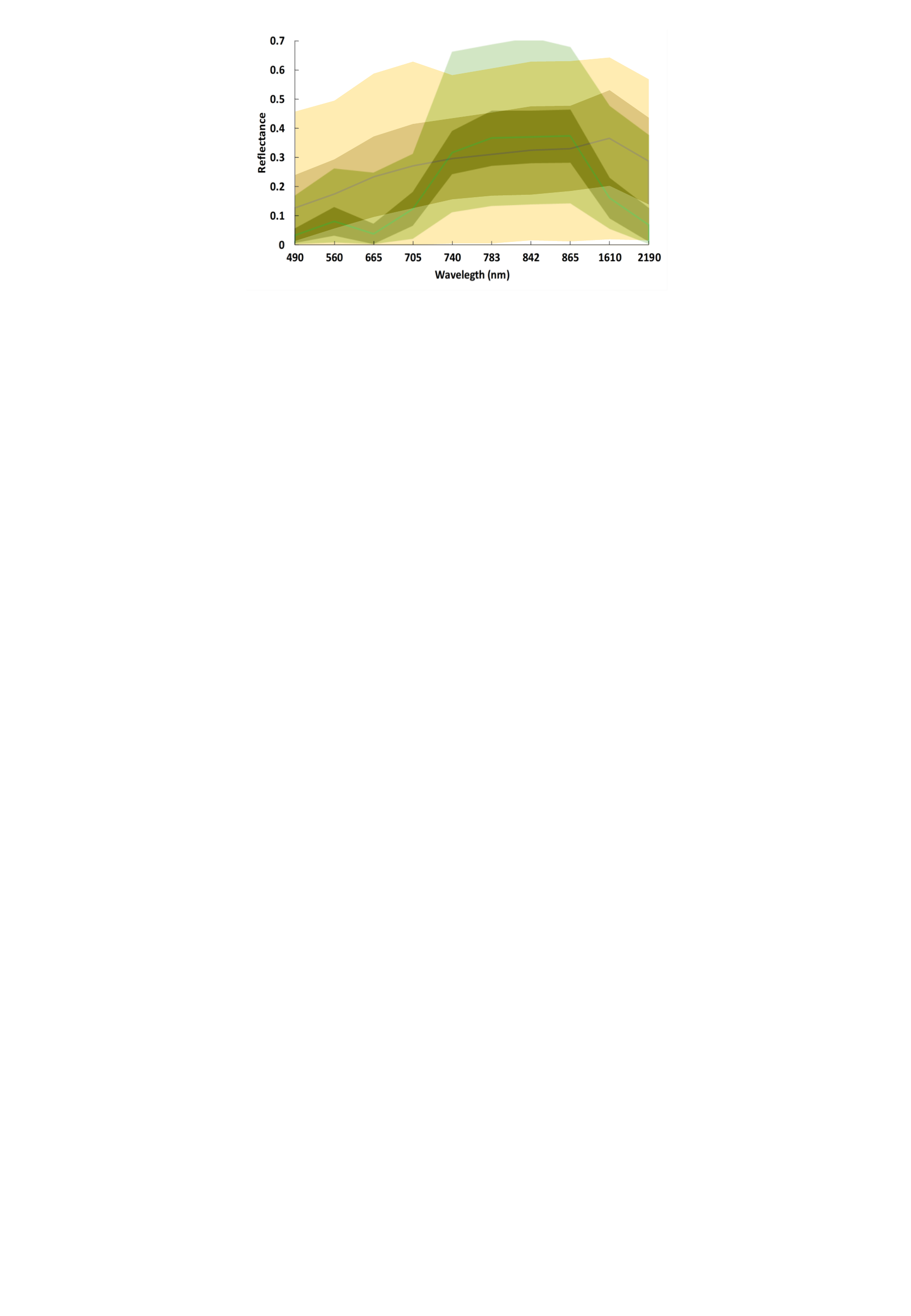}} &
\raisebox{-0.5\height}{\IG[height=3cm,trim={5cm 22.0cm 5cm 0.3cm},clip]{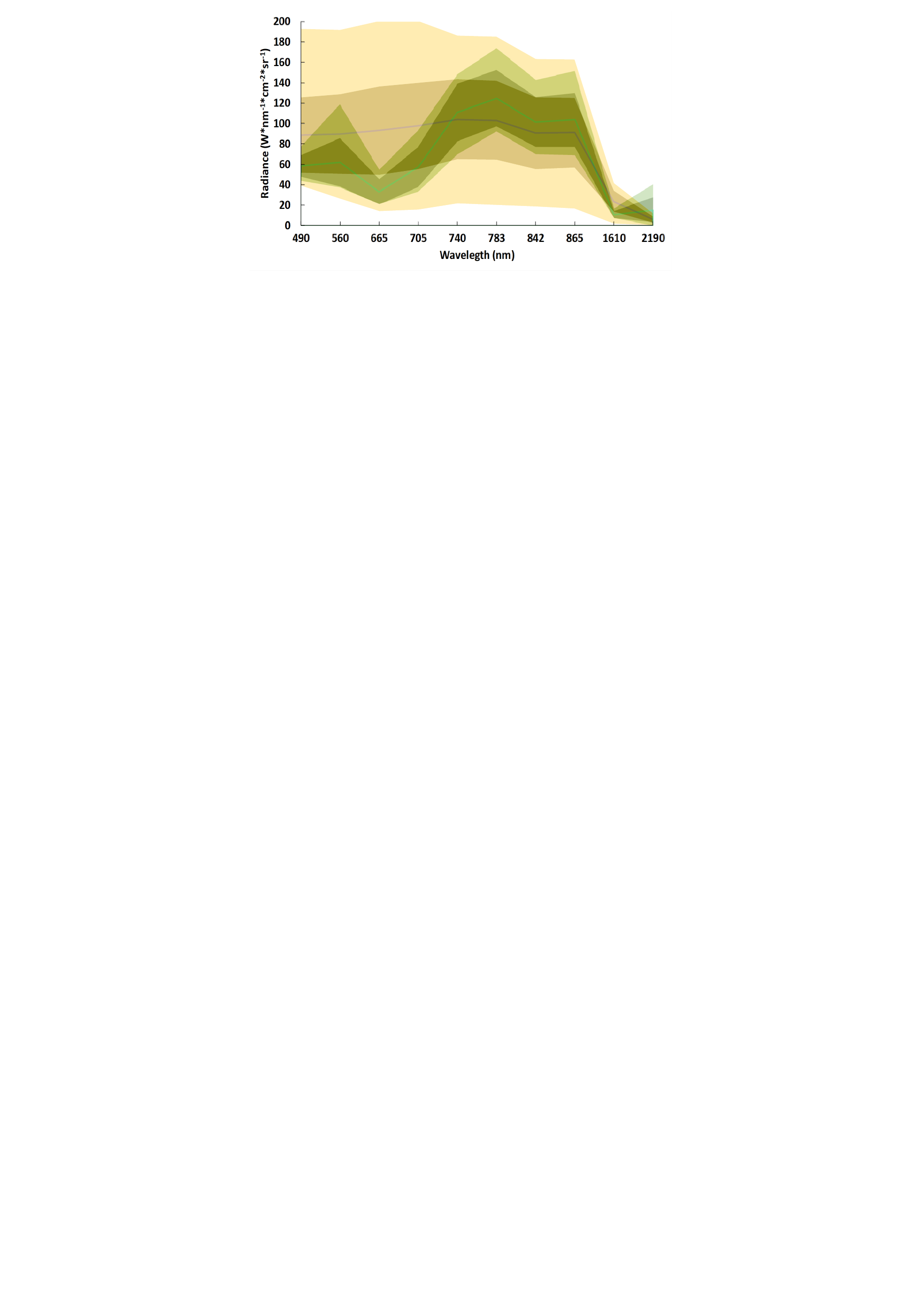}} & \\
\multicolumn{2}{c}{\raisebox{-0.5\height}{\IG[width=20cm,trim={3cm 26cm 3cm 0.4cm},clip]{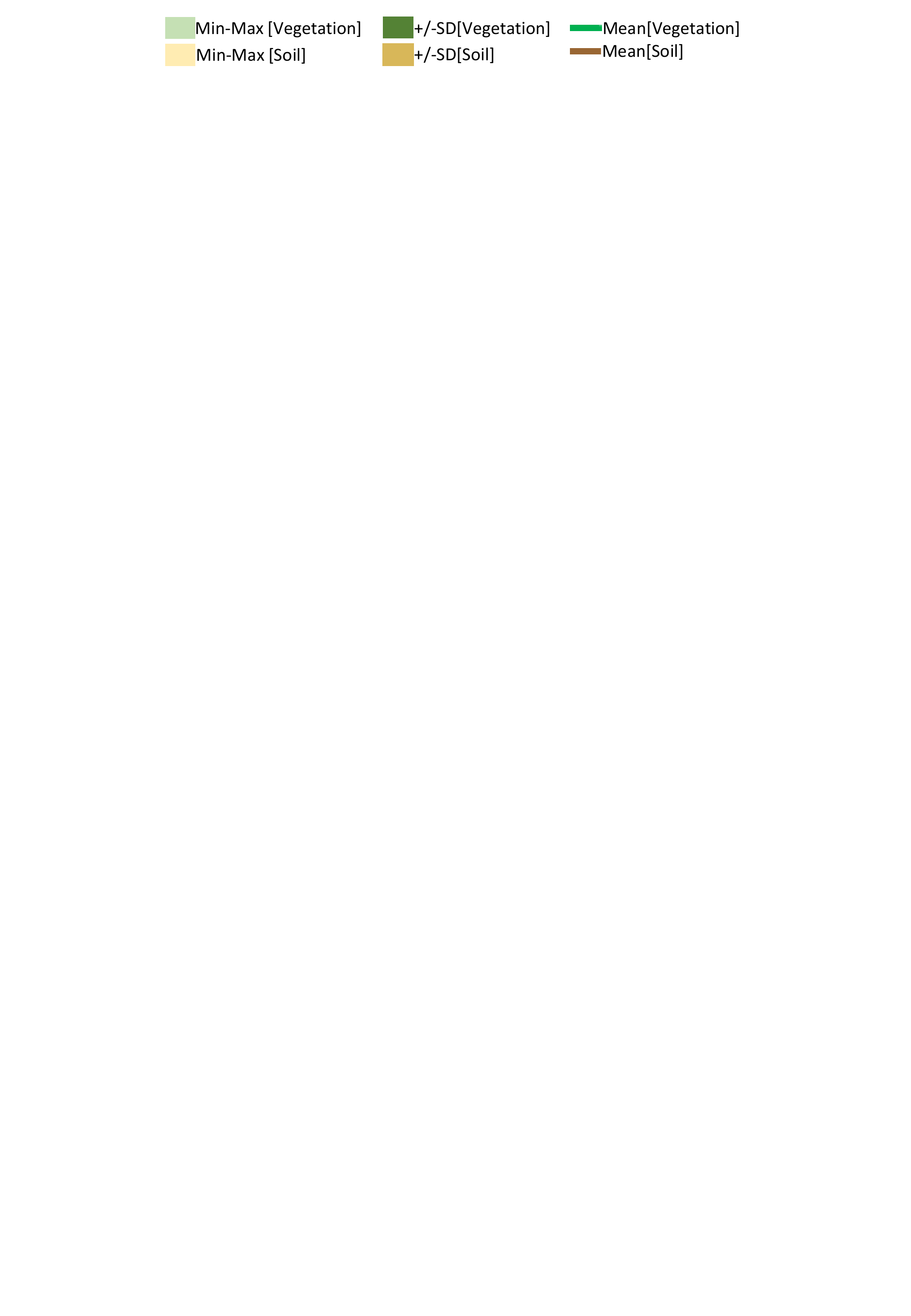}}} \\
\end{tabular}
\caption{General statistics (mean, standard deviation (SD), min-max) for the bare soil spectra collected from S2-L2A product and vegetation spectra simulated with PROSAIL (left) for the 10 and 20 m S2 bands. The same data has been upscaled to TOA radiance with 6S (right).} 
\label{spectraTOC2TOA}
\end{figure*}

Before assessing the validity of the trained GPR and VHGPR models, first inspection of the training data is provided. RTM simulations were run both by PROSAIL, which led to TOC reflectance, and by PROSAIL-6S, which led to upscaled TOA radiance data. Because empirical soil spectra was added to the training data to account for the non-vegetated surfaces, overview statistics of both simulated vegetated spectra and bare soil spectra are shown in Figure \ref{spectraTOC2TOA}. The figures at TOC and TOA scales demonstrate that a large variability of vegetation and soil profiles are covered in the training dataset. This is essential to process a complete S2 image, including all kinds of non-vegetated surfaces. These datasets form the core of the LAI retrieval algorithms applied to S2 BOA and TOA images.

Following the above-described processing scheme (Figure \ref{flowchart}), GPR and VHGPR algorithms were trained with the TOC and TOA training datasets for the development of LAI retrieval models. A first step was to evaluate the theoretical performances of the trained GPR and VHGPR models. A 5-$k$ cross-validation sampling strategy was applied to assess the theoretical goodness-of-fit of the algorithms. Both algorithms showed good and consistent performances at BOA and TOA scale. At BOA scale, GPR obtained a slightly superior accuracy with a $R^2$ of 0.59 (RMSE: 1.08; NRMSE: 15.50\%) than VHGPR, with a $R^2$ of 0.58 (RMSE: 1.09;  NRMSE: 15.68\%). Results were basically identical at the TOA scale yet somewhat poorer than at BOA scale:  GPR with a $R^2$ of 0.54 (RMSE: 1.14; NRMSE: 16.30\%), and VHGPR with a $R^2$ of 0.54 (RMSE: 1.14;  NRMSE: 16.35\%). 
That the TOA retrieval performed somewhat inferior was expected because of the additional variability introduced into the database by means of the coupling with the atmosphere simulations. Yet, the retrieval performances are similar to the TOC dataset, and it reveals the consistency of the LAI retrieval method from TOA radiance data. For both datasets, the added noise and saturation at higher LAI values explain the sub-optimal results, however what matters is the performances against ground validation data.

\begin{figure*}[!t]
\centering
\small
\begin{tabular}{ccc}
& GPR & VHGPR\\
\rotatebox[origin=c]{90}{L2A (BOA)} & 
\raisebox{-0.5\height}{\IG[height=7cm,trim={1cm 11cm 1cm 0cm},clip]{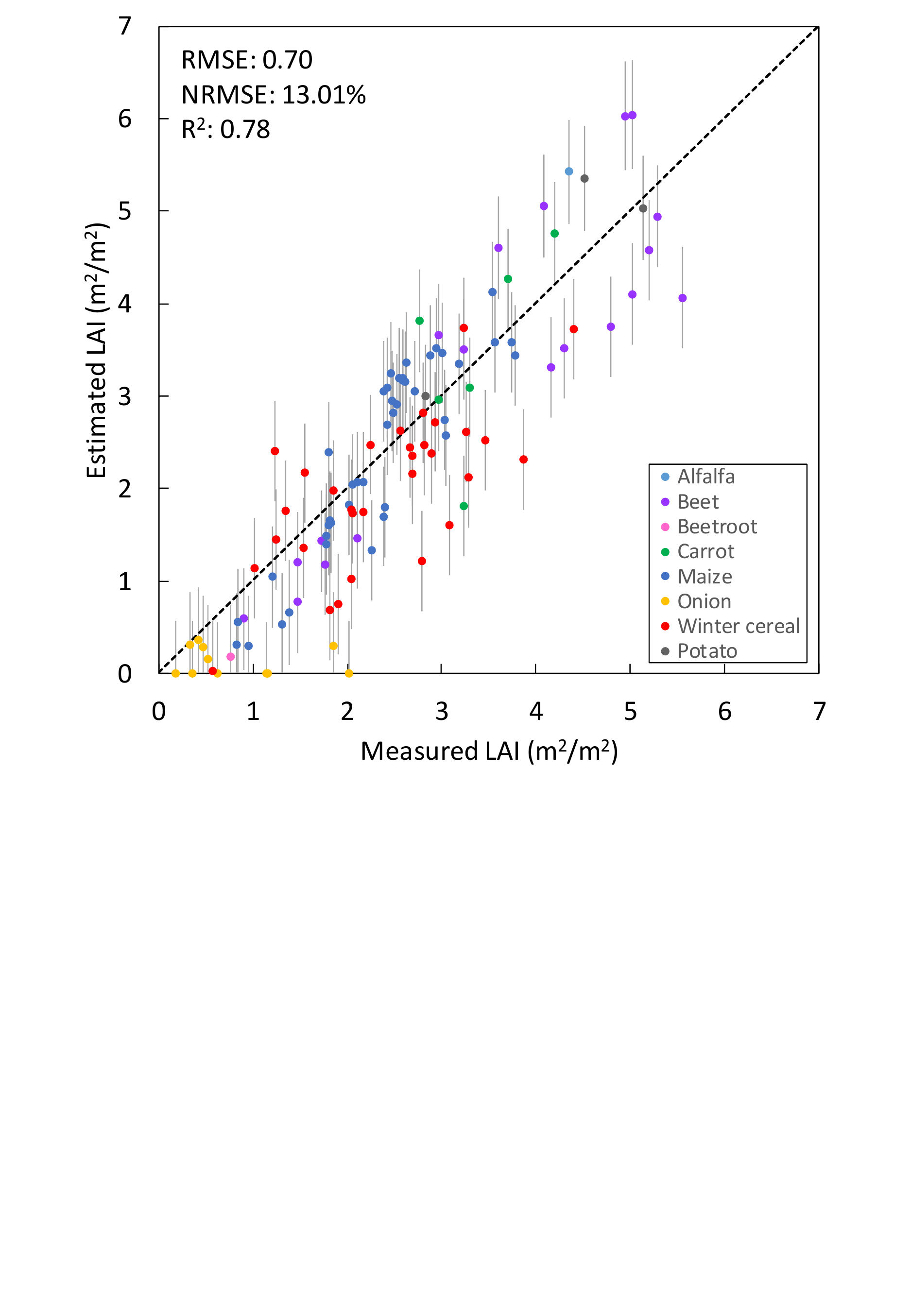}} &
\raisebox{-0.5\height}{\IG[height=7cm,trim={1cm 11cm 1cm 0cm},clip]{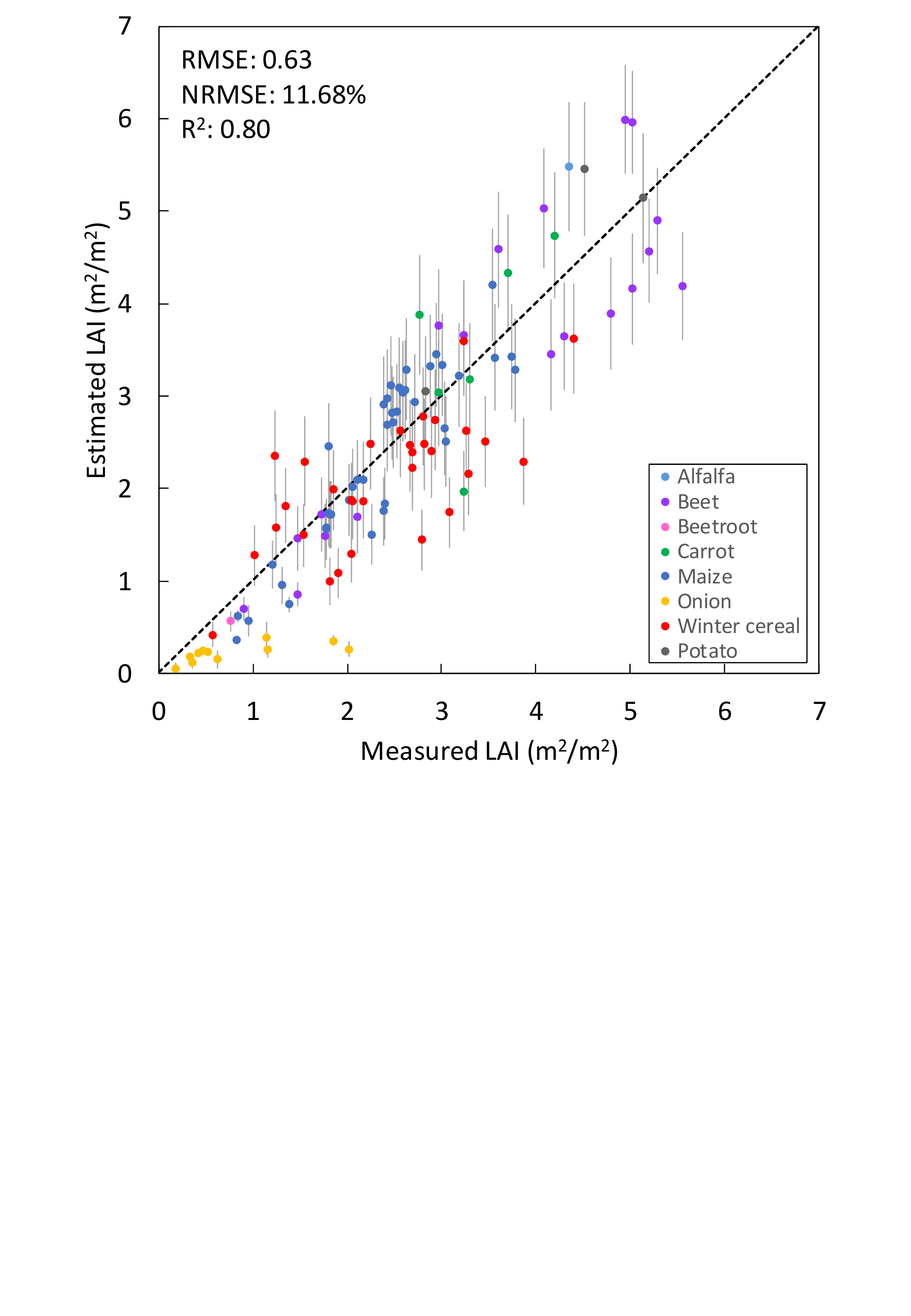}} \\
\rotatebox[origin=c]{90}{L1C (TOA)} & 
\raisebox{-0.5\height}{\IG[height=7cm,trim={1cm 11cm 1cm 0cm},clip]{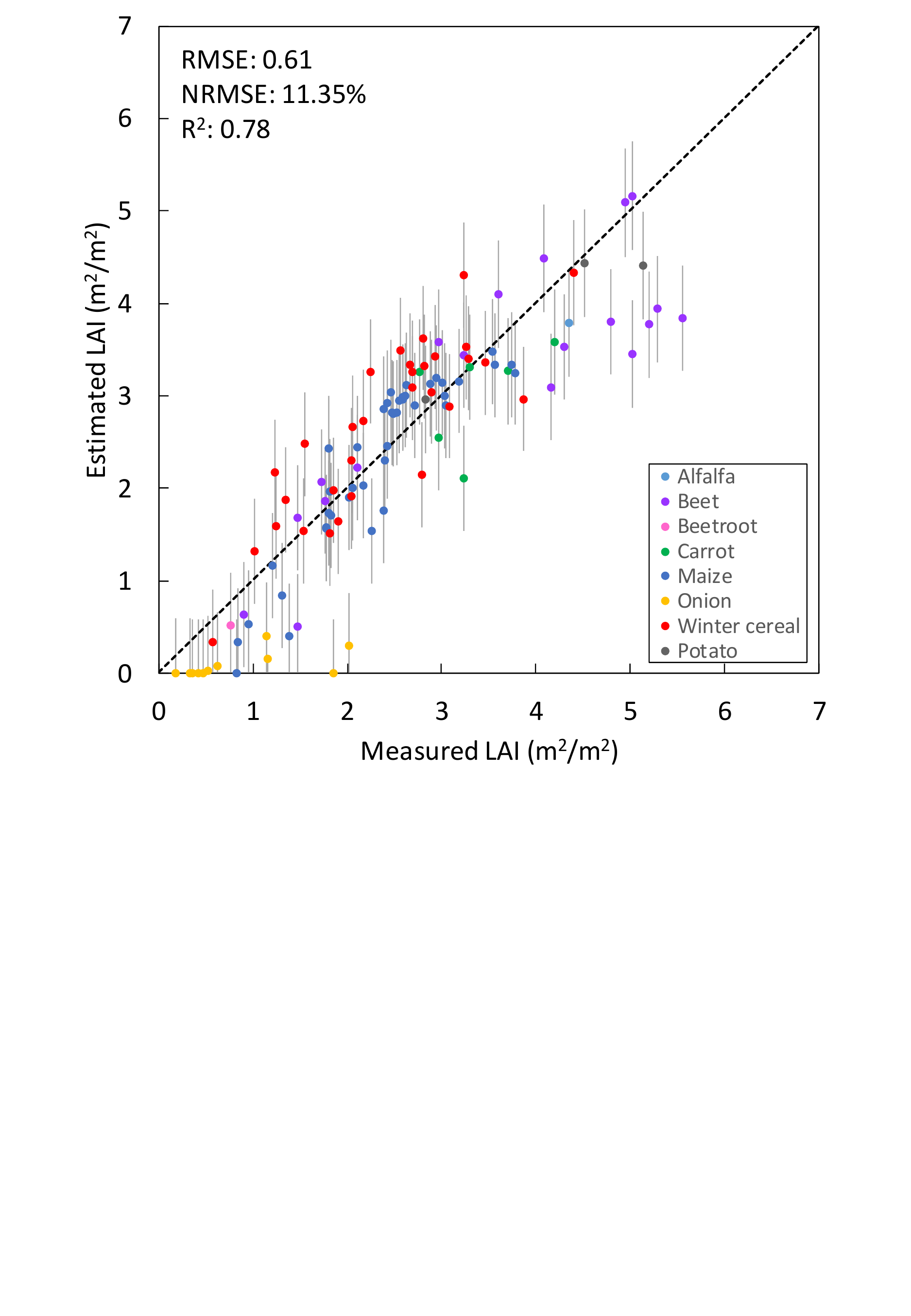}} &
\raisebox{-0.5\height}{\IG[height=7cm,trim={1cm 11cm 1cm 0cm},clip]{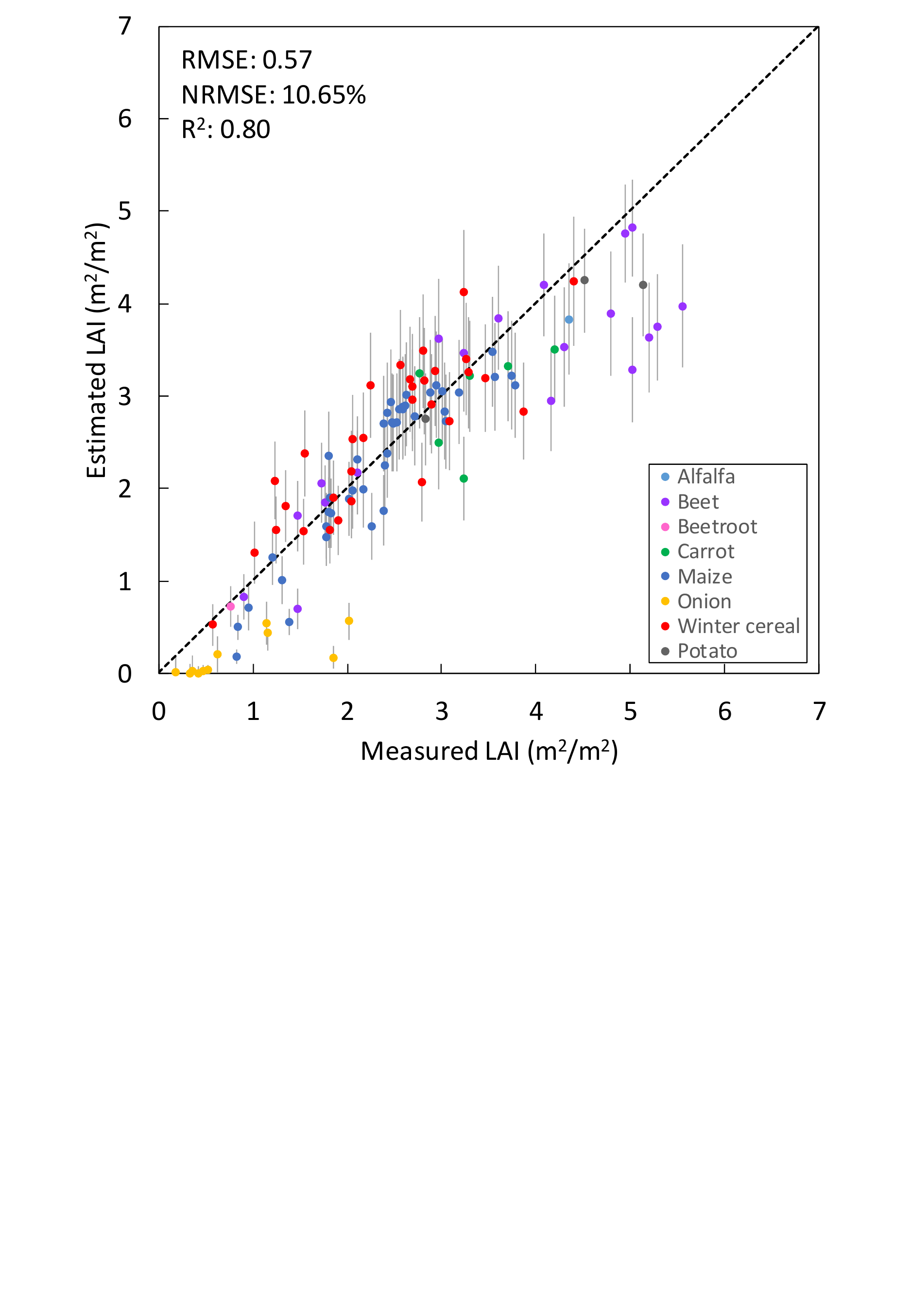}} \\
\end{tabular}
\caption{Measured vs. estimated LAI values along the 1:1-line with associated confidence intervals (1 SD). Ground validation of several crops over Marchfeld site for GPR (\textbf{left}) and VHGPR (\textbf{right}) for retrieval S2-L2A (BOA) data (\textbf{top)} and S2-L1C (TOA) data (\textbf{bottom}) .} 
\label{GV_scat}
\end{figure*}

Regarding validation against field data, the Marchfeld dataset was used. For both BOA and TOA and for GPR and VHGPR, the measured vs. estimated scatter plots are shown in Figure \ref{GV_scat}, along with the goodness-of fit statistics. For both GPR and VHGPR, the validation results do not suggest differences between the BOA and TOA scale for the various crop types. This is of interest, as it suggests that the retrieval models function just as well at both scales. It underlines the possibility of retrieving LAI directly from TOA radiance data. 
Conversely, validation results were slightly superior obtained by VHGPR as opposed to GPR. These results suggest that VHGPR delivers superior accuracies; e.g. the underestimations are smaller as opposed to GPR. An advantage of GPR algorithms is that associated uncertainty estimates (confidence intervals) are provided. As can be observed, all estimates are accompanied with a consistent uncertainty range. Here an additional advantage of VHGPR appeared: it provided lower uncertainties, especially at lower LAI. Hence, VHGPR was used for further mapping applications.
Finally, when inspecting further the estimations for the different crop types, all crops were reasonably well estimated. Only, it can be observed that onion showed a systematic understimation. Most likely for this row crop the influence of bare soil plays a large role at the pixel scale, leading to the underestimation.

\subsection{Mapping LAI from S2 BOA (L2A) and TOA (L1C) data} \label{mapping_results}

\begin{figure*}[!t]
\centering
\small
\begin{tabular}{cccc}
& LAI (m$^{2}$/m$^{2}$) & SD (m$^{2}$/m$^{2}$) & \%CV \\
\rotatebox[origin=c]{90}{L2A (BOA)} & 
\raisebox{-0.5\height}{\IG[height=3cm,trim={0.9cm 0.2cm 1.1cm 0.5cm},clip]{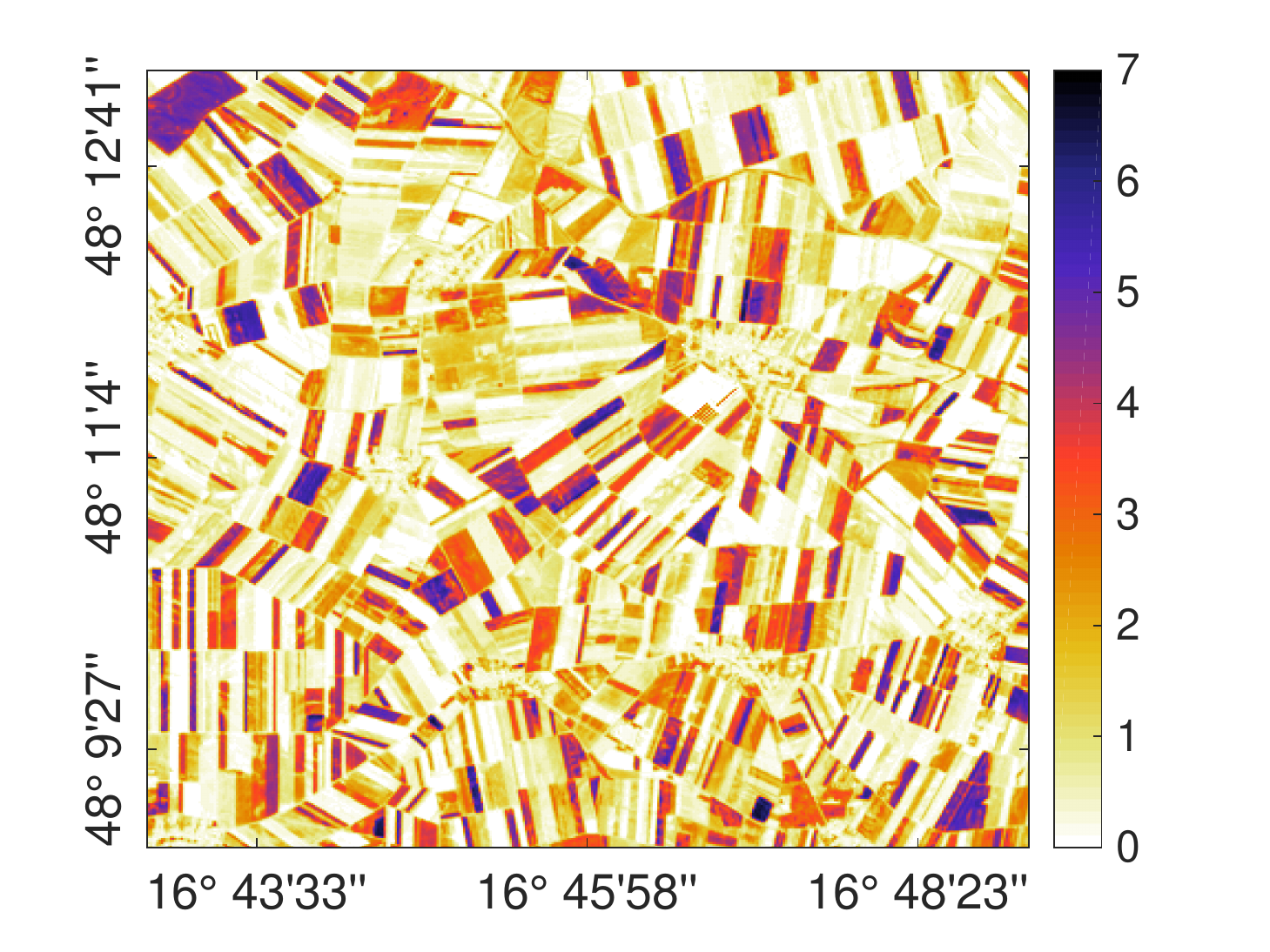}} &
\raisebox{-0.5\height}{\IG[height=3cm,trim={0.9cm 0.2cm 1.1cm 0.5cm},clip]{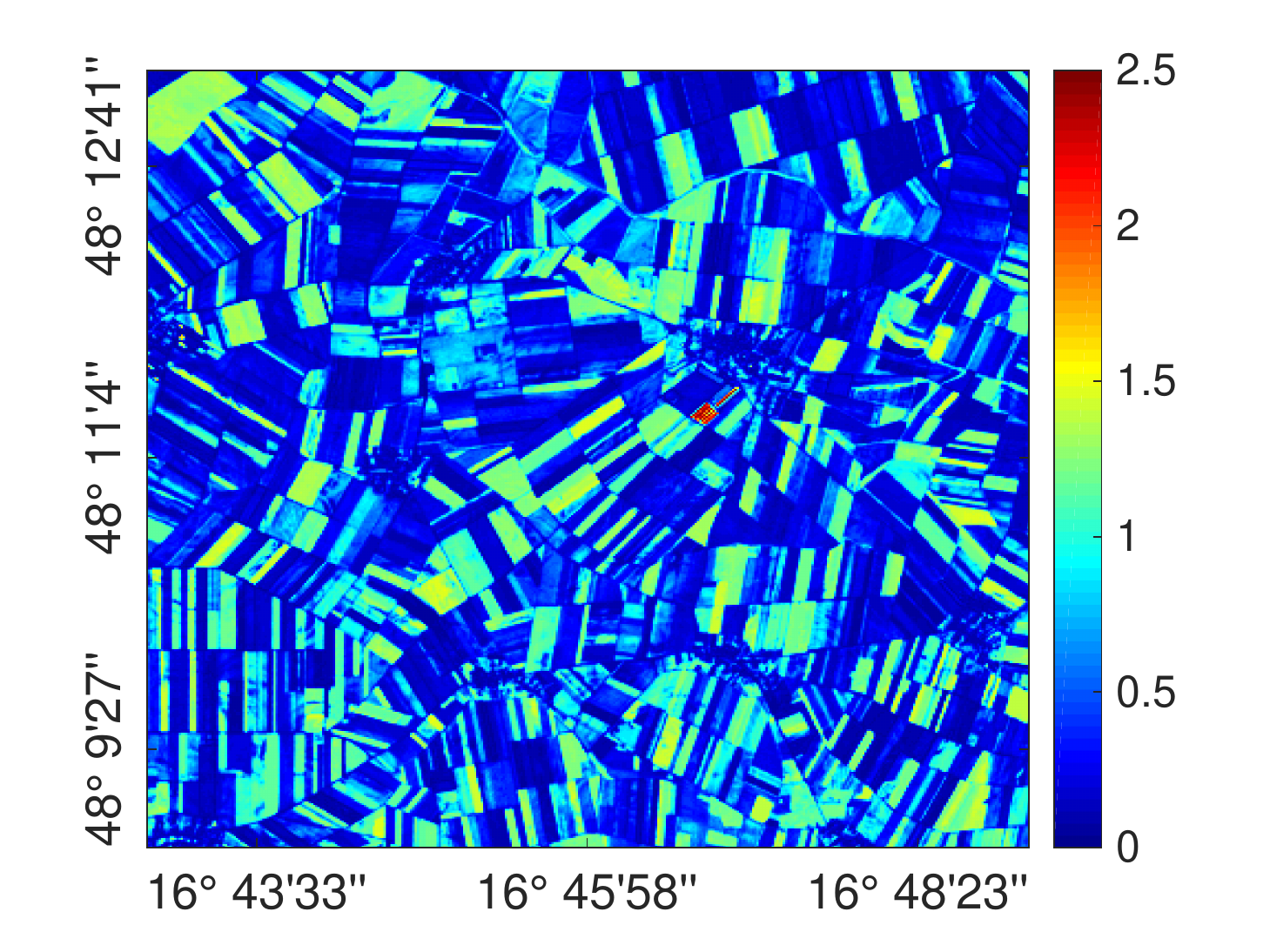}} &
\raisebox{-0.5\height}{\IG[height=3cm,trim={0.9cm 0.2cm 0.9cm 0.5cm},clip]{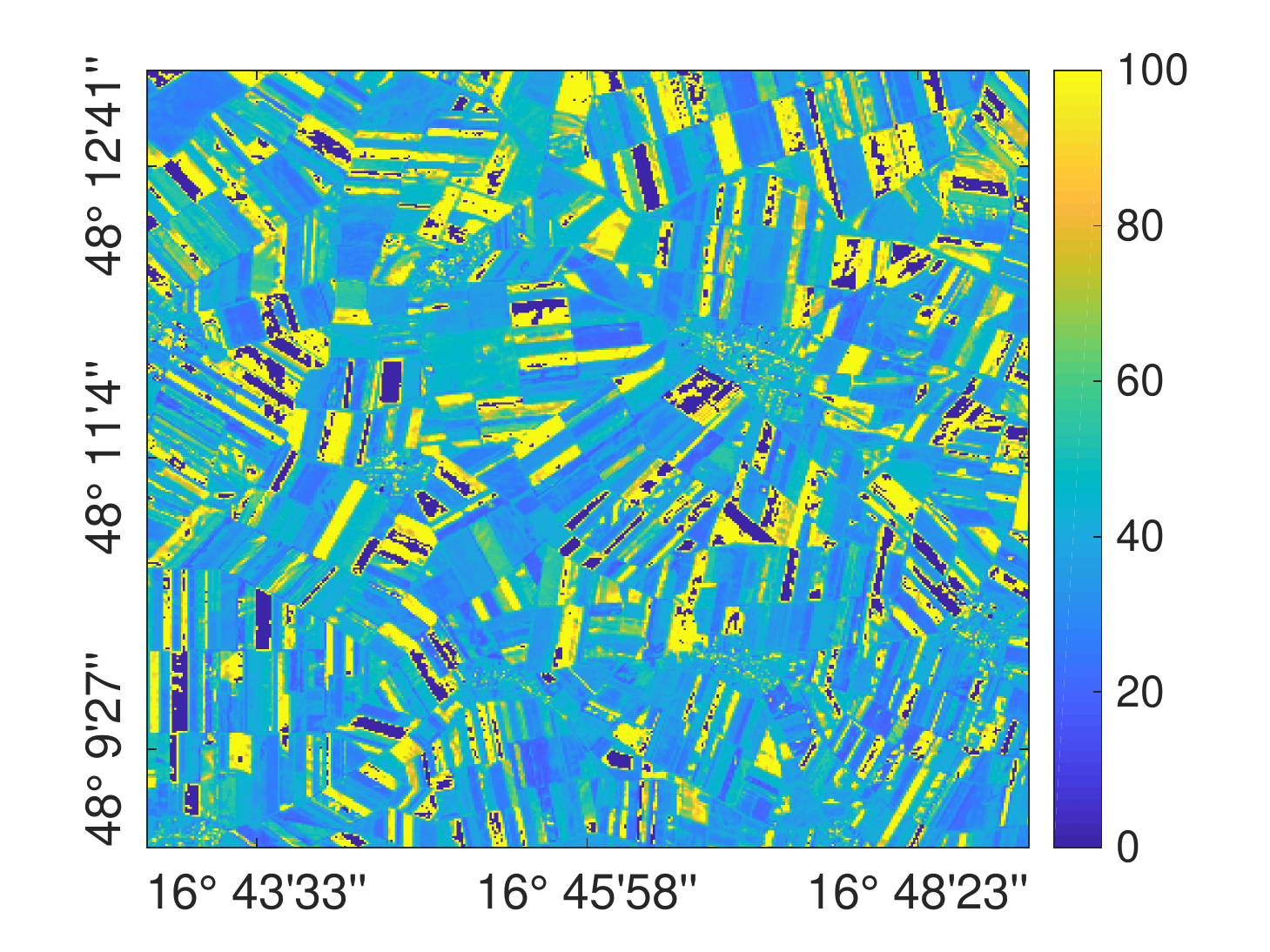}} \\
\rotatebox[origin=c]{90}{L1C (TOA)} & 
\raisebox{-0.5\height}{\IG[height=3cm,trim={0.9cm 0.2cm 1.1cm 0.5cm},clip]{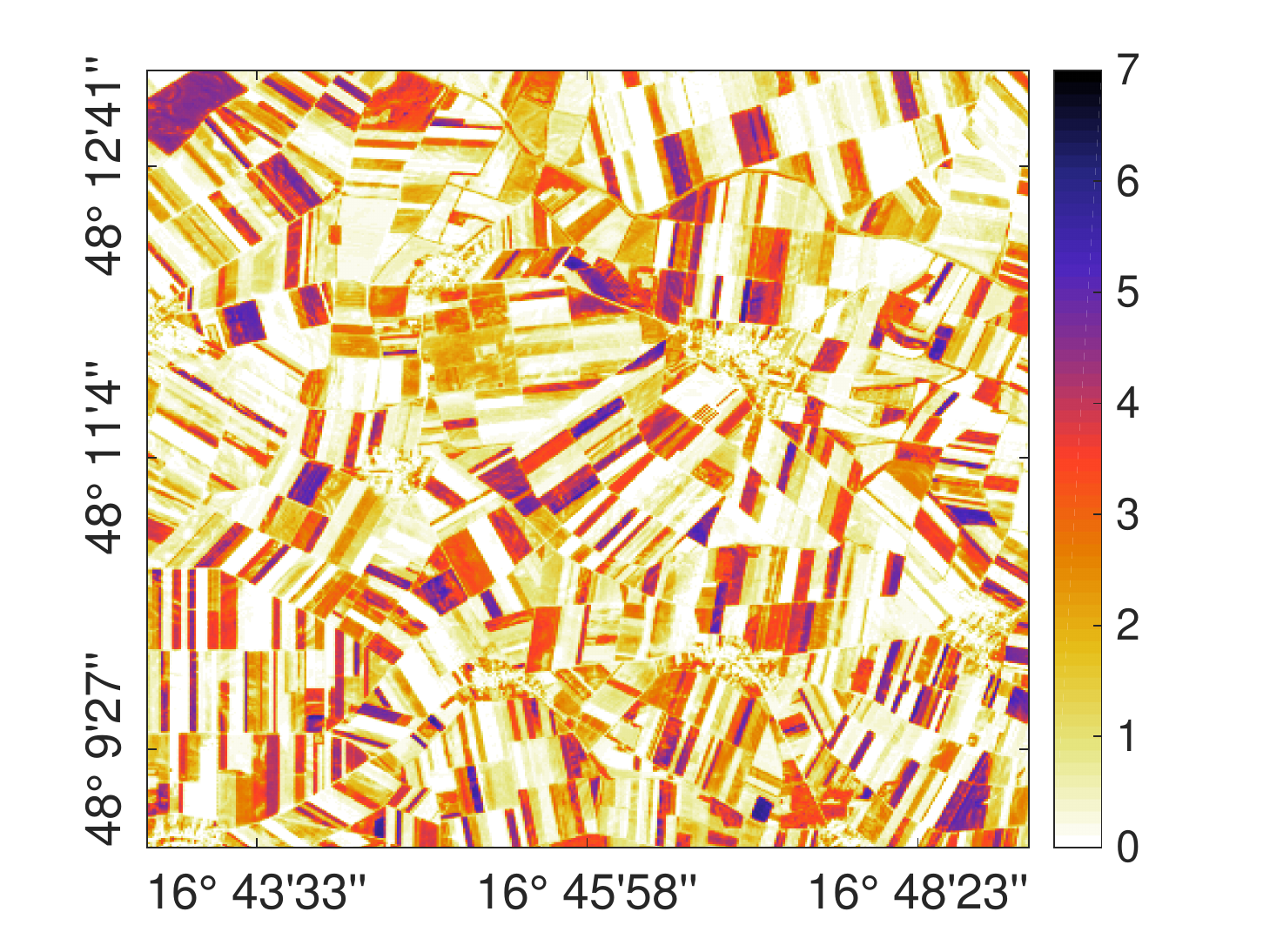}} &
\raisebox{-0.5\height}{\IG[height=3cm,trim={0.9cm 0.2cm 1.1cm 0.5cm},clip]{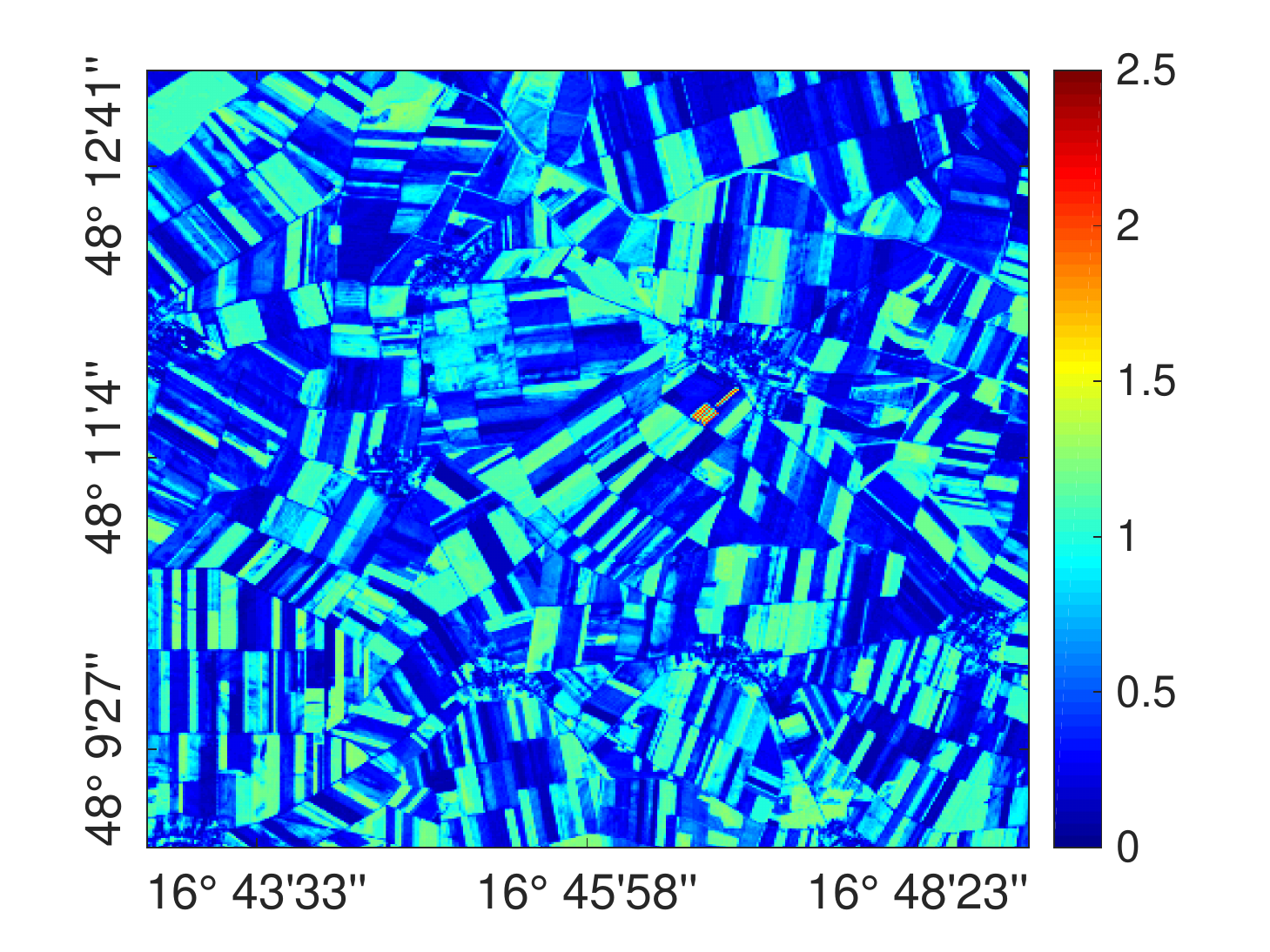}} &
\raisebox{-0.5\height}{\IG[height=3cm,trim={0.9cm 0.2cm 0.9cm 0.5cm},clip]{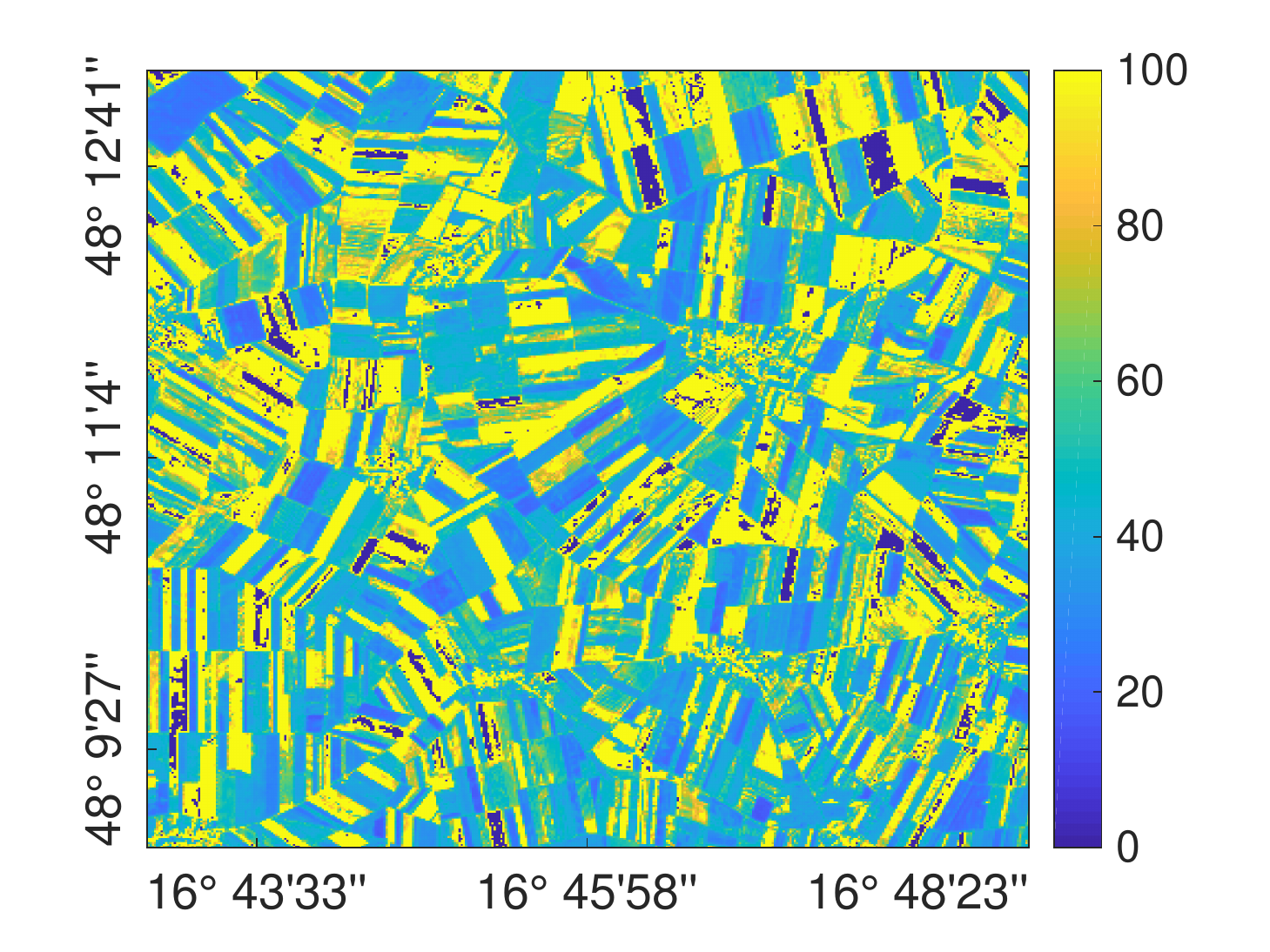}} \\
\rotatebox[origin=c]{90}{Scatterplot} & 
\raisebox{-0.5\height}{\IG[height=3.5cm,trim={1.0cm 0cm 3.0cm 1cm},clip]{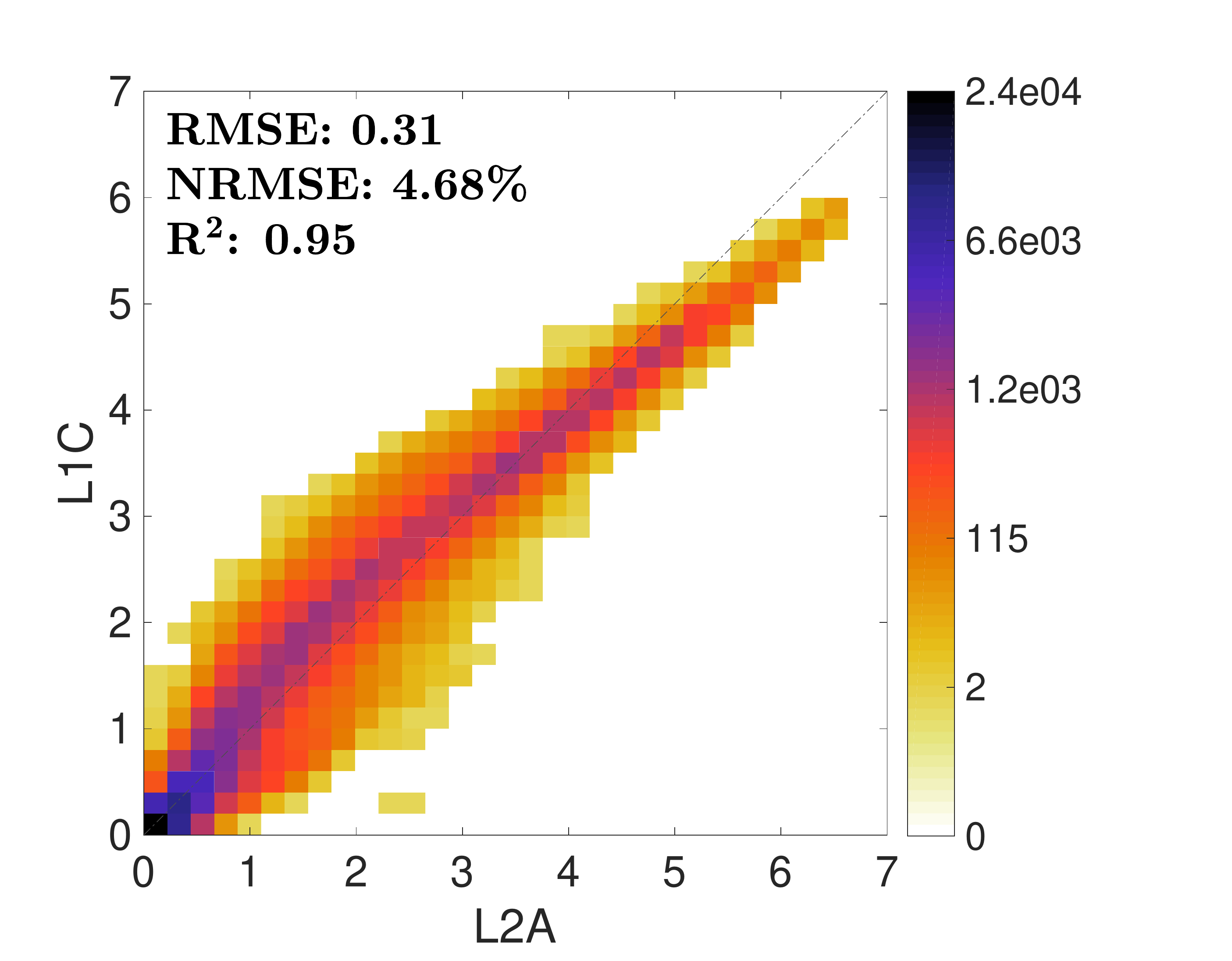}} &
\raisebox{-0.5\height}{\IG[height=3.5cm,trim={0.3cm 0cm 3.4cm 1cm},clip]{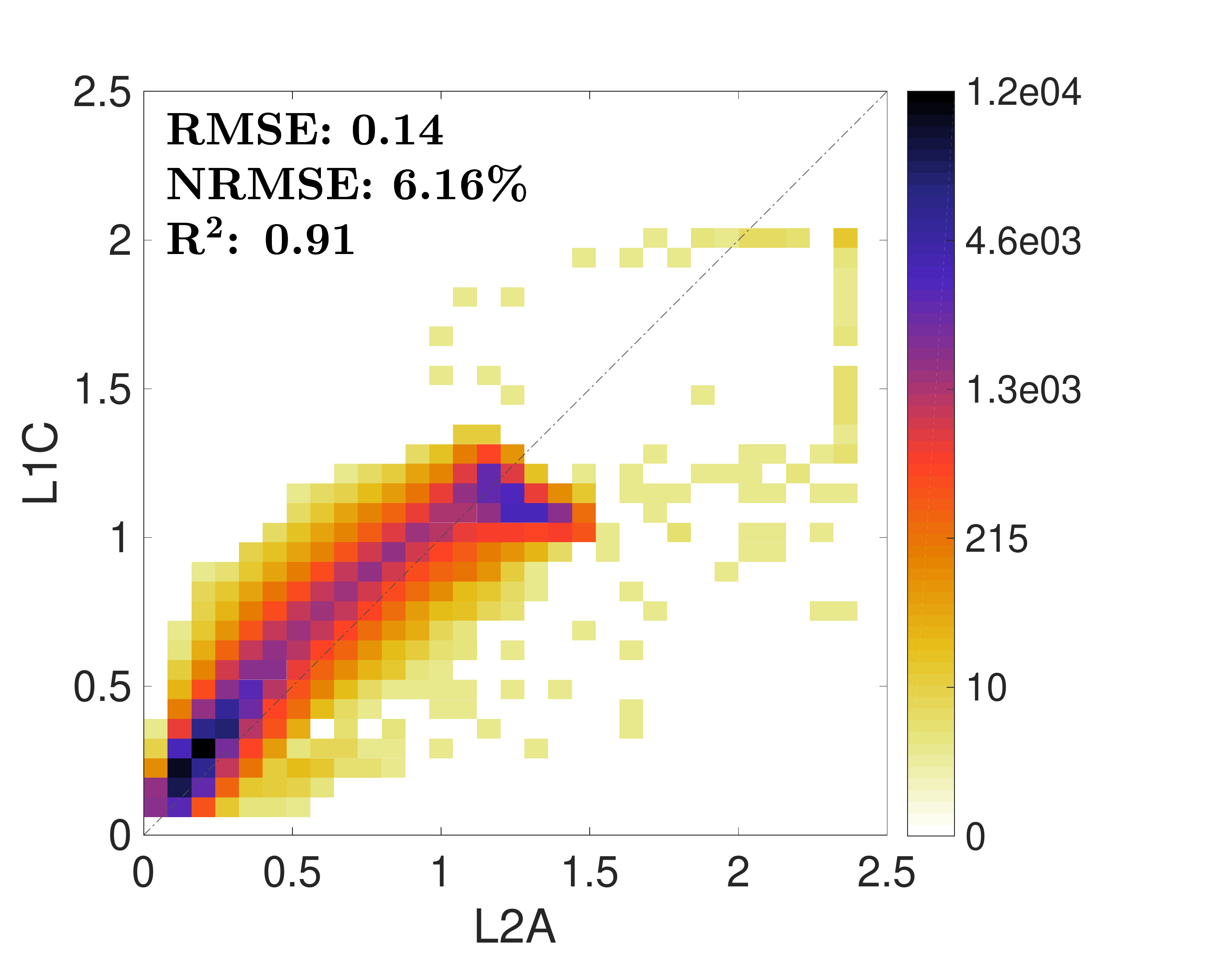}} &
\raisebox{-0.5\height}{\IG[height=3.5cm,trim={0.1cm 0cm 3.6cm 1cm},clip]{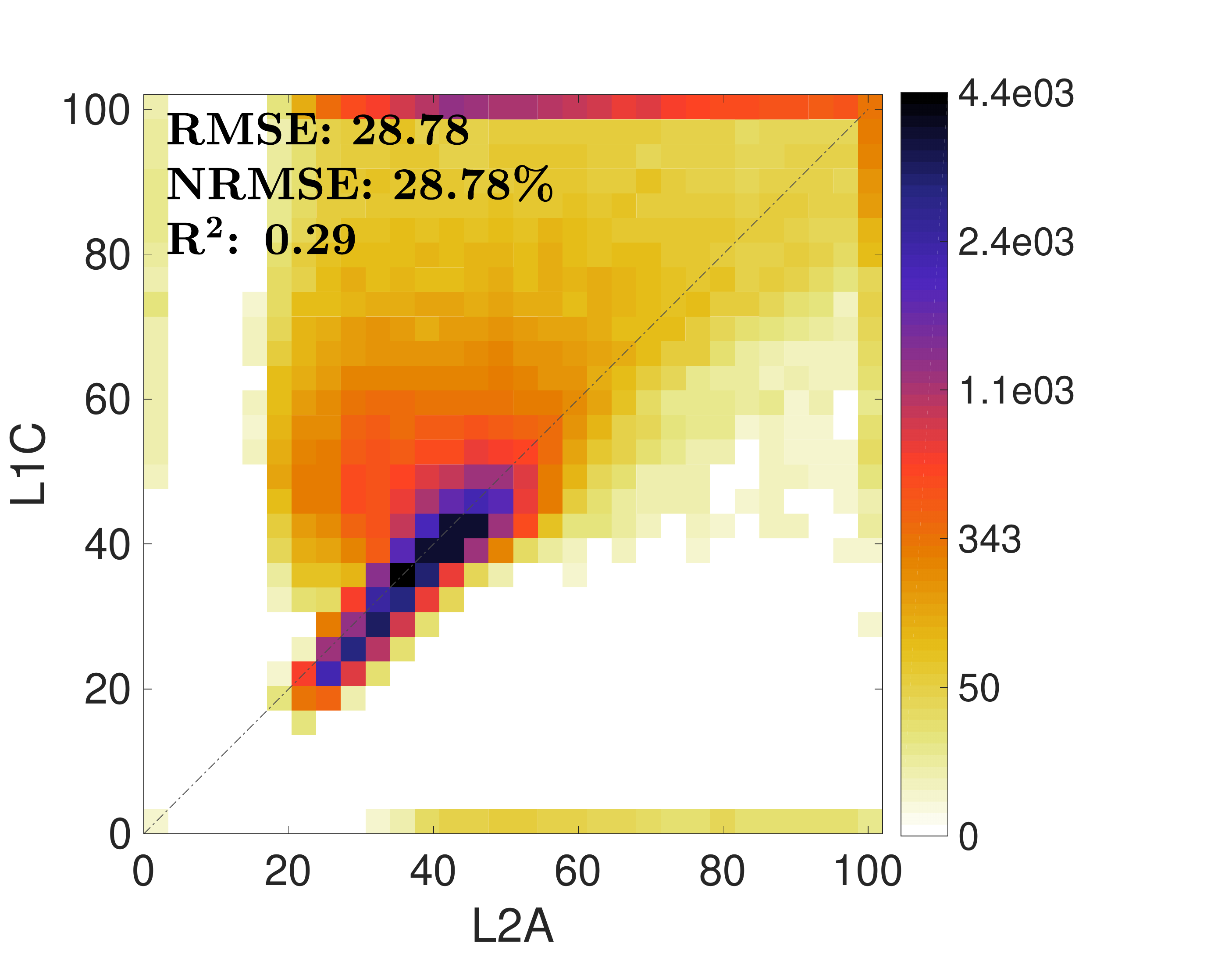}} \\
\end{tabular} 
\caption{LAI map (mean estimates; $\mu$) (left), associated uncertainties (expressed as standard deviation (SD) around the $\mu$) (center), and relative uncertainties (expressed as coefficient of variation (CV = SD/$\mu$ $\times$  100 in \%) (right) as generated by VHGPR algorithm from L2A ({\bf Top}) and L1C ({\bf Middle}) data for Marchfeld test site. Scatter plots of both maps with gridded color density ({\bf Bottom}). In case of \%CV a maximum of 100\% is set. } 
\label{mapsTOATOC_Marchfeld}
\end{figure*}

\begin{figure}[!t]
\centering
\small
\begin{tabular}{cccc}
& LAI (m$^{2}$/m$^{2}$) & SD (m$^{2}$/m$^{2}$) & \%CV \\
\rotatebox[origin=c]{90}{L2A (BOA)} & 
\raisebox{-0.5\height}{\IG[height=3cm,trim={0.9cm 0.2cm 1.1cm 0.5cm},clip]{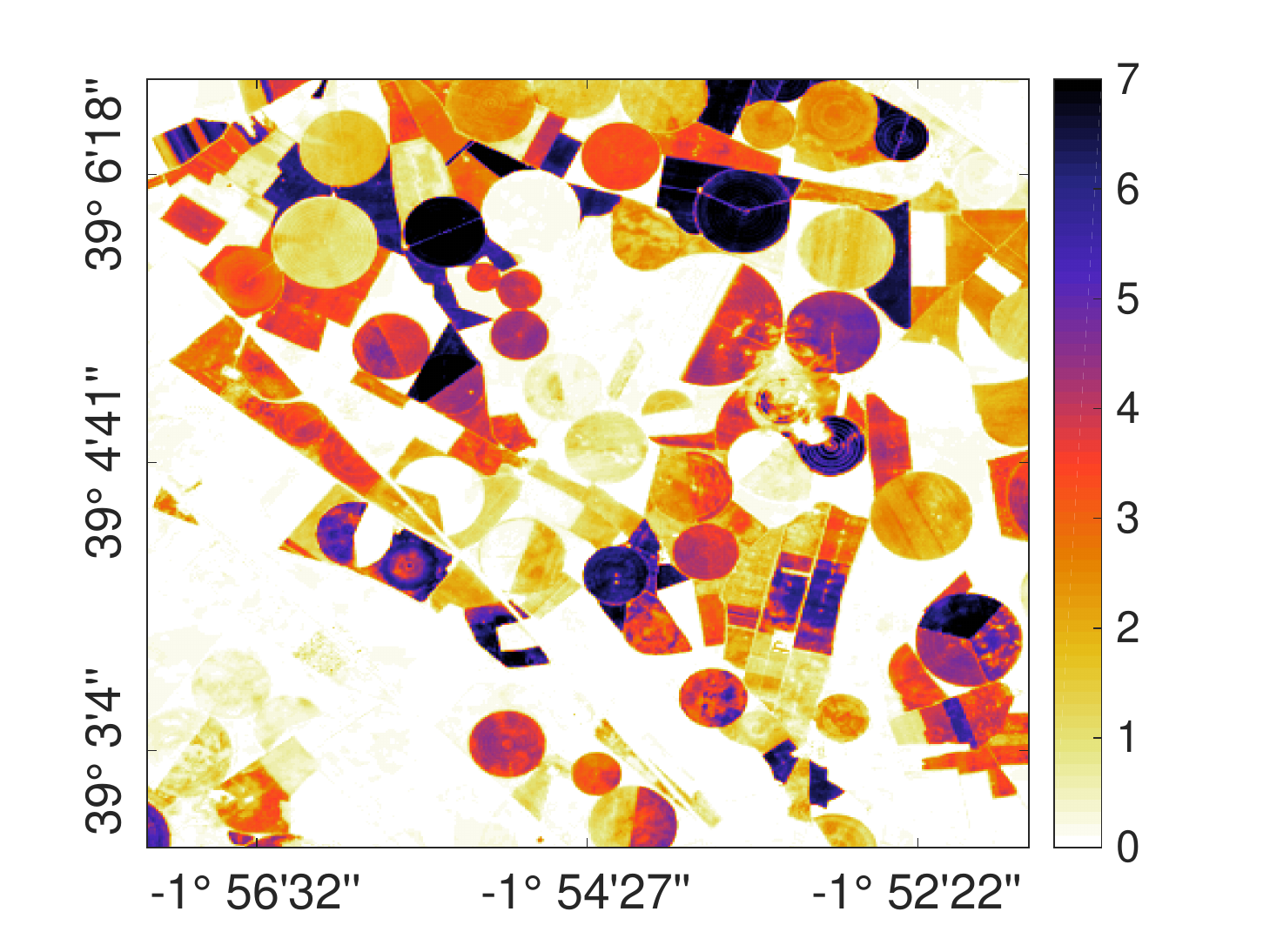}} &
\raisebox{-0.5\height}{\IG[height=3cm,trim={0.9cm 0.2cm 1.1cm 0.5cm},clip]{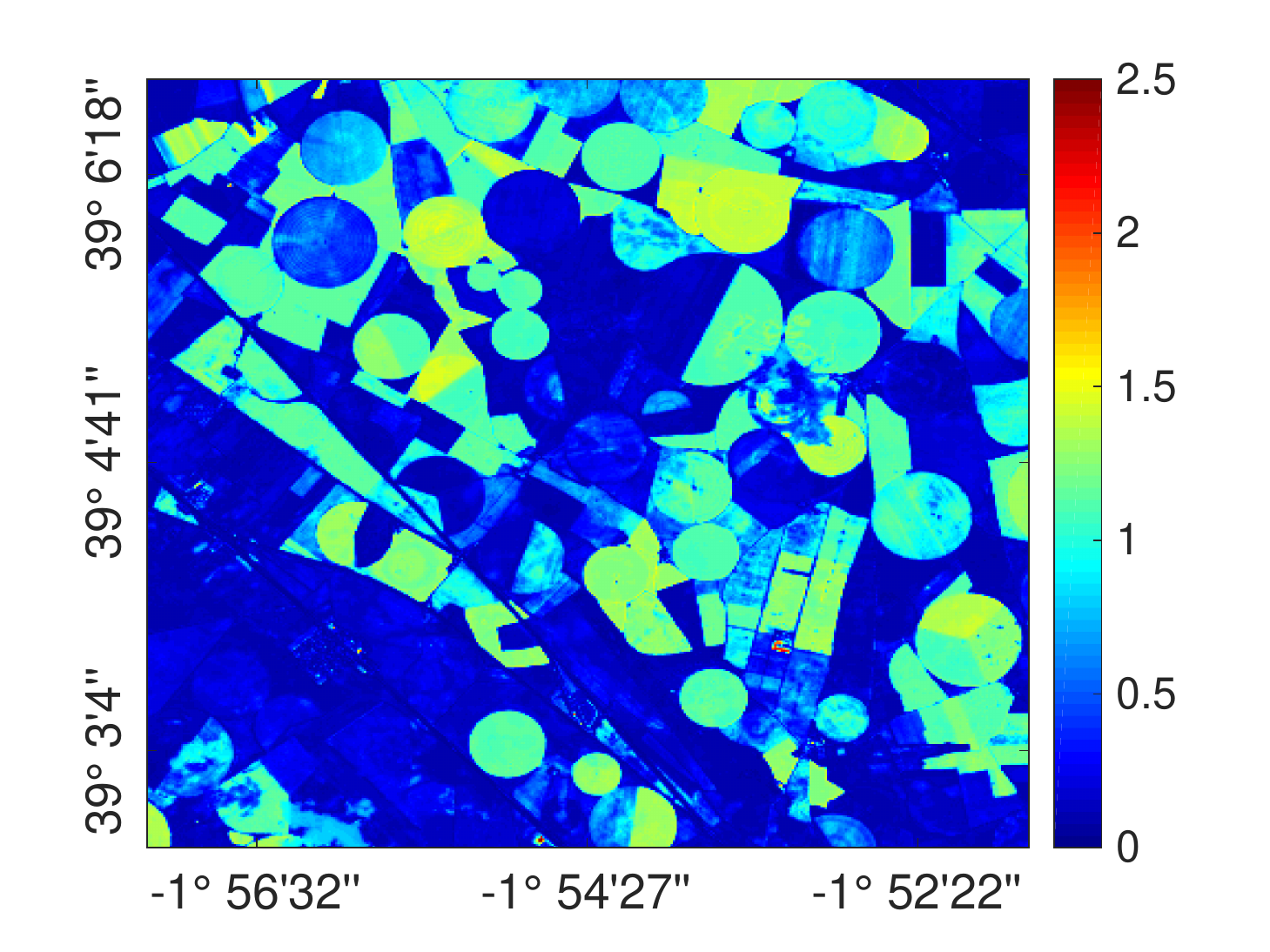}} &
\raisebox{-0.5\height}{\IG[height=3cm,trim={0.9cm 0.2cm 0.9cm 0.5cm},clip]{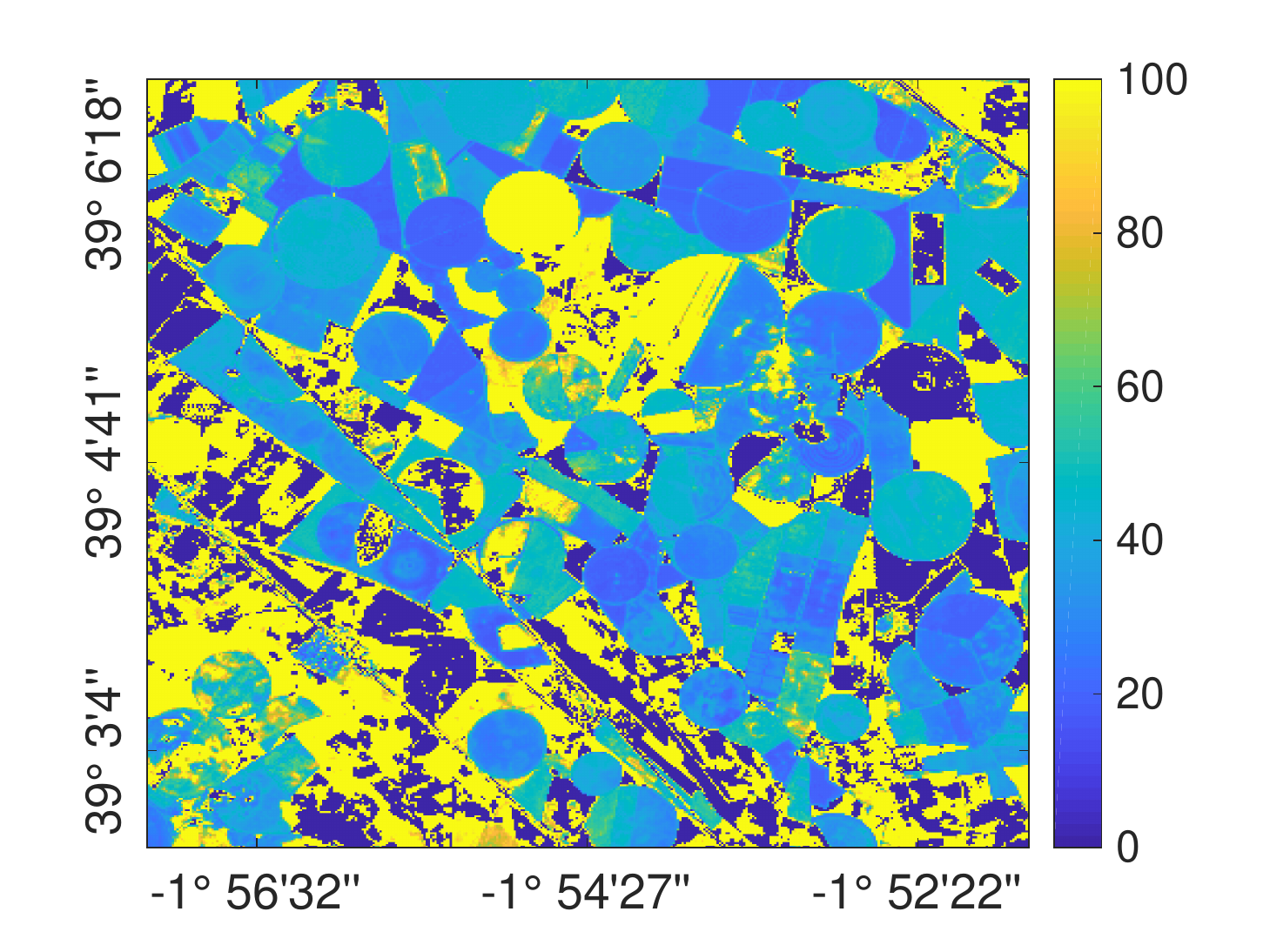}} \\
\rotatebox[origin=c]{90}{L1C (TOA)} & 
\raisebox{-0.5\height}{\IG[height=3cm,trim={0.9cm 0.2cm 1.1cm 0.5cm},clip]{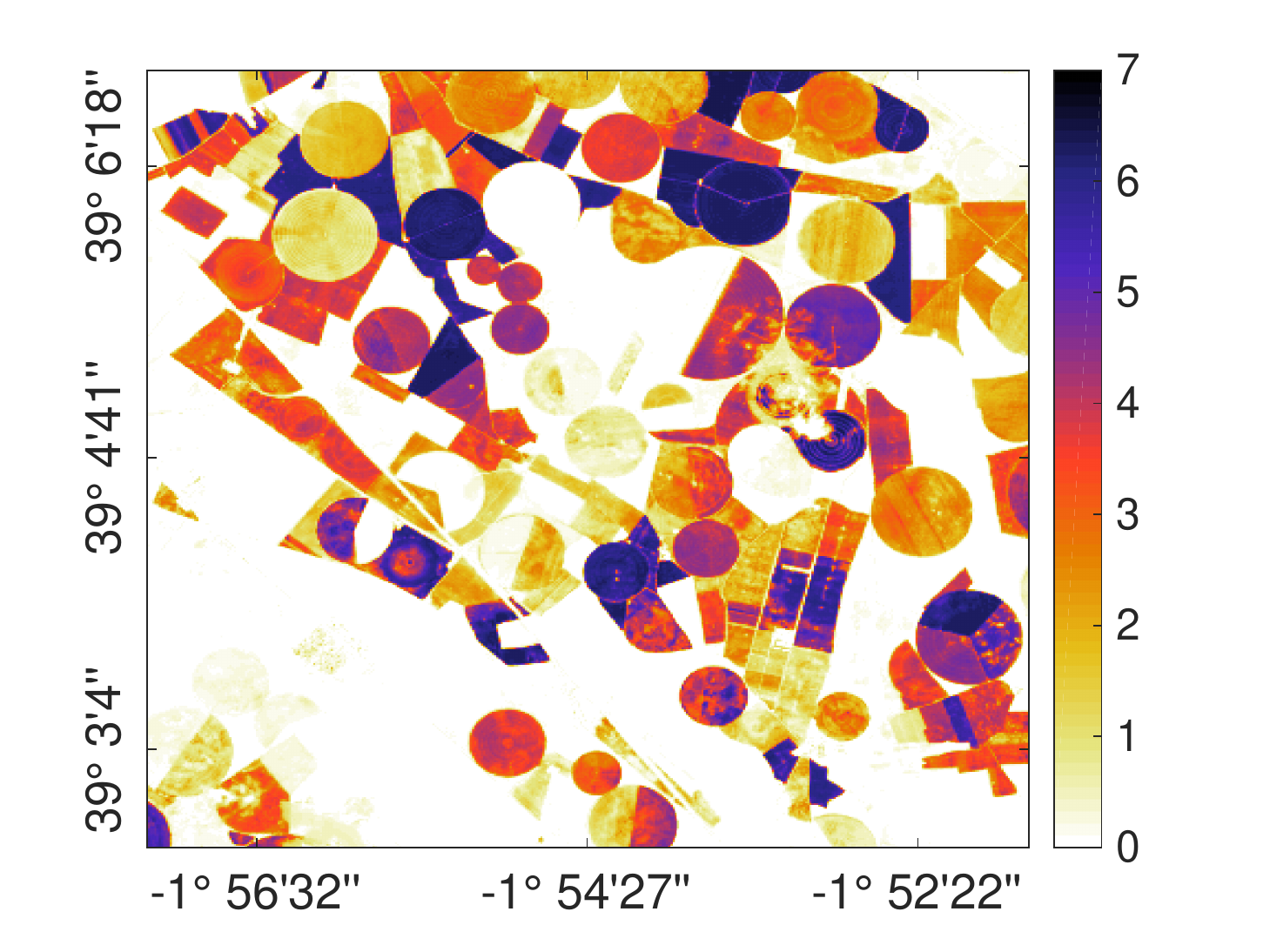}} &
\raisebox{-0.5\height}{\IG[height=3cm,trim={0.9cm 0.2cm 1.1cm 0.5cm},clip]{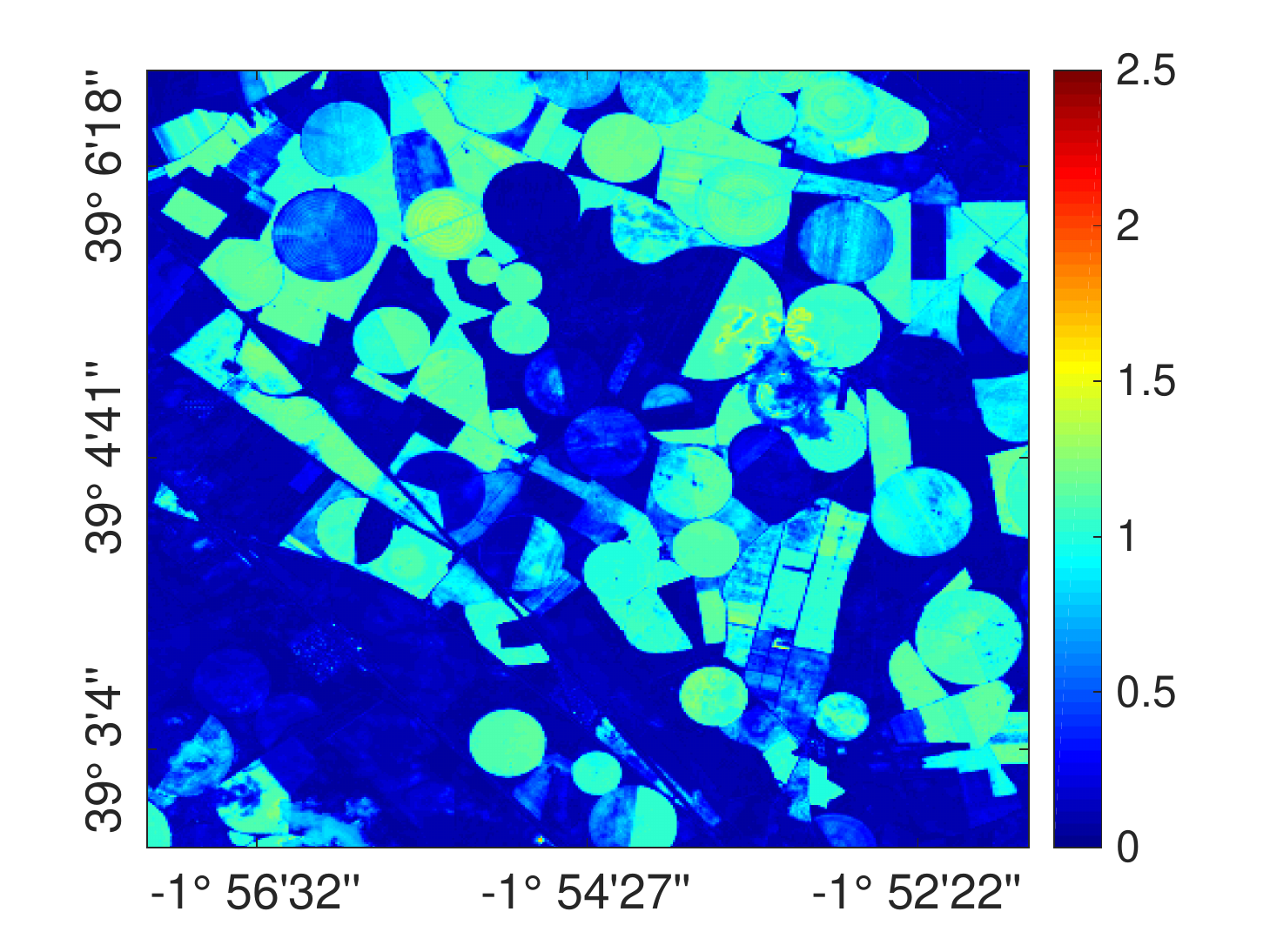}} &
\raisebox{-0.5\height}{\IG[height=3cm,trim={0.9cm 0.2cm 0.9cm 0.5cm},clip]{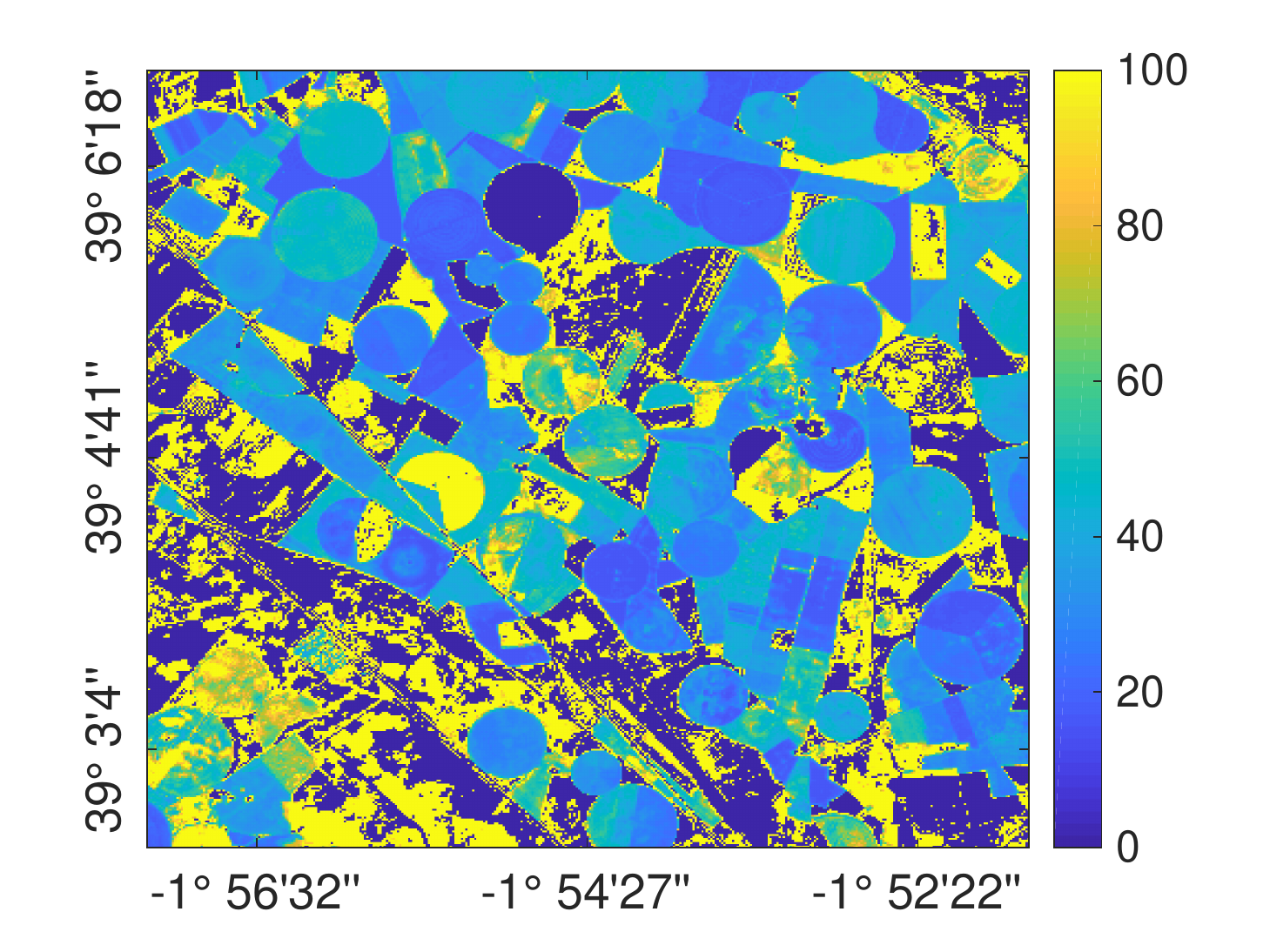}} \\
\rotatebox[origin=c]{90}{Scatterplot} & 
\raisebox{-0.5\height}{\IG[height=3.5cm,trim={1.0cm 0cm 3.0cm 1cm},clip]{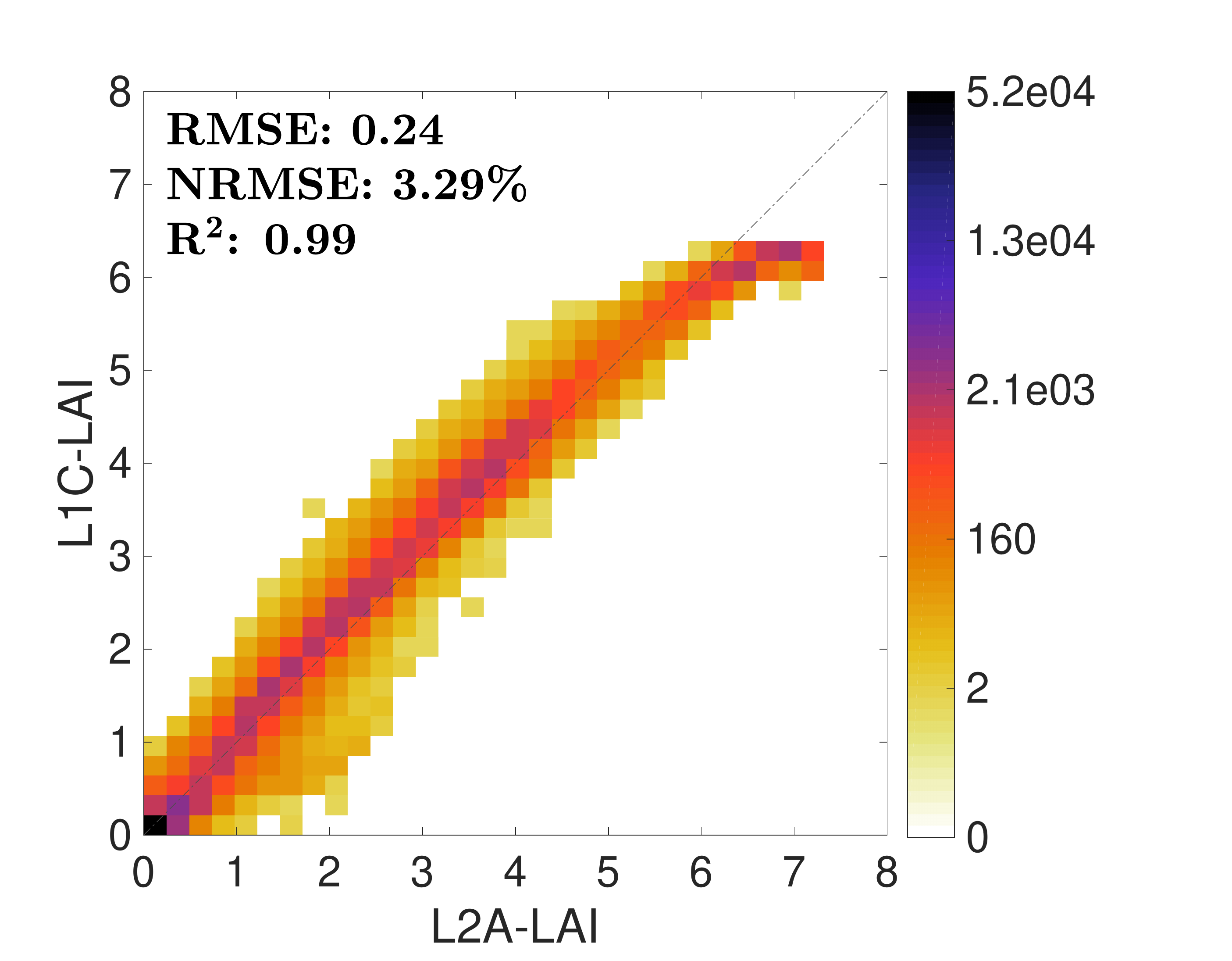}} &
\raisebox{-0.5\height}{\IG[height=3.5cm,trim={0.3cm 0cm 3.4cm 1cm},clip]{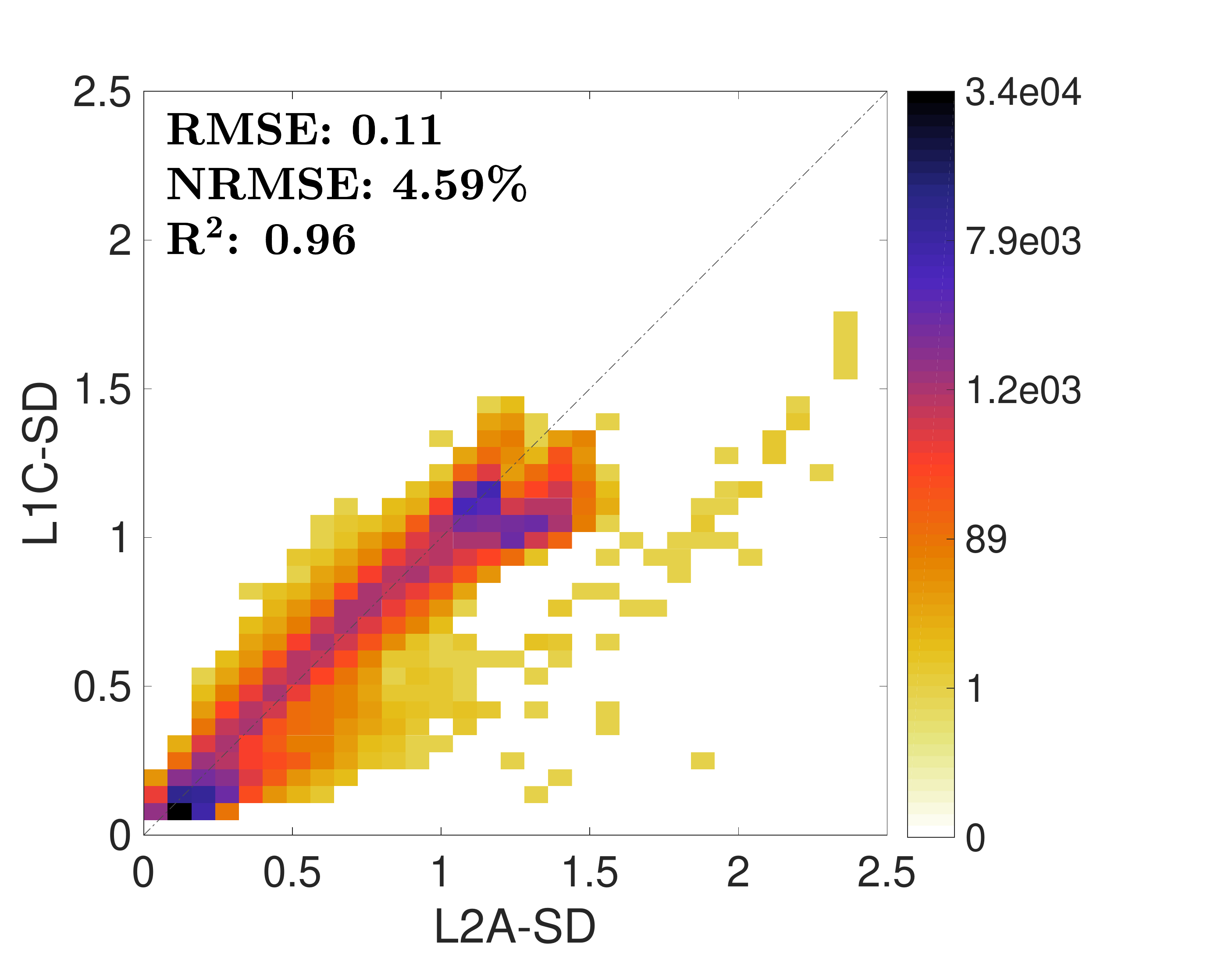}} &
\raisebox{-0.5\height}{\IG[height=3.5cm,trim={0.1cm 0cm 3.6cm 1cm},clip]{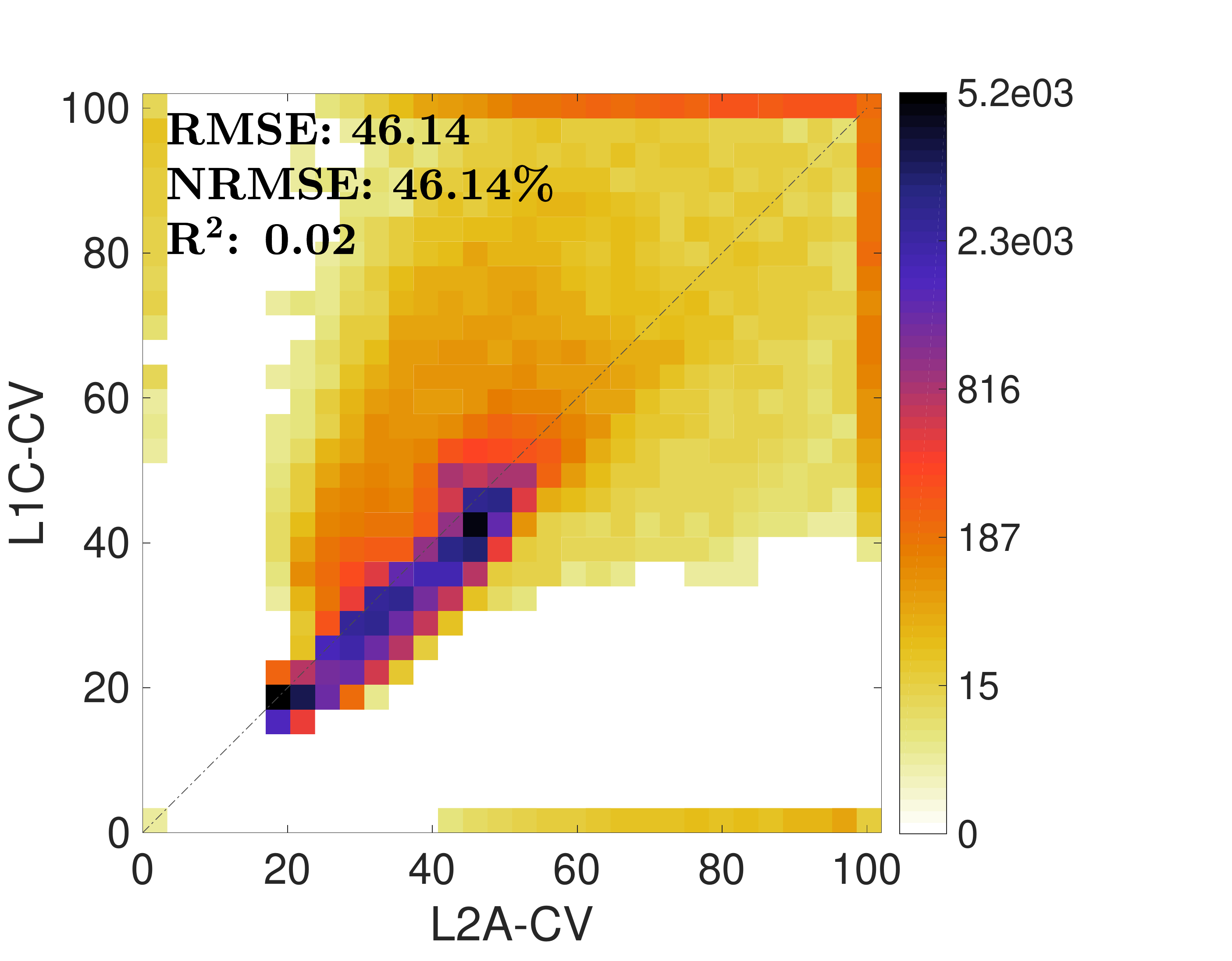}} \\
\end{tabular} 
\caption{LAI map (mean estimates; $\mu$) (left), associated uncertainties (expressed as standard deviation (SD) around the $\mu$) (center), and relative uncertainties (expressed as coefficient of variation (CV = SD/$\mu$ $\times$  100 in \%) (right) as generated by VHGPR algorithm from L2A ({\bf Top}) and L1C ({\bf Middle}) data for Barrax test site. Scatter plots of both maps with gridded color density ({\bf Bottom}). In case of \%CV a maximum of 100\% is set. } 
\label{mapsTOATOC_Barrax}
\end{figure}

Because of the adequate validation of the LAI retrieval algorithms at the S2 BOA and TOA scale, the subsequent step is applying the VHGPR algorithms to the S2 subsets for mapping. The obtained LAI maps as generated for both BOA and TOA data products of the two subsets are shown in Figure \ref{mapsTOATOC_Marchfeld} (Marchfeld) and  Figure \ref{mapsTOATOC_Barrax} (Barrax).

For both test sites, the obtained maps are realistic and represent well the spatial variation of the surface. 
Regarding Marchfeld (Figure \ref{mapsTOATOC_Marchfeld}), a clear distinction can be made between green fields with high LAI and fields with crops that are either senescent or where crop have been harvested with LAI close to zero.  
Regarding Barrax, the irrigated circular agricultural fields can be identified on the maps with their within-field variation. On the other hand, non-vegetated areas are estimated with LAI values close to zero.
Wen comparing the LAI maps extracted at BOA and TOA scale, the spatial distribution and the LAI range appear alike, which suggests the possibility of retrieving LAI directly from TOA radiance data. Probably the high correlation between BOA and TOA  can be better appreciated in the scatter plot for Marchfeld site ($R^2$ = 0.95)  in the bottom of Figure \ref{mapsTOATOC_Marchfeld} and in the scatter plot  for Barrax site ($R^2$ = 0.99) in the bottom of Figure \ref{mapsTOATOC_Barrax}. The similarity between BOA and TOA maps suggests that LAI can be retrieved directly from TOA radiance data, i.e. without the need of an atmospheric correction.

As an advantage of the GPR and VHGPR models, because of developed in a Bayesian framework, apart from the LAI estimates also associated uncertainty maps are provided. Two uncertainty outputs are calculated: absolute uncertainties expressed as standard deviation (SD) around the mean estimate and relative uncertainties (\%CV= SD/mean estimate $\times$ 100). These maps provide additional information about the performance of the retrieval models on a per-pixel basis. Accordingly, the consistency across the BOA and TOA scales can be inspected.

In the associated uncertainties maps, low SD values over vegetated areas indicate high certainties, while high values (light blue and yellow) indicate less certainties. Low values over non-vegetated areas appear because of the close-to-zero LAI values for bare soils or senescent fields. A few areas with high uncertainties can be seen in the maps in red (especially for the Barrax site), belonged to surfaces not included in the LUTs, e.g. man-made surfaces. 
The uncertainty maps can be useful to reveal areas that need more representativeness in the training data set \cite{Verrelst2015}. Generally the BOA and TOA  uncertainty maps are consistent. The BOA map showed some more areas with slightly higher SD values than TOA map, i.e. higher uncertainties. This was expected for BOA due to the errors involved in its complex pre-processing step, as we explained earlier in section \ref{sec:intro}. 
The scatter plot of BOA vs TOA confirms again the consistency between both products  ($R^2$ of 0.91 for Marchfeld and $R^2$ of 0.96 for Barrax). Although not shown, the SD uncertainties obtained from the original GPR model were systematically higher, which confirms that VHGPR leads to higher quality maps as opposed to GPR.

While the SD map is related to the magnitude of LAI, the relative uncertainty as calculated by the coefficient of variation (CV) is probably easier to interpret, as it is a relative estimate expressed in percentage. When inspecting more closely this map for Marchfeld, it can be observed that the BOA map led substantial more parcels with low uncertainties than the TOA map. Nevertheless, the regions with high uncertainties are over parcels with close to zero LAI. The same occurs over Barrax: the irrigated vegetated areas are retrieved with low uncertainties, while the bare soils are retrieved with high uncertainties. This is the result of both very low LAI estimates, which is correct, but they are associated with absolute uncertainties of 1 or higher, as such resulting in high uncertainties. Hence, it is not that the estimates are out-of-range, it is rather that the estimates are accompanied with high uncertainties. 
Typically, adding more bare soil estimates to the training data, accounting for local variability, would further reduce the uncertainties.
The regions with out-of-range values cause to break down the correlation of the scatter plot. Yet, the large majority of pixels fell precisely on the 1:1-line, again suggesting the consistency of both maps. 
As a final remark, although now shown, the CV map as obtained from the GPR model led to a substantial more areas with high values.

\subsection{Comparison with LAI maps obtained from SNAP NN algorithm} \label{comparing_results}

\begin{figure*}[!t]
\centering
\small
\begin{tabular}{cccc}
& SNAP NN & VHGPR vs SNAP NN & VHGPR vs SNAP NN\\
& LAI (m$^{2}$/m$^{2}$) & LAI (m$^{2}$/m$^{2}$) & Relative error (\%)\\
\rotatebox[origin=c]{90}{L2A (BOA)} & 
\raisebox{-0.5\height}{\IG[height=3cm,trim={0.9cm 0.2cm 1.5cm 0.5cm},clip]{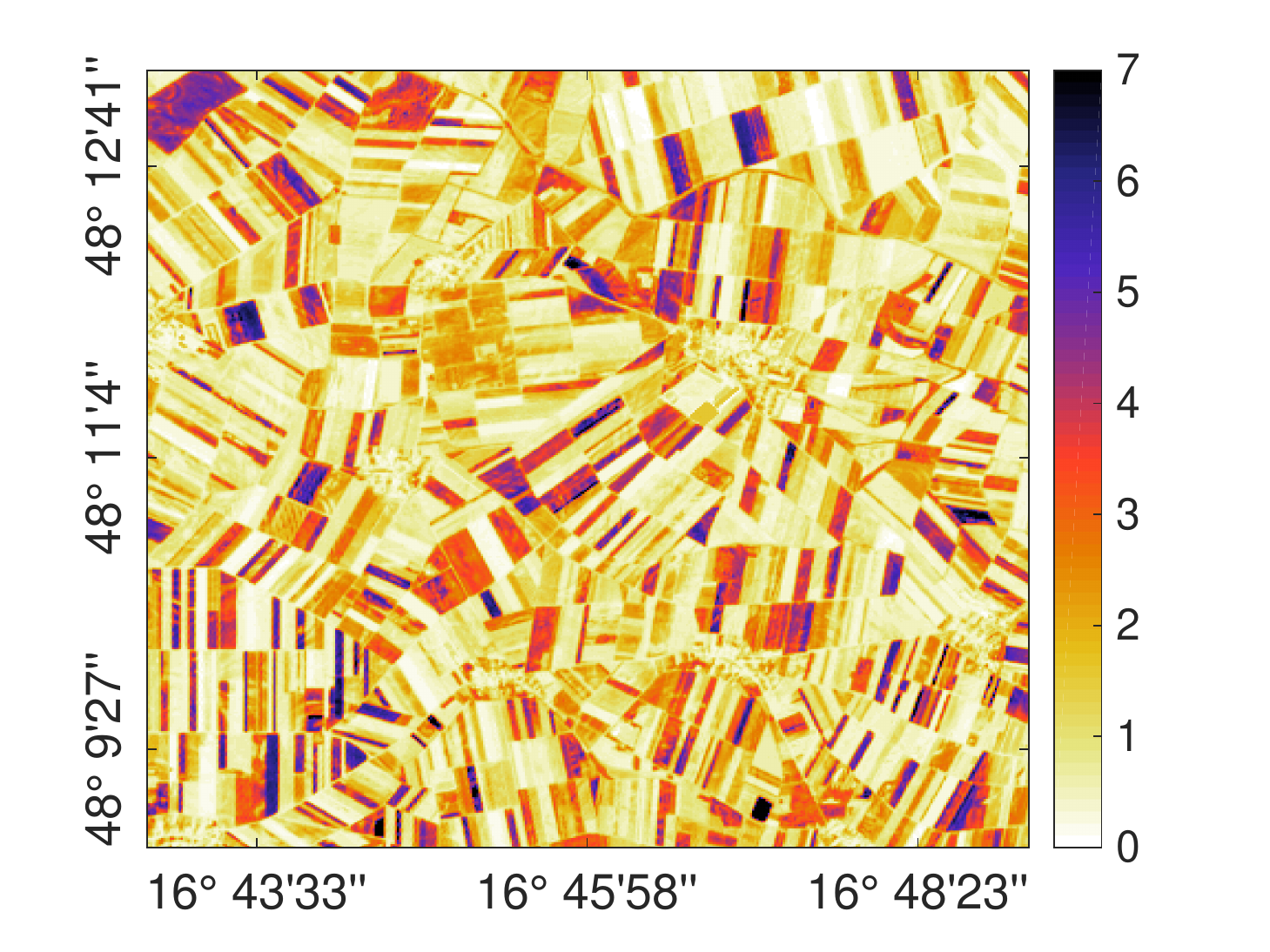}} &
\raisebox{-0.5\height}{\IG[height=3cm,trim={0.8cm 0cm 3.5cm 1.4cm},clip]{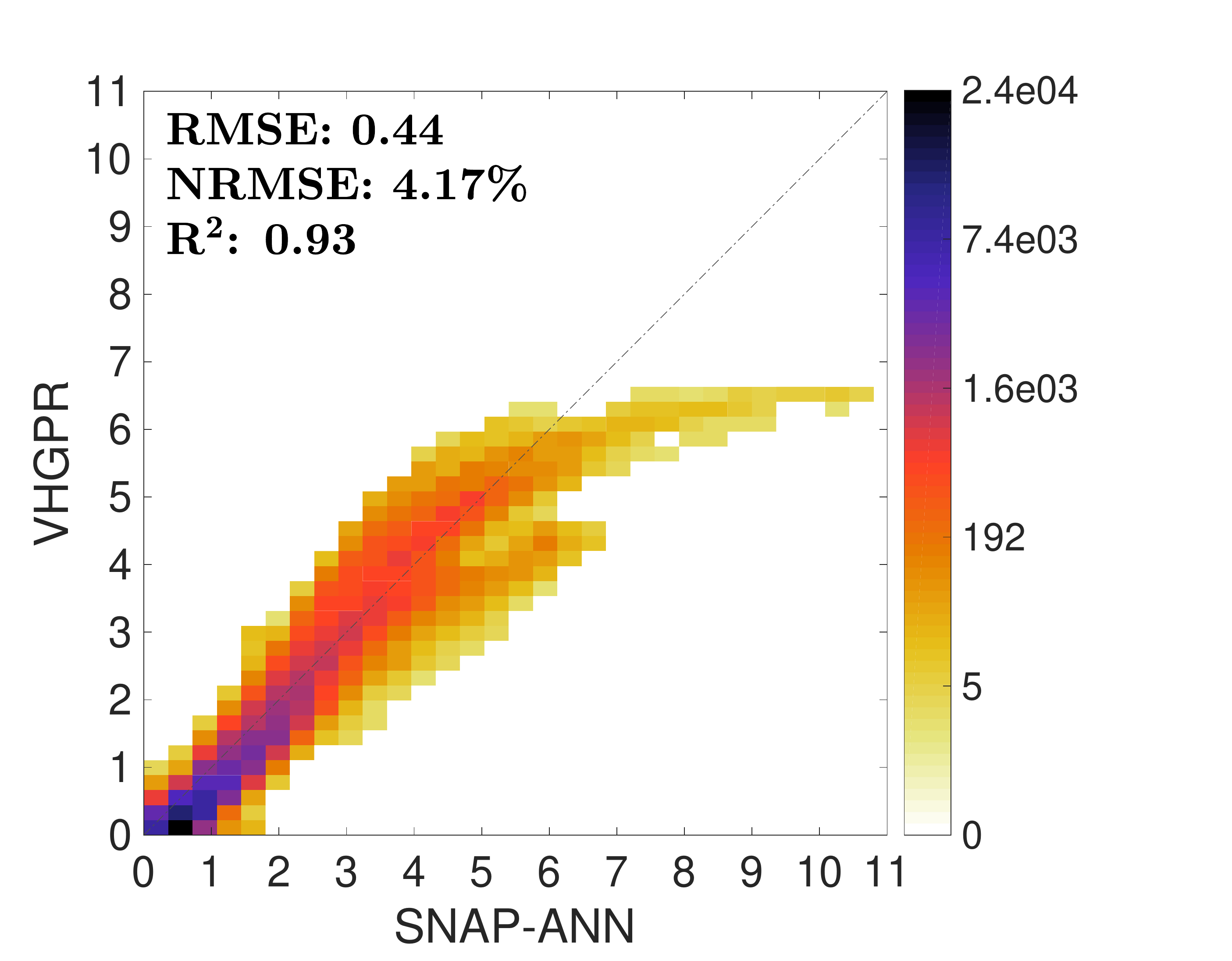}} &
\raisebox{-0.5\height}{\IG[height=3cm,trim={0.9cm 0.2cm 0.8cm 0.7cm},clip]{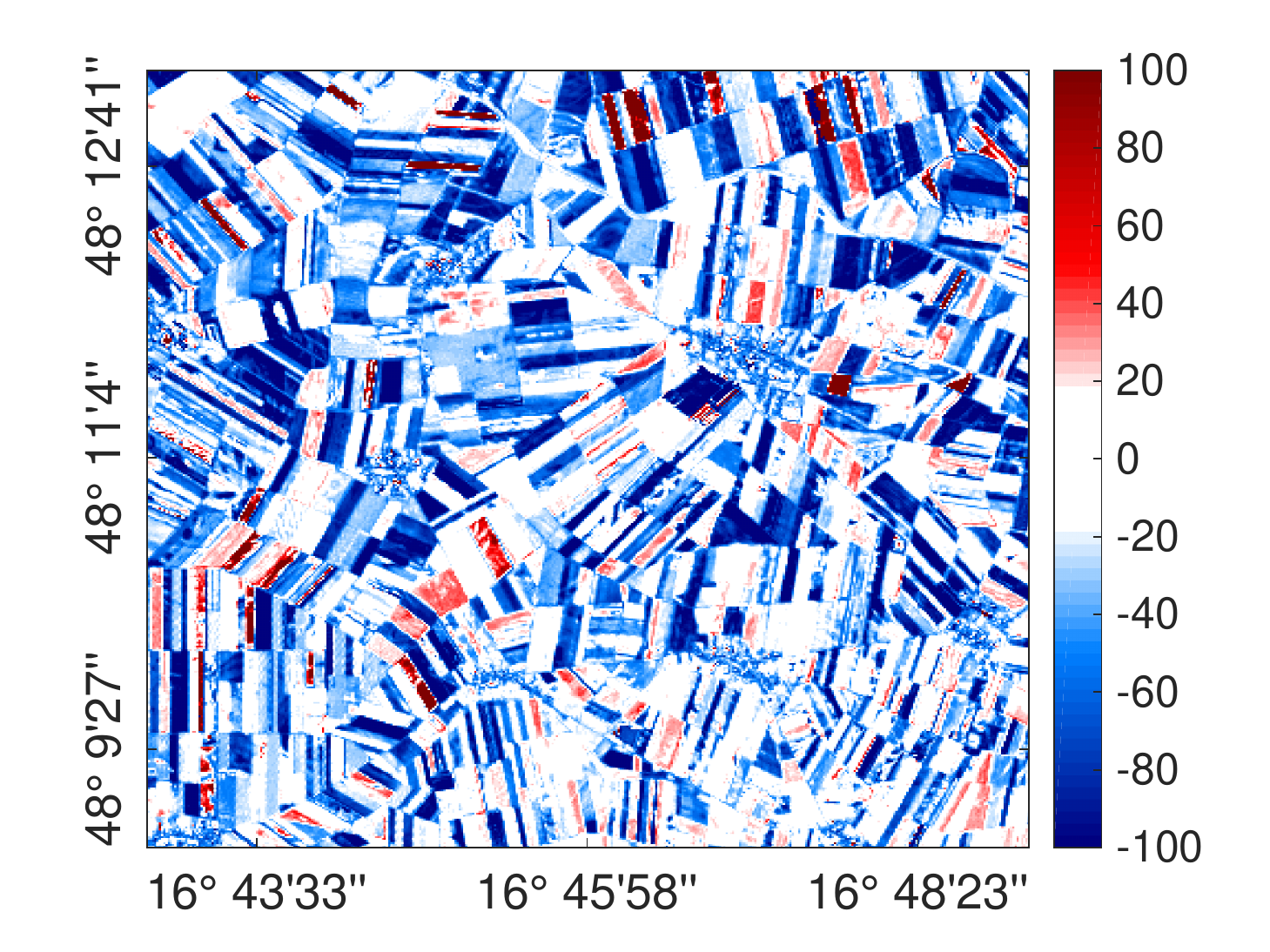}} \\
\rotatebox[origin=c]{90}{L1C (TOA)} & 
\raisebox{-0.5\height}{\IG[height=3cm,trim={0.9cm 0.2cm 1.5cm 0.5cm},clip]{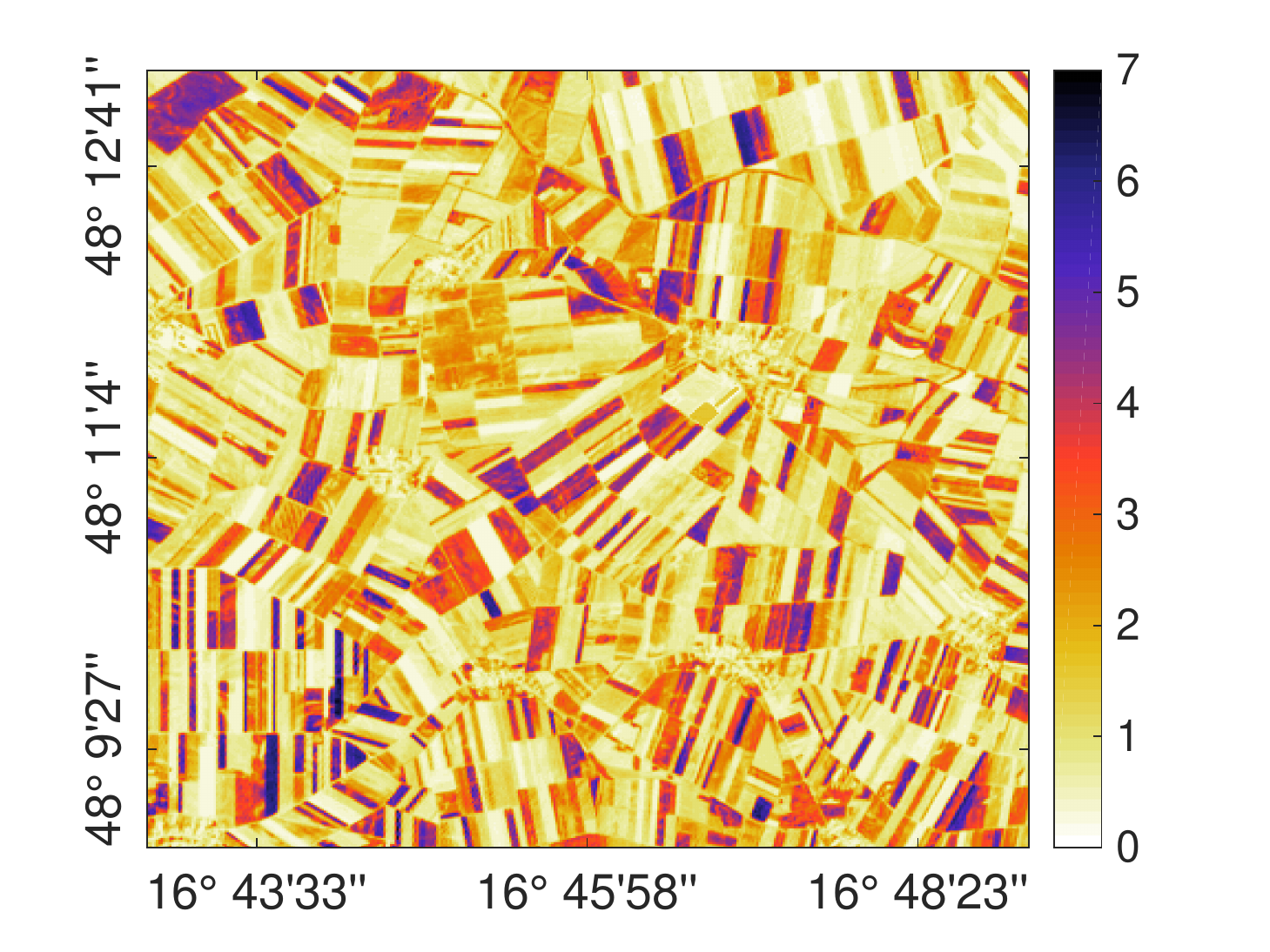}} &
\raisebox{-0.5\height}{\IG[height=3cm,trim={0.8cm 0cm 3.5cm 1.6cm},clip]{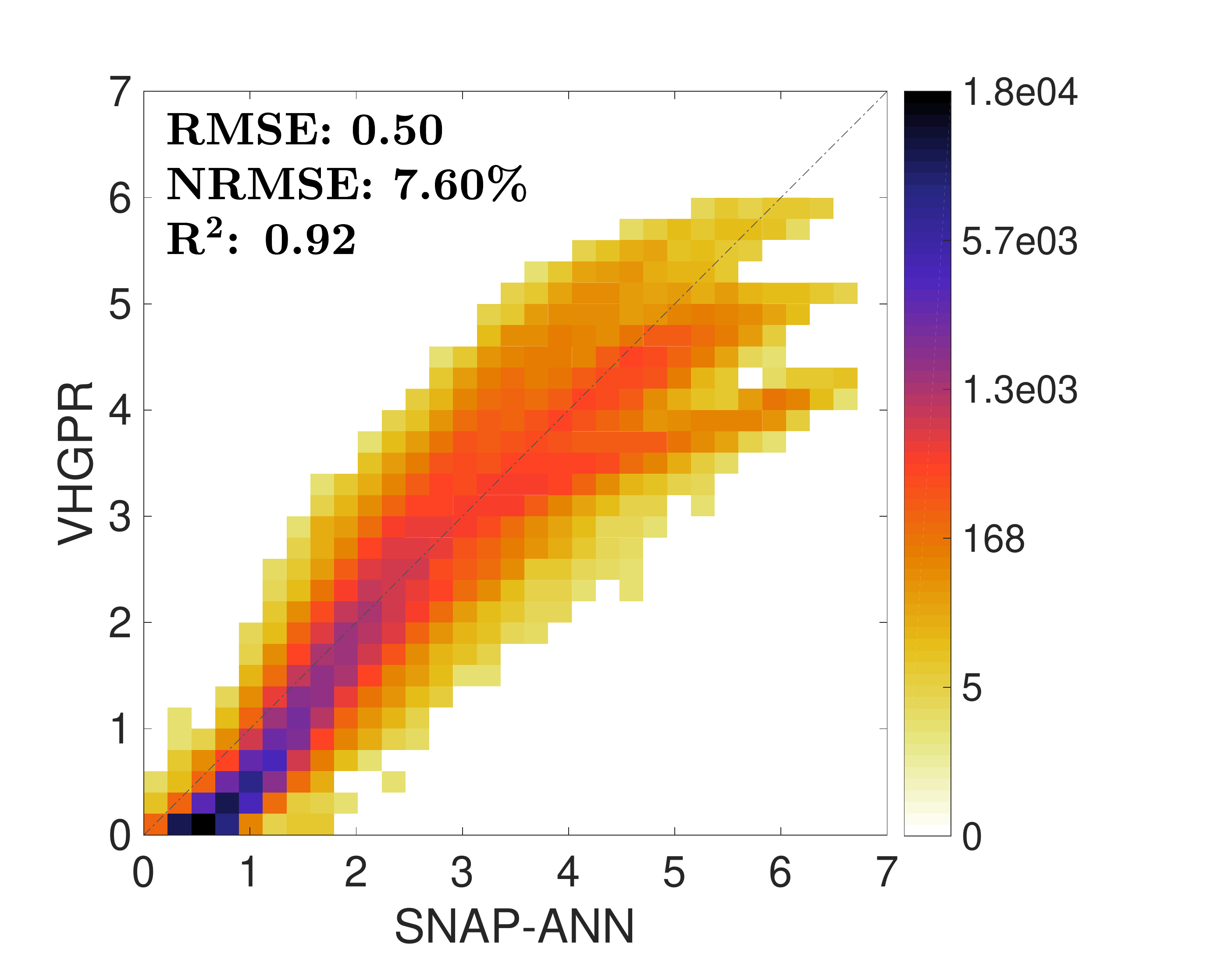}} &
\raisebox{-0.5\height}{\IG[height=3cm,trim={0.9cm 0.2cm 0.8cm 0.7cm},clip]{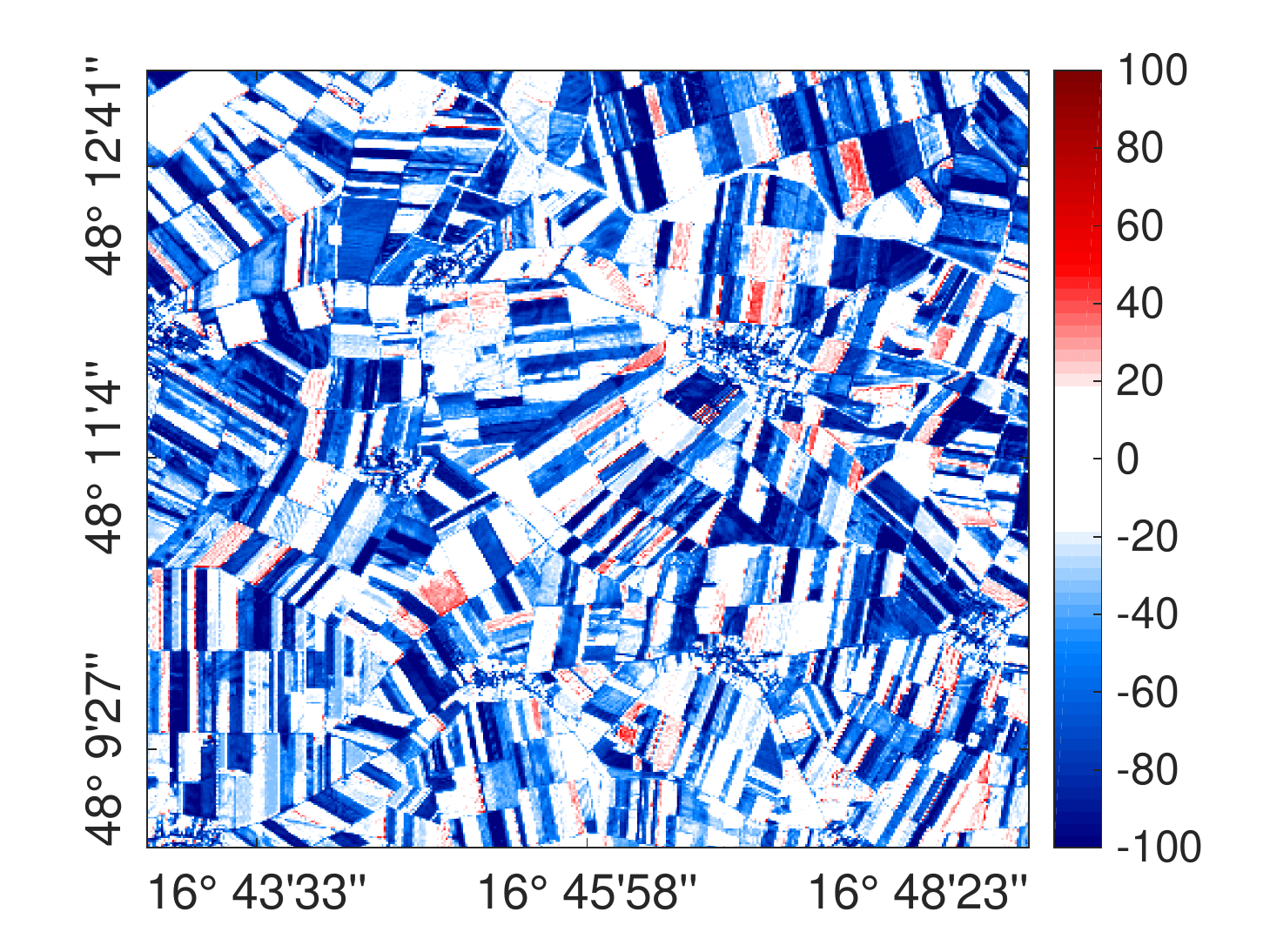}} \\
\end{tabular}
\caption{LAI map obtained with SNAP (left), scatter plot (center) and relative error map (right) between VHGPR (Figure \ref{mapsTOATOC_Marchfeld}) and SNAP NN estimations  from L2A (BOA) data ({\bf Top}) and L1C (TOA) data ({\bf Bottom}) for Marchfeld test site. For visualization purpose, the LAI color bar was limited to a maximum of 7.}
\label{SNAP_figures_Marchfeld}
\end{figure*}

\begin{figure}[!t]
\centering
\small
\begin{tabular}{cccc}
& SNAP NN & VHGPR vs SNAP NN & VHGPR vs SNAP NN\\
& LAI (m$^{2}$/m$^{2}$) & LAI (m$^{2}$/m$^{2}$) & Relative error (\%)\\
\rotatebox[origin=c]{90}{L2A (BOA)} & 
\raisebox{-0.5\height}{\IG[height=3cm,trim={0.9cm 0.2cm 1.5cm 0.5cm},clip]{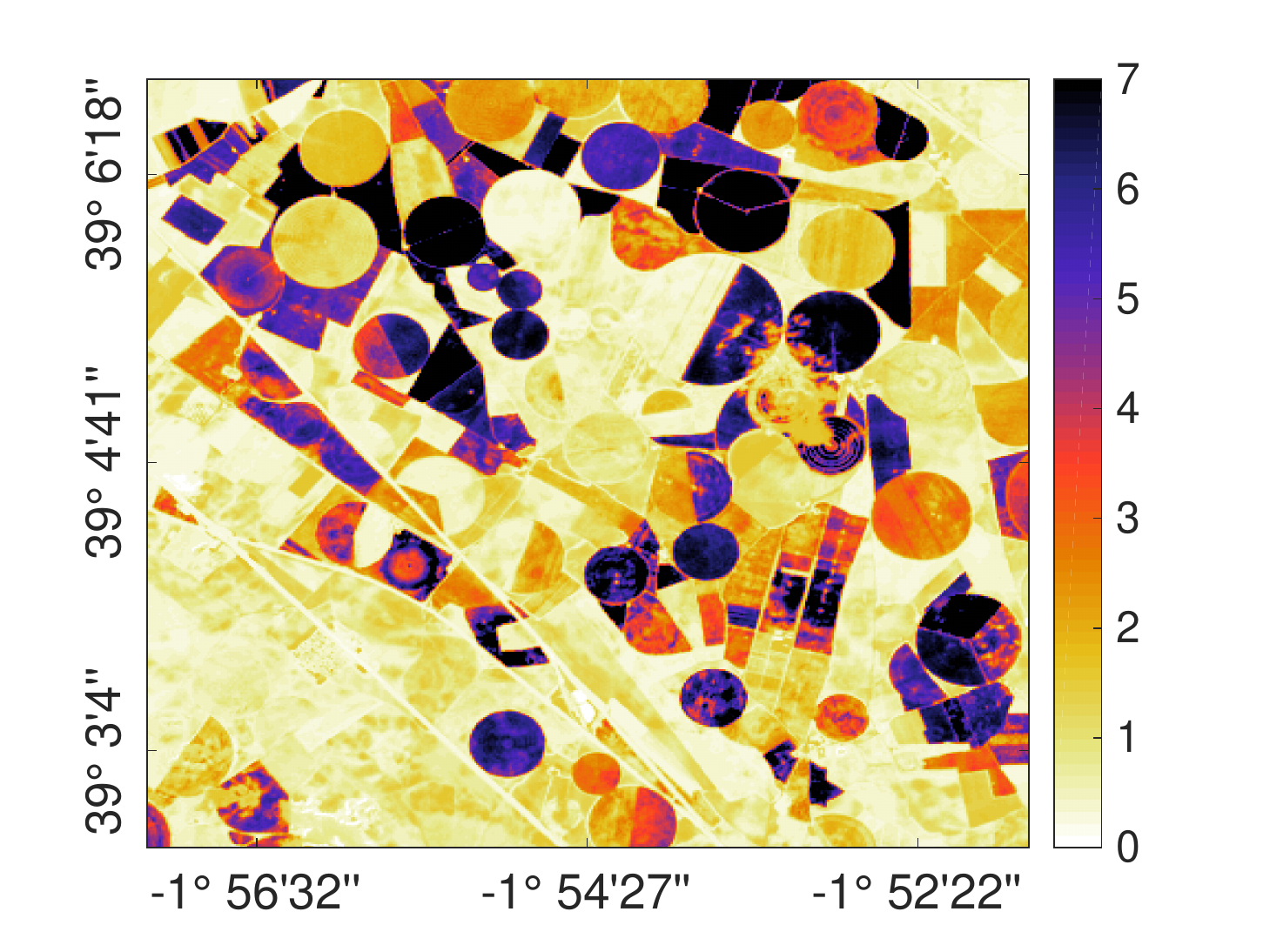}} &
\raisebox{-0.5\height}{\IG[height=3cm,trim={0.8cm 0cm 3.5cm 1.4cm},clip]{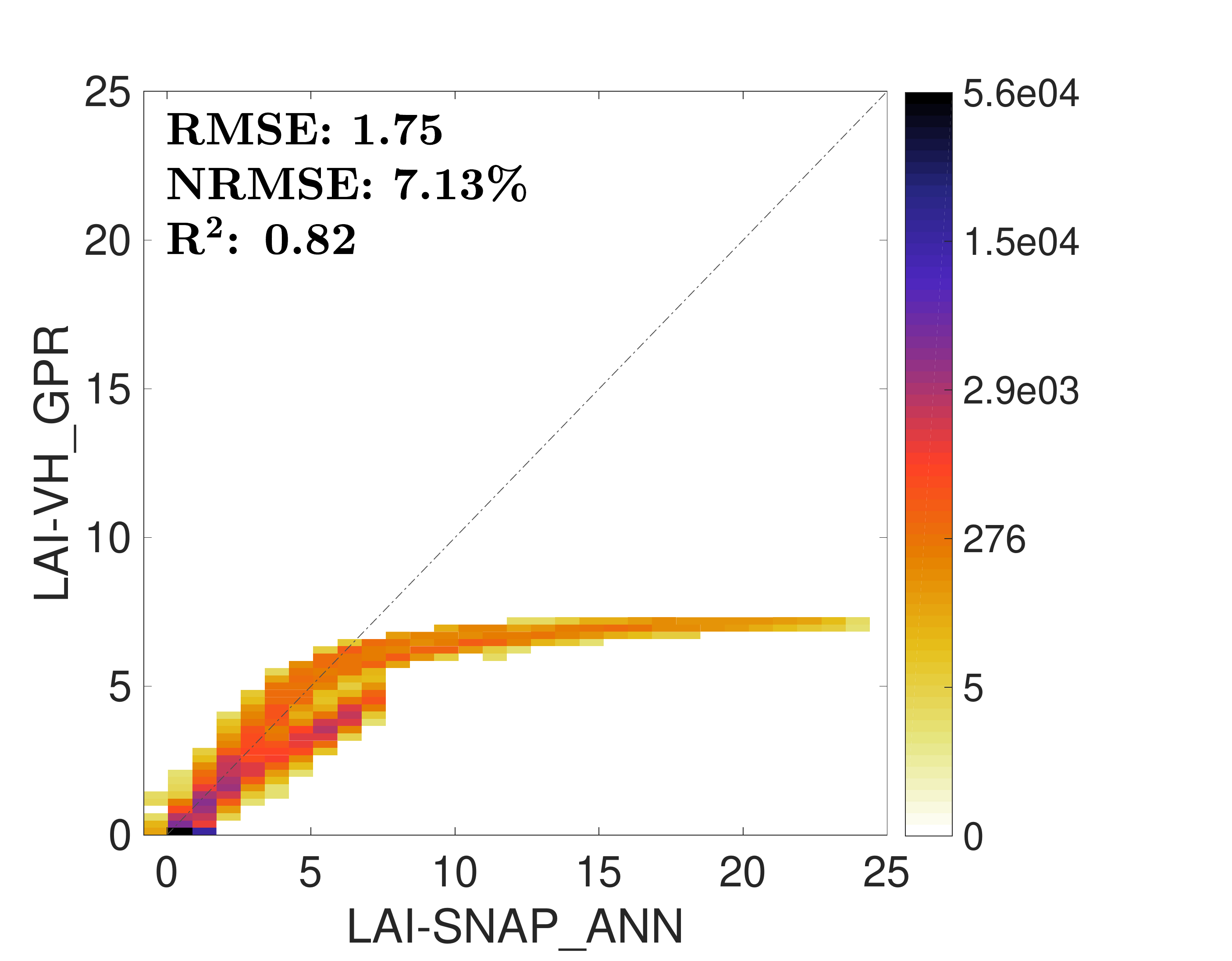}} &
\raisebox{-0.5\height}{\IG[height=3cm,trim={0.9cm 0.2cm 0.8cm 0.7cm},clip]{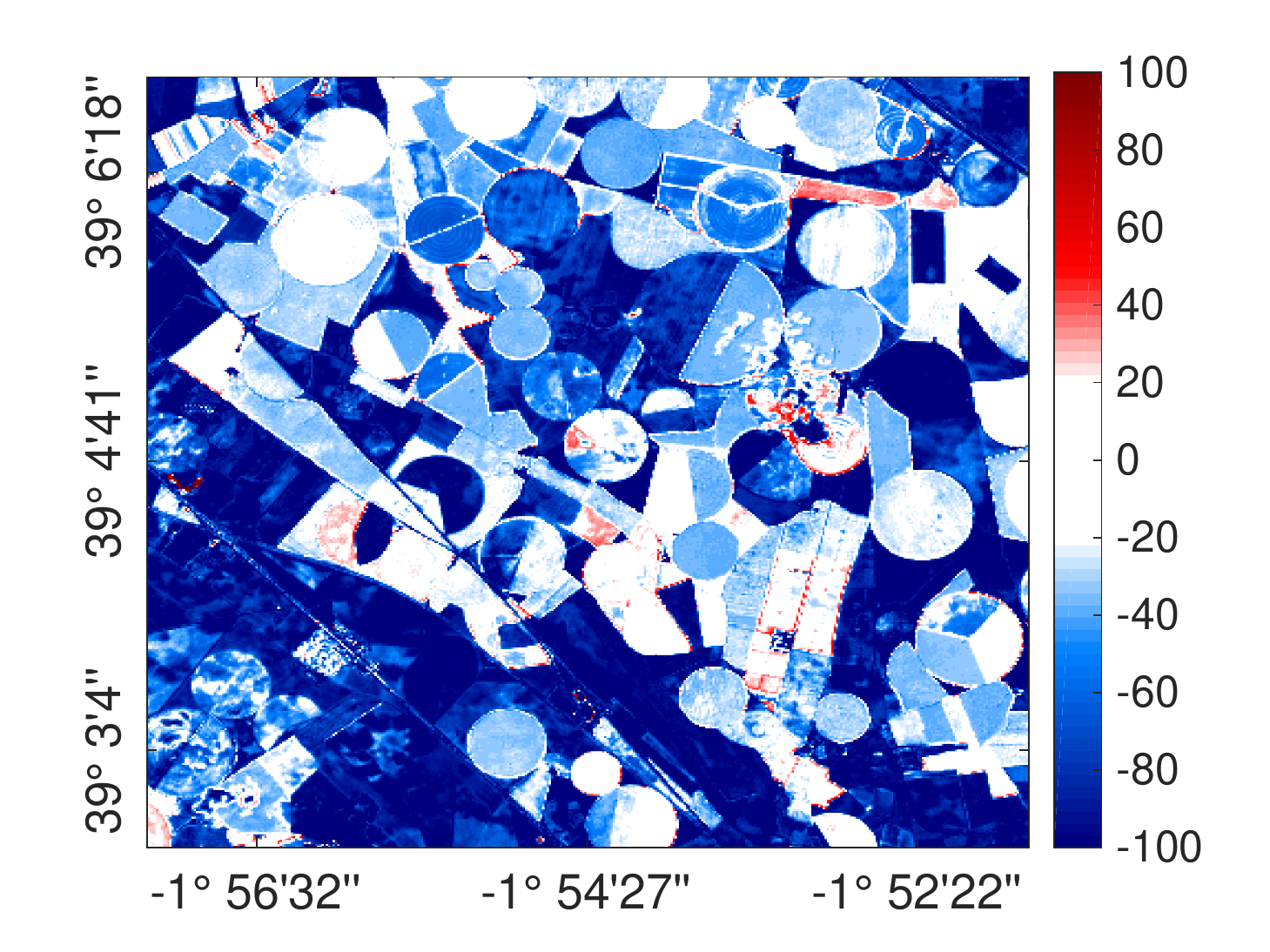}} \\
\rotatebox[origin=c]{90}{L1C (TOA)} & 
\raisebox{-0.5\height}{\IG[height=3cm,trim={0.9cm 0.2cm 1.5cm 0.5cm},clip]{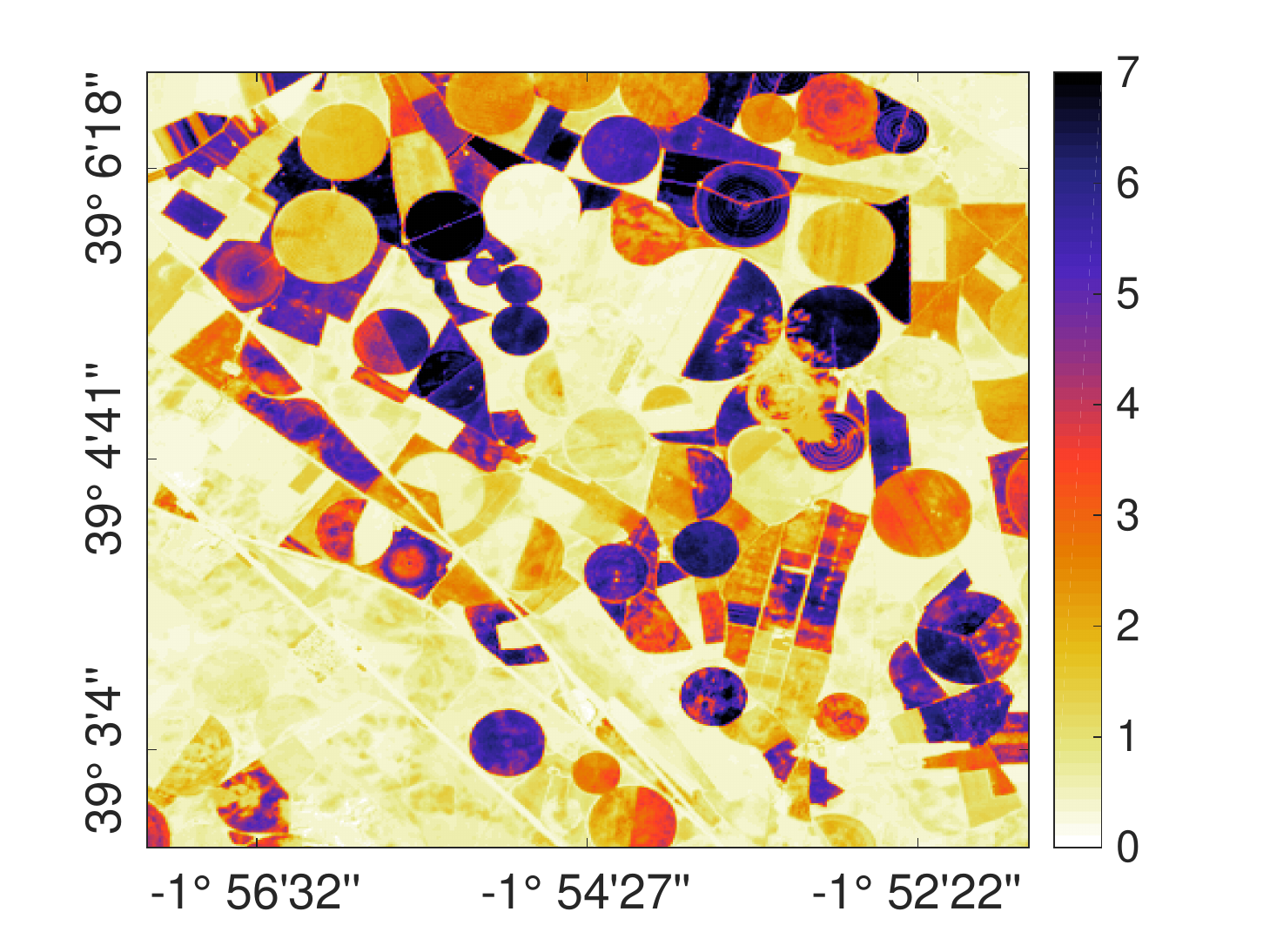}} &
\raisebox{-0.5\height}{\IG[height=3cm,trim={0.8cm 0cm 3.5cm 1.6cm},clip]{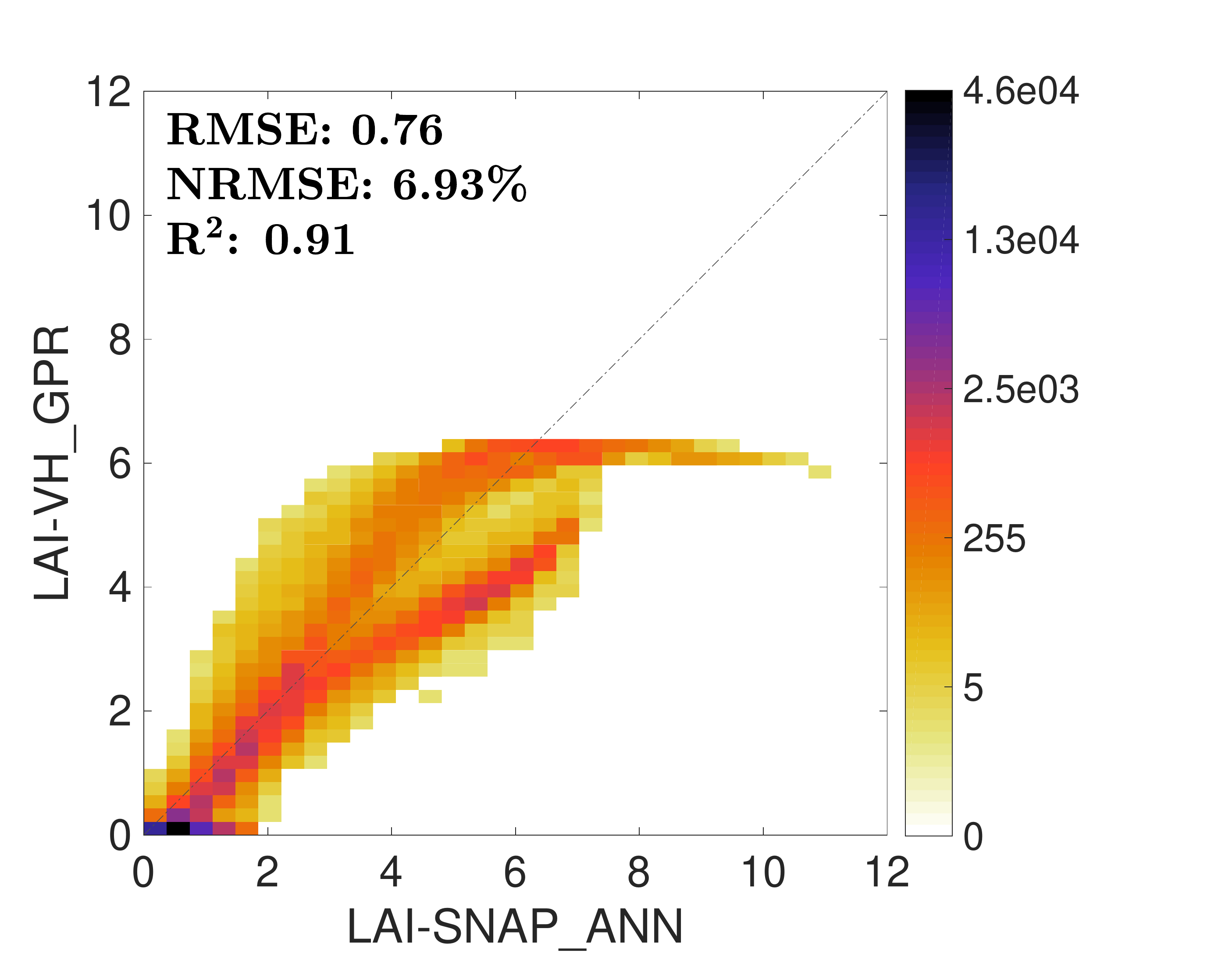}} &
\raisebox{-0.5\height}{\IG[height=3cm,trim={0.9cm 0.2cm 0.8cm 0.7cm},clip]{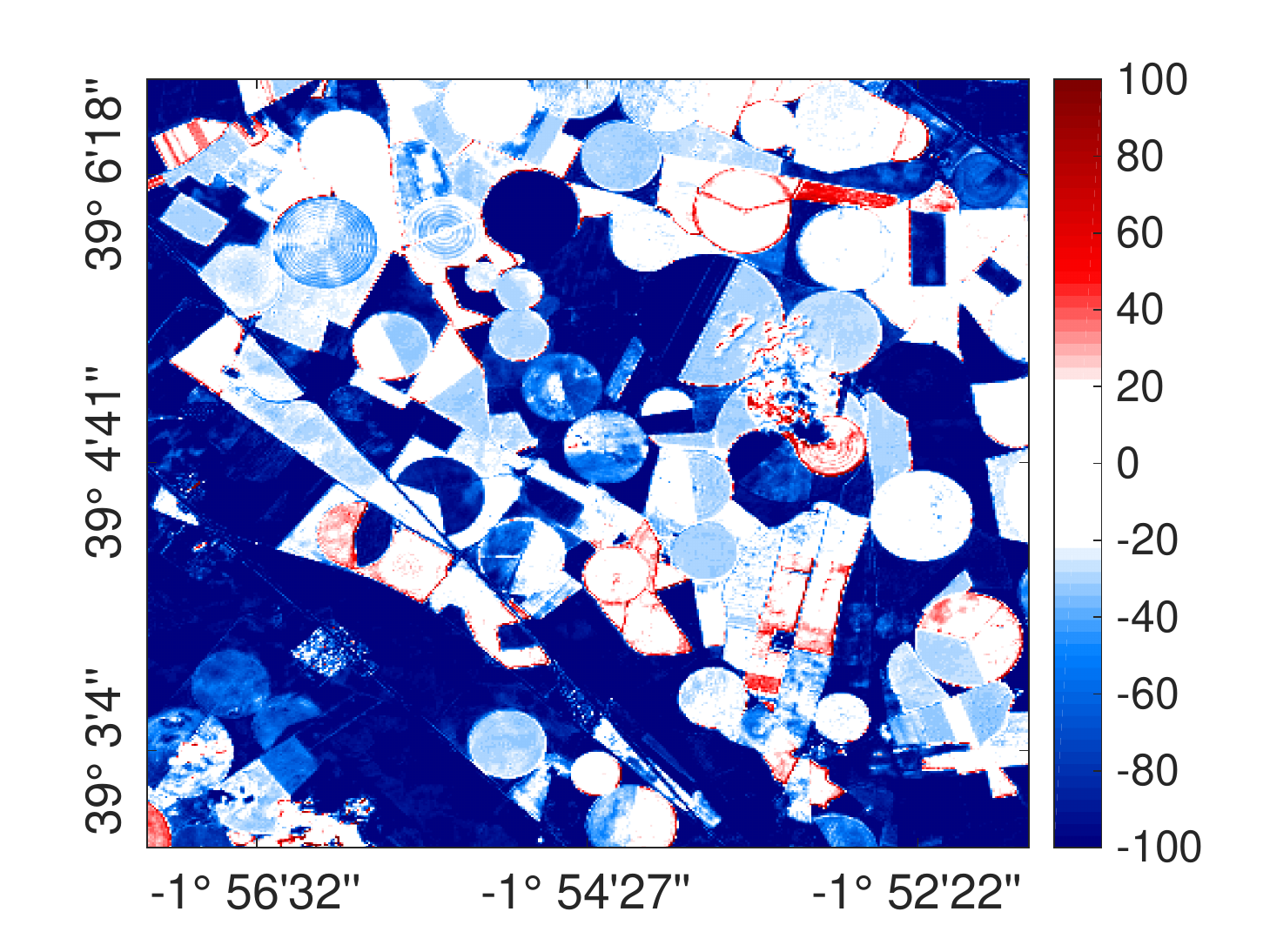}} \\
\end{tabular}
\caption{LAI map obtained with SNAP (left), scatter plot (center) and relative error map (right) between VHGPR (Figure \ref{mapsTOATOC_Barrax}) and SNAP NN estimations  from L2A (BOA) data ({\bf Top}) and L1C (TOA) data ({\bf Bottom}) for Barrax test site. For visualization purpose, the LAI color bar was limited to a maximum of 7.} 
\label{SNAP_figures_Barrax}
\end{figure}

A final step involves comparison against the official LAI product as can be obtained from the ESA's SNAP toolbox. Hence, the SNAP biophysical processor, based on an NN algorithm, was used to generate the LAI maps for both BOA and TOA scale over the Marchfeld and Barrax sites. The obtained LAI maps are provided in Figure \ref{SNAP_figures_Marchfeld} (Marchfeld) and Figure \ref{SNAP_figures_Barrax} (Barrax). 
At a glance, the SNAP map looks similar to the above map as generated by VHGPR; the same patterns are obtained, and LAI estimations seem alike. However, when comparing the BOA map against VHGPR in a scatter plot, it can be observed that SNAP clearly overestimates some vegetated parcels. This is especially the case for Barrax, with unrealistic LAI values up to above 20. See also the scatter plots. Also when the LAI maps are obtained from at sensor radiance (TOA level) the maps look alike. Interestingly, at this level the SNAP NN algorithm does not suffer from extreme overestimation. Despite the overestimation of NN at BOA scale, it should be mentioned that the SNAP biophysical processor also generates quality indicators per pixel. The indicators show when the input reflectances are outside the training definition domain. Also the indicator is flagged when the output value is outside the LAI range, defined by SNAP with a maximum, being 8, a minimum and a tolerance value. Hence, when the estimation is beyond 8, it is flagged as out-of-range.

Probably an easier way to interpret the differences between the SNAP NN and VHGPR algorithms may be by mapping the relative errors. For both BOA and TOA relative error maps are shown in Figure \ref{SNAP_figures_Marchfeld} (Marchfeld) and Figure \ref{SNAP_figures_Barrax} (Barrax). The white areas indicate no change within a 20\% difference. Dark blue and red shades indicate large relative differences. For both the Marchfeld and Barrax site, blue clolors (i.e., underestimation) predominate.  These parcels appear over low LAI values, i.e., parcels with influence of bare soil. When comparing both maps, it appears that the SNAP NN algorithm is not reaching to zero values over senescent or non-vegetated surfaces. It suggests that the NN algorithm is not optimized to estimate 0 values in case vegetation is absent or no longer green.

\section{Discussion}\label{sec:discussion}

Building on the hybrid retrieval processing chain earlier proposed in \cite{Verrelst2019TOC2TOA}, in this work we developed hybrid LAI retrieval algorithms applicable to S2 BOA (L2A) and TOA (L1C) data. These machine learning algorithms can then be applied to S2 images for the operational production of LAI maps. In this context, while demonstrating that the TOA retrieval approach can be developed using a hybrid method based on coupled vegetation-atmosphere RTMs and novel MLRAs is one thing, what was left to be done was assessing  the robustness and the maturity of the retrieval models. 
This was done here as follows. First the retrieval algorithms were validated against ground measurements, then the mapping performances were tested over two European agricultural sites, and finally the maps were compared against the official S2 LAI product.
The obtained validation results and consistency of the generated maps confirmed that biophysical variables such as LAI can be meaningfully retrieved directly from at-sensor radiance data. Here we discuss the multiple aspects of the retrieval algorithm, starting with the chosen variable. 

The reason to focus on LAI retrieval from TOA radiance data was motivated by an earlier conducted leaf-canopy-atmosphere global sensitivity analysis (GSA) \cite{Verrelst2019TOC2TOA}. According to those GSA results, LAI is the most dominant variable that drives the variability of TOA radiance along the visible-SWIR spectral range, followed by leaf chlorophyll and water content  \cite{Verrelst2019TOC2TOA}. 
The good validation results obtained at the TOA scale were to be expected; the TOA-based GSA results indicated that -- outside the water absorption regions and the blue region -- the vegetation variables drive TOA radiance much more than atmospheric variables  \cite{Verrelst2019TOC2TOA}.
Moreover, the used S2 bands are conveniently located in the spectral regions where LAI plays a dominant role, especially in the red and SWIR bands (B11 at 1610 nm and B12 at 2190 nm). The S2 band settings exploit efficiently the spectral information because according to the GSA results it is in these two SWIR bands where LAI is mostly driving, with about 80\% of the total sensitivity ($S_{Ti}$) \cite{Verrelst2019TOC2TOA}. At the same time, the influence of the atmosphere in the visible part can introduce some error in both the TOA and BOA retrieval. While in principle at the BOA scale atmospheric correction takes care of it, in TOA retrieval, the idea is that the influence of the atmosphere in the visible is accounted for directly in the retrieval algorithm. 

In an attempt to figure out the role of S2 band positions, we revised the used S2 bands with respect to LAI retrieval performance. Related studies used similar S2 band settings in LAI algorithm development but with one or two bands less. For instance \cite{Verrelst2019TOC2TOA} used 9 bands excluding B8, and \cite{upreti2019} used 8 bands by excluding B2 and B8. To clarify the contribution of the bands, we conducted similar tests using 8, 9 and 10 bands in the LAI retrieval algorithms. Validation results showed insignificant differences in the retrieval performance (results not shown), which suggests that using one band more or less will not impact the results. 
Another aspect is that our validation results at BOA level conducted with GPR (RMSE: 0.70) and VHGPR (RMSE: 0.63) are consistent with the related work of \cite{upreti2019}, who found a similar performance using least-squares linear regression (RMSE: 0.68) and GPR with active learning (RMSE: 1.31).
In comparison to the latter case, we obtained superior validation results without implementing an active learning strategy. At the same time, the reason for a good validation may as well lie in the good quality of the validation data; with over 100 ground measurements covering multiple crop types it is rich dataset. 
It is also worth mentioning that we did not encounter related studies that provide LAI validation results at the TOA scale, implying that the conducted BOA vs TOA intercomparison is probably the more informative. To sum up, the consistent validation results and also the similar estimations for all crop types at each scale support that the pursued hybrid retrieval method works at both TOA and BOA scales.

Apart from the targeted variable and selected bands, a key factor of the hybrid algorithm is the specific machine learning algorithm used, i.e. GPR and VHGPR. Both flavors of GP models achieved consistent performances in LAI estimation with both BOA and TOA data. The small but systematic superiority by VHGPR over GPR can be explained by its more flexible nature accounting for signal-to-noise relations. Unlike GPR, VHGPR does not assume that the noise is independent of the signal. 
When observing the information of uncertainties offered by these methods, we can appreciate how VHGPR yields consistently lower uncertainties than GPR at low LAI values. 
We hypothesize that VHGPR adjusts better to the noise conditions at low levels of the variable due to its heteroscedastic characteristic.  
Although VHGPR takes somewhat longer training time than GPR (roughly twice as much, see~\cite{Lazaro-GredillaT11,GRSM_2016}), mapping run-time are on the same order for both algorithms, 
which suggests that training time should not be an obstacle to prefer VHGPR over GPR for achieving lower uncertainties and higher accuracy.

Having grip on the data and retrieval algorithm, the essence of this work was to explore how well LAI can be retrieved directly from TOA data.  
Although the obtained results at BOA and TOA scale are consistent, it is not escaping our attention that the TOA validation results slightly outperformed those of BOA. It can be argued that retrieving from BOA data is probably a more complex approach due to the multiple steps involved in the processing to convert the L1C product into L2A.  In addition to the uncertainties introduced by the hybrid LAI retrieval algorithm, the atmospheric correction carried out with the Sen2Cor procedure can introduce residual errors in the L2A product that are propagated in the biophysical variables retrieval. 
Among the elements that can affect the atmospheric correction accuracy of the Sen2Cor algorithm involve the atmospheric effects (i.e., Rayleigh and aerosol scattering effects) on the TOA data across the spectrum, particularly in the visible bands \cite{Martins2017}. 
In this respect, \cite{Sola2018,Doxani2018} compared the role of atmospheric correction, alternative approaches of the atmospheric correction (6S, ACOLITE and Sen2Cor). The authors reported diverging performances with respect to location, spectral band and land cover. These sensitivities suggest that atmospheric correction is not that straightforward, and can easily influence the resulting BOA reflectance. It has been argued that more research is required to quantify the impact of different atmospheric correction algorithms on biophysical parameters retrieval \cite{Djamai2018}.

The here proposed alternative approach, i.e. developing retrieval models directly at the TOA scale, seems to simplify the problem. Yet likewise challenges appear, although probably less than at BOA scale. First of all, a bright, cloud-free atmosphere is assumed, as was the case in the two demonstration sites. It must be remarked that the TOA models have not yet been tested in case of hazy conditions, so a bright sky is a requirement. 
Notwithstanding, the observed moderate discrepancies between BOA and TOA against the ground validation require further explanation; they may be attributed to the use of different RTMs for the atmospheric simulation and for the TOA data preprocessing. While  6S is used to generate the training data set, the S2 TOA reflectance-to-radiance conversion in SNAP is done with routines based on libRadtran \cite{Verrelst2019TOC2TOA} using the solar irradiance spectrum from Thuillier \cite{thuillier2003}. 
Further work should ensure consistency among the atmospheric models used to simulate the training dataset as well as the preprocessing of real TOA data. This can be easily done in the ALG-TOC2TOA packages. 
For instance, new TOA LUTs can be generated with different atmospheric RTMs for comparison purposes. PROSAIL can be kept as the vegetation model that has proven to be suitable for generating the training dataset, while distinct atmospheric models can be applied to perform the atmospheric correction of the TOA product, e.g. 6S, libRadtran, MODTRAN. 
Another research task would be to enable TOC2TOA to generate TOA reflectance spectra instead of TOA radiance spectra so that retrieval can use directly the provided S2 TOA reflectance product.

Once the VHGPR model is trained, it can be applied to any S2 image for LAI mapping, such as here demonstrated over two agricultural sites. The consistency among the TOA and BOA scale and the low deviation shown in the uncertainty maps confirm the good performance of the VHGPR models. The uncertainty maps are a valuable addition as opposed to other retrieval methods: it immediately provides information about the performance and thus portability of the VHGPR algorithm \cite{Verrelst2013b}. It implies that the model can be applied to any S2 image, as long as the uncertainties stay within a certain threshold. On the other hand, the uncertainty maps could also be used as a spatial mask to show only the pixels that meet a minimum level of confidence \cite{Verrelst2015b}. 
For example, the pixels below 20\% of relative certainties can be discarded, as such meeting the recommendation of the GCOS for the minimum LAI quality \cite{gcos2011}. 
Moreover, the models were performing more consistent than the SNAP NN algorithm. As opposed to the NN-derived maps, more close-to-zero values were achieved for non-vegetated surfaces, while overestimations over dense vegetation did not occur. In the SNAP NN algorithm, estimations above an LAI of 8 are flagged by additional quality layers as out-of-range, which can can thus also be masked out. When comparing both approaches, the VHGPR-produced uncertainty estimates seem more valuable than threshold-based quality flags as an independent and quantitative source of information about the model performance \cite{Verrelst2013b,Verrelst2015b}. 

It must also be remarked that, although the LAI models perform  well over agricultural areas, it remains necessary to test and validate them over other vegetation types typically present in a S2 image, such as forest and more heterogeneous, natural areas. The estimations will most likely be poorer, as the models are not optimized for these vegetation types. Complementary training data is therefore required, e.g. coming from RTMs that are better equipped to simulate more heterogeneous vegetation types such as forest or shrubland. 
At the same time, the maps show higher levels of uncertainty over non-vegetation surfaces (such as bare soil, man-made surfaces) than over vegetation zones. The reference soil spectra extracted directly from the images would be limiting when moving beyond local conditions. 
In order to be more generically applicable, instead of extracting the soil spectra directly from the images, an idea is to implement reference soil data set that represent a large variation of soil types, moisture, roughness and geometrical configuration \cite{Jacquemoud1992,LiuWeidong2002}. 
From this database a small subset of field measured spectra could be implemented that does not significantly increase the size of the LUT but that represents well the variability of the soil making the retrieval model more global \cite{weiss2016}. 
The diversity of soil properties can be further increased by applying the concept of brightness. The brightness coefficient can multiply the reflectance values of the soil in order to reach a better soil representation \cite{Verger2011}. In case the soil spectra is too smooth, the implemented noise model should also be applied to these soil spectra. Additionally, the models are currently not trained for water bodies and man-made surfaces. It would be either a matter of adding such spectra (e.g., USGS database \cite{usgs} or ECOSTRESS spectral library \cite{MEERDINK2019111196}) to the training dataset or otherwise masking out those non-vegetated surfaces.

Finally, in view of moving beyond the here presented results and with ambition to obtain consistent LAI retrievals across the globe,  optimally trained and robust models are required. To achieve this, a fine tuning of PROSAIL and 6S was conducted to ensure that the LUTs represent the most variable conditions possible for a cloudless agricultural scene. For the vegetation variables, Gaussian distributions applied to LAI and LCC were appropriate to represent the vegetation conditions in spring and summer. This allowed to concentrate the maximum information on the most characteristic LAI and LCC values for those dates. However, the lower representativeness at the extremes of the ranges may contribute to the underestimation and overestimation observed in the extreme LAI values. This configuration focused for a period of the year might limit the global application of the model. In order to mitigate these limitations, a common approach is to distribute better the variable along the range, even if this generates an increment in the size of the LUT that complicates the (VH)GPR training process and affects the efficiency of the retrieval \cite{Verrelst2016AL}. 
Regarding the development of the LUTs for training, several modifications can be introduced that may lead to improved or more robust models.  For instance, various PROSAIL variables were kept fixed, such as the hot-spot and the sun-target sensor geometry in order to facilitate the coupling between the vegetation and atmospheric RTMs. 
To deal with the growth of the dataset, reduction strategies such as active learning \cite{Pasolli2012} can be used, which has been demonstrated to be superior to random sampling, improving retrieval accuracy with lower sampling rates \cite{Verrelst2016AL,upreti2019}. 
Another option is to use sample distributions that reflect reality more, e.g., normal or log-normal distributions for key variables \cite{Verrelst2019TOC2TOA}.
Global processing applications of the methods presented here should take into account the variability in the illumination conditions along the orbit and acquisition time. Rather than being an independent variable, the illumination conditions should be taken as an additional source of information for the retrieval, which could be included as an additional band for training the statistical model. Several of these LUT optimization aspects have been studied and implemented in the SNAP NN algorithm, which led to a large and optimized LUT (41472 simulations with 2/3 used for training) \cite{weiss2016}. However, it must be remarked that the validation of the NN model against the same Marchfeld field data yielded only a slightly better performance at the BOA scale ($R^2$ of 0.83) \cite{Vuolo2016} than reported in this study  ($R^2$ of 0.80). The here presented (VH)GPR algorithms made use of about 1000 samples, which is considerably less than what  is used in the SNAP NN algorithm. Hence, LUT size is not a key factor for the (VH)GPR algorithm; in fact large training datasets make the model unnecessary heavy and slow. What matters is the quality of the training data, which eventually boils down to seek for an optimized threshold between realism, diversity and keeping it manageable.  

As a closing remark, although in this work the focus was on LAI retrieval from S2, it must be emphasized that essentially any hybrid retrieval algorithm can be developed for any kind of TOA radiance data with the developed ALG-ARTMO software framework. Future work will be dedicated to the retrieval of other vegetation and atmosphere variables and from other sensors, such as Sentinel-3. The software framework can be freely downloaded at \url{http://artmotoolbox.com/}. Code snippets and demos for both GPR, VHGPR and other machine learning regression algorithms is available from \url{https://isp.uv.es/soft_regression.html}.

\section{Conclusions}\label{sec:conclude}

This study aimed to develop LAI retrieval algorithms directly from Sentinel-2 (S2) top-of-atmosphere radiance (TOA) radiance data  (L1C product). 
To do so, a hybrid machine learning regression approach was developed by making use of simulations from leaf-canopy-atmosphere radiative transfer models (RTMs). The coupled  PROSAIL-6S RTMs were used to simulate a look-up table (LUT) of TOA radiance data and associated input variables. This LUT was then used to train the Bayesian algorithms Gaussian processes regression (GPR) and variational heteroscedastic GPR (VHGPR).  Similarly, PROSAIL simulations were used to train GPR and VHGPR models for LAI retrieval from S2 images at BOA level (L2A product) for comparison purposes. The LAI products were adequately validated at TOA and BOA scale against a field dataset acquired at the agricultural site Marchfeld (Austria). VHGPR was further used for LAI mapping because of delivering superior accuracies and lower uncertainties. Obtained LAI maps over the agricultural sites Marchfeld and Barrax (Spain) from a S2 BOA and TOA subset were alike. A similar degree of consistency was obtained when comparing the obtained LAI maps against the SNAP LAI products at BOA and TOA scale. Associated uncertainty maps supported the spatial consistency. 

Altogether, this study demonstrated that hybrid LAI retrieval algorithms can be developed from TOA radiance data given a cloud-free sky -- thus without the need of atmospheric correction. It is expected that for a diversity of vegetation properties hybrid retrieval algorithms will be developed operated directly from TOA radiance data. The development of such hybrid retrieval algorithms can be easily achieved within the freely downloadable ALG-ARTMO software framework.


\end{document}